 \documentclass[final,3p,times]{elsarticle}

\usepackage{graphicx} 
\usepackage{tabularx}
\usepackage{amsmath}
\usepackage{amssymb}
\usepackage{subfigure}
\usepackage{subfigure}
\usepackage{comment}
\usepackage{color}

\usepackage{amssymb,amsmath,bm}

\definecolor{lightgray}{gray}{0.75}

\usepackage{fancyhdr}

\usepackage{lipsum}
\usepackage[dvipsnames]{xcolor}

\usepackage{hyperref}

\usepackage{cleveref}

\newcommand\myshade{85}
\colorlet{mylinkcolor}{violet}
\colorlet{mycitecolor}{YellowOrange}
\colorlet{myurlcolor}{Aquamarine}

\hypersetup{
  colorlinks=true,
  linkcolor  = mylinkcolor!\myshade!black,
  citecolor  = mycitecolor!\myshade!black,
  urlcolor   = myurlcolor!\myshade!black
}

\journal{CMA report}

\begin{document}

\begin{frontmatter}

\title{ \LARGE Review on development of a scalable high-order nonhydrostatic multi-moment constrained finite volume dynamical core}

\author[cnwp]{Xingliang Li} 
\author[xjtu]{Chungang Chen}
\author[cnwp]{Xueshun Shen \corref{cor}}
\author[titech]{Feng Xiao}

\address[cnwp]{Center of Numerical Weather Prediction of NMC, China Meteorological Administration,  46 Zhongguancun South St., Beijing 100081, China }
\address[xjtu]{State Key Laboratory for Strength and Vibration of Mechanical Structures \& School of Human Settlement and Civil Engineering, Xi'an Jiaotong University, 28 Xianning West Road, Xifan, Shaanxi, 710049, China}
\address[titech]{ Department of Mechanical Engineering, Tokyo Institute of Technology, Tokyo 226-8502, Japan }

\cortext[cor]{Corresponding Address: Center of Numerical Weather Prediction, China Meteorological Administration,  46 Zhongguancun South St., Beijing 100081, China. Email address: shenxs@cma.gov.cn.}

\begin{abstract}
This report summarizes the major progresses recently archived in the Center of Numerical Weather Prediction at China Meteorological Administration to develop the dynamic core for  next-generation atmospherical model for both numerical weather prediction and climate simulation. The numerical framework is based on a general formulation, so-called multi-moment constrained finite volume (MCV) method, which is  well-balanced among solution quality (accuracy and robustness), algorithmic simplicity,  computational efficiency and flexibility for model configuration. Rigorous numerical conservation is guaranteed by a constraint of finite volume formulation in flux form. A local high-order limiting projection is also devised to remove spurious oscillations and noises in numerical solutions, which allows the numerical model working well alone without artificial diffusion or filter. The resulted numerical schemes are very simple, efficient and easy to implement for both structured and unstructured grids, which provide a promising plateform of great practical significance. We have implemented the MCV method to shallow water equations on various spherical grids, including Yin-Yang overset grid, cubed sphere grid and geodesic icosahedral grid, non-hydrostatic compressible atmosherical model under complex topographic boundary conditions. In addition,  the moist dynamics simulation like moist thermal bubble has been validated by using direct microphysical feedback. We have also constructed a prototype of 3D global non-hydrostatic compressible atmosherical model on cubed sphere grid with an explicit/implicit-hybrid time integration scheme,  which can be used as the base to develop global atmospheric GCM. All the MCV models have been verified with widely used benchmark tests. The numerical results show that the present MCV models have solution quality competitive to other exiting high order models. Parallelization of the MCV shallow water model on cubed sphere grid reveals its suitability for large scale parallel processing with desirable scalability.  
\end{abstract}

\end{frontmatter}

\clearpage
\tableofcontents
\clearpage

\section{Introduction}

\subsection {The GRAPES system in CMA}
The GRAPES \index{grapes}(Global and Regional Assimilation and PrEdiction System) global model has been used as an operation system for numerical weather prediction (NWP) since June 1, 2016,  which was seen as a milestone in the development of numerical models for NWP and climate simulations in China. The GRAPES project started in 2001, and continuously supported by CMA. As one of the major outputs of the project, the dynamic core of the GRAPES system has been built with some advanced numerics commonly used for atmospheric modeling in the past decades.  Based on the LAT-LON grid, the Semi-implicit Semi-Lagrangian (SISL) scheme  was adopted in the dynamical core \citep{cxy08} with the central difference of the pressure gradient on the staggering C-grid and the vertical Charney-Phillips variable distribution where the vertical velocity $w$ and the potential temperature are staggered with the horizontal velocity and mass ($\Pi$). The height-based terrain following coordinate is used to represent the topographic effects as lower boundary conditions. The Helmholtz equation is solved by the GCR (Generalized Conjugate Residual) algorithm to obtain the dimensionless pressure perturbation $\Pi'$ at the new time level,  which is then used to advance the prognostic variables $(u,v,\hat{w},\theta')$ where the base hydrostatic state of thermodynamic variables has been removed.

As can be expected from the numerical framework of its dynamic core, some intrinsic drawbacks exist in the GRAPES system. For example, with the grid resolution increased, the ratio of grid spacings between the equatorial  and polar regions becomes very significant, which might not only cause heavily uneven  errors in numerical solutions but also be a big barrier to implementing large scale simulations on modern supercomputers unless extra efforts would be made. Another important issue is that in the present GRAPES system numerical conservation is not automatically guaranteed. As the continuity equation is reformulated in terms of the dimensionless pressure $\Pi$, it provides an ad hoc remedy to fix the global mass if needed. Currently the GRAPES system, with well equipped physical packages and data-assimilation modules, performs reasonably well for operational use. Further development  demands for some new and innovative reformulations in its numerical framework as well. 

\subsection{Recent trends in the field}

With an increased amount of supercomputing resources available to present-day modelers, it is observed as a noticeable trend that finer grid resolutions are available for global models due to the rapid progress in computer hardware, and mesoscale dynamics can be solved on a global base, which requires the global models to be built in the nonhydrostatic framework as well \citep{putman07,sa08,wa08,gassmann11,uj12jcp}. Another requirement for an atmospheric model to be able to efficiently use next-generation large-scale supercomputers, which usually consist of tens to hundreds of thousands of processors with distributed-memory nodes, is that the dynamical core should be highly scalable for large-scale parallel processing. Some recent studies demonstrate that the dynamical cores based on high-order schemes with local spectral reconstructions are superior in parallel scalability and overcome the barrier that prevents the spherical-harmonic spectral transform-based method and finite-volume method from efficient implementations on supercomputers toward exa-flops computation \citep{dennis05,dennis12}. 

It necessitates the development of the numerical discretization schemes that are capable of facilitating excellent parallel efficiency on supercomputers. Researches on these numerics have been so far carried out in such a direction by using the spectral element (SE) method \citep{thomas05,ihl02,GR04,ftt04,taylor10} or the discontinuous Galerkin (DG) method \citep{lnt07,giraldo08,rg09,nair09,blaise12} , mixed finite element method\citep{cs12jcp,staniforth13,ct14jcp,msc14,cm14arxiv} and multi-moment method (see below).

More closely related to real-case applications, some next generation dynamic cores like UK's GungHo, NGGPS of USA ( Note that the incremental performance improvement of MPAS with additional computer processors -its scalability- was better than FV3 although FV3 outperformed MPAS on tests of more than 100,000 processors ) and KIAPS of Korea projects come to reality. They can address these challenges and have computationally desirable local properties such as compact computational stencils, high on-processor operations, and minimal communication footprints.

Another type of high-order schemes among this kind of numerics can be devised by using the multi-moment concept \index{multi-moment} \citep{yabe91,yabe01,xiao04,xiao06,ii05,xiao07,ii07,chen08}, where we make use of different kinds of discrete quantities, so-called \textit{moments} in our context, to describe the physical field, such as the pointwise value, derivatives and volume integrated value. These moments are locally defined over each mesh cell, which allows us to build high order local reconstructions. Different moments can be then updated by different formulations that may have different forms but should be consistent with the original conservation law. For example, the point value can be updated by a point-wise Riemann solver or semi-implicit semi-Lagrangian (SISL) solver for hyperbolic type equations, while the volume integrated average value is computed by a finite volume formulation to ensure the rigorous numerical conservation. In our previous studies, two global multi-moment finite volume shallow water models have been reported using either a point-wise Riemann solver \citep{chen08} or SISL method \citep{li08}. In these schemes, different  moments, i.e. cell-integrated average and point value, are directly treated as the prognostic variables.

A more efficient alternative to the multi-moment finite volume formulation is to define the unknowns (or the degree of freedoms, DOFs) as the values at the points located within each grid cell and to use the time evolution equations of different moments as the constraint conditions to derive the governing equations for updating the unknown point values.  The resulting scheme is so-called multi-moment constrained finite volume (MCV) \index{MCV} method \citep{ii09} where the moments are not directly used as the predicted unknowns (DOFs), but the constraint conditions. The numerical conservation is exactly guaranteed through a constraint on the cell-integrated average which is cast in a finite volume formulation of flux form. In the present multi-moment constrained method, all predicted unknowns are the nodal  values at the solution points and the volume integration is not involved in the computation, which makes the numerical formulation very efficient, especially when the physical source term and metric term are included in the governing equations.
Our experiences also show that an MCV scheme can use larger CFL number for computational stability compared to other high-order schemes of the same order, and the location of the solution points can be determined in a more flexible manner, which is of particular importance in atmospheric modeling.   We have  implemented the MCV method to develop global shallow water models on icosahedral grid \citep{ii10}, hexagonal geodesic grid \citep{chen12} and Yin-Yang grid \citep{li12,li15}.  We recently have also presented the global shallow water models based on the third-order MCV scheme and the three quasi-uniform spherical grids such as Yin-Yang overset grid, cubed-sphere grid and geodesic icosahedral grid in \cite{chen14} and  developed the MCV nonhydrostatic atmospheric dynamics \citep{li13}, which demonstrated its potential as the fundamental numerics for new type dynamical cores.

In the rest part of this report, we will describe in section 2 the basic idea and formulations of the MCV scheme with as a limiting projections which effectively eliminate spurious oscillation while retain the high-order accuracy of the numerical solutions.   In section3, implementations of multi-moment finite volume method and MCV to shallow water equations in spherical geometry will be presented in different spherical grids with numerical results of the representative benchmarks. The  nonhydrostatic atmospheric dynamic model will be introduced in section 4. We present in section 5 the prototype of the three dimensional global nonhydrostatic dynamic core using the cubed sphere grid. Parallelization and scalability performance  are presented in section 6. 
\section{Multi-moment constrained finite volume formulations}

\label{sec:multiform}

\subsection{The one dimensional formulation}\label{sec:multiform-1d}
We firstly present the basic idea and the solution procedure of the third order MCV (MCV3) scheme to solve the following hyperbolic conservation law  
\begin{equation}
{{\partial q} \over {\partial t}}+{{\partial f} \over {\partial x}} =0,  
\label{1dge}
\end{equation}
where  $q$  is the state variable, and  $f(q)$ the flux function. 

The computational domain is divided into $I$ non-overlapping cells or elements $\Omega_i=[x_{i-{1\over2}},x_{i+{1\over2}}]$, $i=1,2,\cdots,I$, and three solution points $x_1$, $x_2$ and  $x_3$ are set over  $\Omega_i$ where the solution $q_{ik}$, $k=1,2,3$, is defined and updated at every time step.  The locations of the solution points can be flexibly determined in an MCV scheme, and we used $x_{i1}=x_{i-{1\over2}}$,  $x_{i2}=(x_{i-{1\over2}}+x_{i+{1\over2}})/2$ and $x_{i3}=x_{i+{1\over2}}$ for simplicity. 

Given ${q}_{ik} (k=1,2,3)$, a cell-wise Lagrange interpolation can be constructed as,  
\begin{equation} 
{\mathcal Q}_i(x)=\sum^3_{k=1}{q}_{ik}\phi_{ik}(x).  
\label{lagintp_q3}
\end{equation}
where 
\begin{equation}
\phi_{ik}=\prod_{\alpha=1,\alpha \neq k}^3 \frac{x-x_{i\alpha}}{x_{ik}-x_{i\alpha}}
\end{equation} 
is the basis function of the Lagrange interpolation. 

The equations to update ${q}_{ik} (k=1,2,3)$ are derived from the time evolution equations of two types of moments, which are defined as follows,
\begin{itemize}
\item The volume-integrated average (VIA moment), 
\begin{align}\label{via-def}
   \overline{{q}}_i(t)\equiv \frac{1}{\Delta x_i} \int_{x_{i-{1\over2}}}^{x_{i+{1\over2}}}{q}(x,t)dx, 
\end{align} 
where $\Delta x_i=x_{i+{1\over2}}-x_{i-{1\over2}}$.

\item The point values (PV moment) at the two ends, 
\begin{align}\label{pv-def}
   {q}_{i1}(t)\equiv{q}(x_{i-{1\over2}},t) \quad {\rm and} \quad  {q}_{i3}(t)\equiv{q}(x_{i+{1\over2}},t).
\end{align} 
\end{itemize}

The time evolution equation of VIA moment is obtained by integrating the conservation law  \eqref{1dge}, yielding a finite volume formulation, 
\begin{equation}
\frac{{d  \overline{{q}}_i}} {d t}=  - \frac{1}{\Delta x_i} \left ( {\hat f}_{i+{1\over2}}-{\hat f}_{i-{1\over2}}\right ),
\label{conservation-mcv3-1d}
\end{equation}
where ${\hat f}_{i-{1\over2}}  $ and  ${\hat f}_{i+{1\over2}}  $ are the numerical fluxes consistent to $ f(x,t)$ at the two ends of cell $i$. The present MCV scheme has a $C^0$ continuity, and the point values at the cell boundaries are shared by the two neighboring cells, i.e. $q_{i3}=q_{(i+1)1}=q_{i+{1\over2}}$. It allows us to directly obtain the numerical flux by ${\hat f}_{i+\frac{1}{2}}  =f(q_{i+{1\over2}}(t)  )$.

The PV-moment values at the two ends of cell $i$ are predicted by solving the conservation law \eqref{1dge} point-wisely, 
\begin{equation} \label{pv_1d}
\frac{d q_{i1}  }{dt} =-\hat{f}_{xi-{1\over2}}    \ {\rm and} \    \frac{d q_{i3}  }{dt} =-\hat{f}_{xi+{1\over2}}  ,
\end{equation}
where $\hat{f}_{xi+{1\over2}}  $  is a numerical approximation to the derivative of the flux function, $\partial {f}/\partial x$, at cell boundary $x_{i+{1\over2}}$, which needs to be computed from the approximate Riemann solver as discussed later. 

From \eqref{lagintp_q3}, we know immediately
\begin{equation}\label{q2}
 q_{i2}= \frac{3}{2}\overline{q}_i- \frac{1}{4}\left( q_{i1}+q_{i3} \right).  
\end{equation}
With the multi-moment constraint conditions \eqref{conservation-mcv3-1d} and \eqref{pv_1d},  relation \eqref{q2} directly leads to the time evolution equation of $q_{i2}$, 
\begin{equation}\label{eq-q2}
\frac{dq_{i2}}{dt} = -\frac{3}{2\Delta x_i} \left ( {\hat f}_{i+{1\over2}}  -{\hat f}_{i-{1\over2}}  \right ) +\frac{1}{4}  \left( \hat{f}_{x i-\frac{1}{2}}   + \hat{f}_{x i+\frac{1}{2}}   \right). \\
\end{equation}

Eqs. \eqref{pv_1d} and \eqref{eq-q2} make up of the formulation to update the point value solutions, $q_{ik}$, $k=1,2,3$.
We re-write them into the following vector-matrix form,
\begin{equation}\label{eq:conrel3}
\left( \begin{array}{c}
\displaystyle \frac{d}{dt}(q_{i1})\\
\displaystyle \frac{d}{dt}(q_{i2})\\
\displaystyle \frac{d}{dt}(q_{i3})
\end{array}\right)
=\left(\begin{array}{ccccc}
0&0&-1&0\\
\frac{3}{2\Delta x_i}&-\frac{3}{2\Delta x_i}&\frac{1}{4}&\frac{1}{4}  \\
0&0&0&-1
\end{array}\right) \left(\begin{array}{c}
\hat{f}_{i-\frac{1}{2}}  \\
\hat{f}_{i+\frac{1}{2}}  \\
\hat{f}_{x i-\frac{1}{2}}  \\
\hat{f}_{x i+\frac{1}{2}}  
\end{array}\right).
\end{equation}

Denoting the elements of matrix 
\begin{equation}\label{def-matrix-3}
\mathbf{M}=\left(\begin{array}{ccccc}
0&0&-1&0\\
\frac{3}{2\Delta x_i}&-\frac{3}{2\Delta x_i}&\frac{1}{4}&\frac{1}{4}  \\
0&0&0&-1
\end{array}\right) 
\end{equation}
by $\mathcal{M}_{k\alpha}$, $k=1,2,3$ and $\alpha=1,\cdots,4$, and the components of 
\[ \mathbf{F}= \left(\hat{f}_{i-\frac{1}{2}}  , \hat{f}_{i+\frac{1}{2}}  , \hat{f}_{x i-\frac{1}{2}}  , 
\hat{f}_{x i+\frac{1}{2}}   \right)^T \]  by $\mathcal{F}_{\alpha}$, we recast \eqref{eq:conrel3} into a component form,
\begin{equation}\label{semi-d-component-1d}
\displaystyle \frac{d}{dt}(q_{ik})=\sum_{\alpha=1}^4\mathcal{M}_{k \alpha} \mathcal{F}_{\alpha}, \ {\rm for} \ k=1,2,3.
\end{equation}

As a matter of fact, the equation for $k=1$ duplicates that for $k=3$ in \eqref{semi-d-component-1d} since $q_{i3}=q_{(i+1)1}$ in the present MCV scheme. So, we need only solve either $k=1,2$ or $k=2,3$ in \eqref{semi-d-component-1d}. Bear this in mind, we still use  \eqref{semi-d-component-1d} as a more general expression which includes the case where the solution points don't include the cell boundaries, such as the Gauss quadrature points. 

Recall that in the present MCV3 scheme the piecewisely reconstructed interpolations have a $C^0$ continuity at cell boundary, i.e. ${\mathcal Q}_{i} (x_{i+\frac{1}{2}})={\mathcal Q}_{i+1} (x_{i+\frac{1}{2}})=q_{i+\frac{1}{2}}$, the numerical flux is directly computed by 
\begin{equation} \label{f-1d}
{\hat f}_{i+\frac{1}{2}}=f \left( q_{i+\frac{1}{2}} \right).
\end{equation}

What left now is how to calculate the gradient of the flux function at cell boundaries, $\hat{f}_{x i+\frac{1}{2}}$. 
To this end, we firstly find the left-side and right-side values for the gradient of state variable at the cell boundary 
$x_{i+1/2}$ by Lagrange interpolations cell-wisely constructed over the two neighboring cells,  
\begin{equation}\label{qx-1d}
\begin{split}
 &q^L_{xi+\frac{1}{2}}=\frac{d}{d x} {\mathcal Q}_{i} (x_{i+\frac{1}{2}}) = \frac{q_{i1}-4q_{i2}+3q_{i3}}{\Delta x_{i}}, \\
 &q^R_{xi+\frac{1}{2}}=\frac{d}{d x} {\mathcal Q}_{i+1} (x_{i+\frac{1}{2}}) = \frac{-3q_{(i+1)1}+4q_{(i+1)2}-q_{(i+1)3}}{\Delta x_{i+1}}.
\end{split}
\end{equation}

The left-side and right-side values for the first-order derivative of flux function are then computed by 
\begin{equation} \label{fx-1d}
\begin{split}
 &f^L_{xi+\frac{1}{2}}=f_{x } \left( q_{i+\frac{1}{2}},q^L_{x i+\frac{1}{2}} \right), \\
 &f^R_{xi+\frac{1}{2}}=f_{x } \left( q_{i+\frac{1}{2}},q^R_{x i+\frac{1}{2}} \right).
\end{split}
\end{equation}

The numerical approximation to the  first-order derivative of the flux function at cell boundary, $x_{i+1/2}$,  is obtained by solving the (derivative) Riemann
problems,
\begin{equation} \label{flux-1d}
 \hat{f}_{x i+\frac{1}{2}} = {\rm Riemann} \left( f^L_{xi+\frac{1}{2}},f^R_{xi+\frac{1}{2}} \right).
\end{equation}

For computational efficiency, a local Lax-Friedrich (LLF) approximate Riemann solver \citep{so89} is adopted in this paper, which reads
\begin{align} \label {llf-1d}
  \hat{f}_{x i+\frac{1}{2}}=\frac{1}{2}\left( f^L_{xi+\frac{1}{2}}+ f^R_{xi+\frac{1}{2}}\right)
   -|\lambda_{\rm max}| \left(q^R_{xi+\frac{1}{2}}-q^L_{xi+\frac{1}{2}}\right),
\end{align}
where $|\lambda_{\rm max}|$ is the local maximum of the characteristic speed $\lambda=df/dq$, which is evaluated from the point values of neighboring cells, i.e. $|\lambda_{\rm max}|=\max (|\lambda_{ik}|,|\lambda_{(i+1)k}|)$, $k=1,2,3$ . 

We have described  the spatial discretization procedure so far, and get the semi-discretized equations \eqref{semi-d-component-1d} which are ordinary differential equations with respect to time. The third order TVD Runge-Kutta time integration method \citep{sh88} is applied to solve  \eqref{semi-d-component-1d}. Consider a semi-discretized equation,  
\begin{equation}
\frac{d q}{d t}=\mathcal{L}\left(q \right)
\end{equation}
where $\mathcal{L}(q)$ stands for the spatial approximations discussed above. The third order TVD
Runge-Kutta time integration method to update the solution $q^n$ at time level $t^n$ to $q^{n+1}$ at time level $t^{n+1}$  reads 
\begin{align}\label{rk3}
   q^{(1)}&=q^{n}+\Delta t
   \mathcal{L}(q^{n}) \nonumber \\
   q^{(2)}&=\frac{3}{4}q^{n}+\frac{1}{4}q^{(1)}+\frac{1}{4}\Delta t
   \mathcal{L}(q^{(1)}) \\
   q^{n+1}&=\frac{1}{3}q^{n}+\frac{2}{3}q^{(2)}+\frac{2}{3}\Delta t
   \mathcal{L}(q^{(2)}),\nonumber
\end{align}
where $\Delta t=t^{n+1}-t^n$ is the time integration interval. 

We summarize the solution procedure of the third order explicit MCV scheme as follows, 
\begin{enumerate}[{Step}.1]
\item Given solution, $q_{ik}$, at the current time step, make the Lagrange interpolation \eqref{lagintp_q3};
\item Calculate the left-side and right-side values of the first order derivatives (gradients) of state variable and flux function by \eqref{qx-1d} and \eqref{fx-1d} respectively; 
\item Compute the numerical flux from \eqref{f-1d} and the first order derivative of flux function from the Riemann solver \eqref{flux-1d};
\item  Update  solution $q_{ik}$ using Runge-Kutta scheme \eqref{rk3} to integrate the semi-discretized equations \eqref{semi-d-component-1d};
\item  Proceed to next time level.
\end{enumerate}
 
\subsection {Some remarks} \label{remark}
We make some remarks on the important features of the MCV schemes in comparison with other methods. 

\begin{enumerate}[{Remark}.1]
\item Different from other existing methods, the derivation of an MCV scheme is based on a set of prognostic equations for different moments which can be chosen by various principles, not limited to the Galerkin formulation for example in the DG method. 

\item Using VIA as one of the moments leads to a finite volume formulation that guarantees the numerical conservation, which is substantially different from the SE method.  It should also be noted that the
constraint conditions can be chosen in view of not only
numerical accuracy and efficiency but also underlying
physics. 

\item In the present MCV scheme, the solutions at the cell boundaries are shared by the neighboring cells. So, the DOFs of solutions required in the third-order MCV 
scheme is $2I+1$ for a 1D computational domain of
$I$ cells. One derivative Riemann solver is required at each cell
boundary. Given the solution at each cell boundary updated every time
step, the numerical flux needed in the constraint \eqref{conservation-mcv3-1d}  on the
VIA moment can be directly computed. 
 
\item It is shown by Fourier analysis and numerical experiments that replacing the equidistance
solution points in the present scheme by Gauss-Legendre or Gauss-Chebyshev points
doesn't noticeably change the numerical result, whereas the constraint conditions in an MCV scheme matters substantially the property of the numerical schemes, such as accuracy
and stable CFL number \citep{xiao13}.  It reveals that the solution points can be located within
each cell in a more flexible manner. 

\item  We use equally spaced solution points in the present MCV
formulation, which simplifies the formula for updating
the unknowns and is more attractive for real applications.
 All solutions to be updated are nodal values, which
particularly benefits the computations of source terms
and metric terms.

\item The equations to update
the nodal values are derived from a set of constraint
conditions without the involvement of
the inner product in a weak-form formulation like the
Galerkin method. So the mass matrix of the equations to update the unknowns  are always in a diagonal form no matter whether an orthogonal basis function is used or not.

\item A Fourier analysis with the linear advection equation shows  that the spectral radius (the largest eigenvalue of the spatial discretization operator) 
of the third  MCV  scheme is only half of a $\it P_2$ DG scheme, which reveals that the largest allowable CFL number for computational
stability of  the MCV scheme can be twice as large as the DG scheme. This conclusion is further confirmed by numerical experiments with the third-order TVD Runge-Kutta time integration scheme.

\item A more general formulation to construct high order numerical schemes by
using multi-moment constraint conditions on the flux function reconstruction, so called multi-moment constrained flux reconstruction (MMC-FR), is presented in \cite{xiao13} where the MCV schemes of different orders can be derived as the special cases. 
\end{enumerate}

\begin{figure}[htbp]
\begin{center}
{\includegraphics[width=0.4\textwidth]{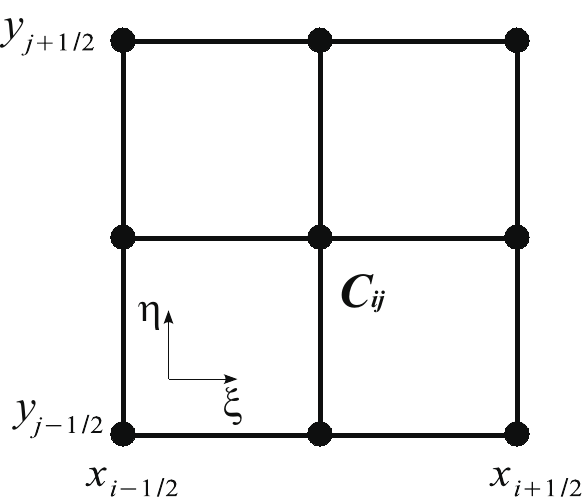}}
\end{center}
\caption{Solution points of the third-order MCV scheme on a 2D quadrilateral grid cell. }\label{structured-2d}
\end{figure}

\subsection{Extension to multi-dimensions}

The MCV formulation can be easily extended to multi-dimensions for both structured and unstructured grids. 

We consider a scalar conservation law in two dimensional domain, $(x,y)\in \mathcal D$, 
\begin{align}\label{eq-scalar-2d}
   \frac{\partial {q}}{\partial t}+\frac{\partial {f}}{\partial x}+\frac{\partial {g}}{\partial y}=0, 
\end{align}
where $q$ is the state variable, and $f$ and $g$ the flux functions in the $x$ and $y$ directions, respectively. 

The quadrilateral element is used in structured grids, such as the Yin-Yang and cubed-sphere grids shown later. In a structured grid, the computational domain of is partitioned into mesh cells $\mathcal{C}_{ij}$ indexed by cell number $i$ and $j$ in $x$ and $y$ directions respectively, such that
\begin{equation}\label{eq:domain}
   {\mathcal D} = \bigcup_{i,j=1}^{I,J} {\mathcal C}_{ij}
\end{equation}
where the mesh cell spans as ${\mathcal C}_{ij} = [x_{i-1/2},x_{i+1/2}]\otimes[y_{j-1/2},y_{j+1/2}]$ with $\Delta x_i = x_{i+1/2}-x_{i-1/2}$ and $\Delta y_j =y_{j+1/2}-y_{j-1/2}$, and $I$ and $J$ the total numbers of mesh cells in $x$ and $y$ directions.

As shown in Fig. \ref{structured-2d}, the solution points for mesh cell $\mathcal{C}_{ij}$ are denoted by $x_{ijkl}$, $(k,l=1,2,3)$, where the point-value solutions $q_{ijkl}$ are defined as the predicted unknowns. Same as in the one dimensional case,  we use the equi-distanced point configuration including the cell boundary points in the two dimensional case. The values at boundary points are shared by the adjacent cells and the physical field is $C^0$ continuous.   For simplicity, we drop the cell indices $i$ and $j$ from now on and focus our discussions on the local control volume
 where the solution points are equally spaced over $[x_{i-1/2},x_{i+1/2}]$ and $[y_{j-1/2},y_{j+1/2}]$ respectively by
\begin{align*}
  &\xi_1=x_{i-\frac{1}{2}}, \quad  \xi_2=(x_{i-\frac{1}{2}}+x_{i+\frac{1}{2}})/2, \quad \xi_3=x_{i+\frac{1}{2}};\\
  &\eta_1=y_{j-\frac{1}{2}}, \quad  \eta_2=(y_{j-\frac{1}{2}}+y_{j+\frac{1}{2}})/2, \quad \eta_3=y_{j+\frac{1}{2}}.
\end{align*}     

The 1D building block described in \ref{sec:multiform-1d} is straightforwardly applied  in $x$ direction along the $l$th line segment $\overline{\xi_1\xi_3}\times \eta_l$, $l=1,2,3$, and   in $y$ direction along the $k$th line segment $\xi_k \times \overline{\eta_1\eta_3}$, $k=1,2,3$.  We obtain the semi-discretized equations for the point-value solutions within cell  ${\mathcal C}_{ij}$
 by
\begin{equation}\label{semi-d-component-2d}
\displaystyle \frac{d}{dt}(q_{kl})=\sum_{\alpha=1}^4\mathcal{M}^{(x)}_{k \alpha} \mathcal{F}_{\alpha}+\sum_{\beta=1}^4\mathcal{M}^{(y)}_{l \beta} \mathcal{G}_{\beta}, \ {\rm for} \ k,l=1,2,3.
\end{equation}

In \eqref{semi-d-component-2d}, $\mathcal{M}^{(x)}_{k\alpha}$ denotes the elements of matrix 
\begin{equation}\label{def-matrix-3x}
\mathbf{M}^{(x)}=\left(\begin{array}{ccccc}
0&0&-1&0\\
\frac{3}{2\Delta x_i}&-\frac{3}{2\Delta x_i}&\frac{1}{4}&\frac{1}{4}  \\
0&0&0&-1
\end{array}\right),  
\end{equation}
 and $\mathcal{F}_{\alpha}$ the components of 
\[ \mathbf{F}= \left(\hat{f}_{i-\frac{1}{2}l}, \hat{f}_{i+\frac{1}{2}l}, \hat{f}_{x i-\frac{1}{2}l}, 
\hat{f}_{x i+\frac{1}{2}l} \right)^T. \]  
A similar formulation is obtained in $y$ direction with 
$\mathcal{M}^{(y)}_{l\beta}$ being the elements of matrix 
\begin{equation}\label{def-matrix-3y}
\mathbf{M}^{(y)}=\left(\begin{array}{ccccc}
0&0&-1&0\\
\frac{3}{2\Delta y_j}&-\frac{3}{2\Delta y_j}&\frac{1}{4}&\frac{1}{4}  \\
0&0&0&-1
\end{array}\right),  
\end{equation}
 and $\mathcal{G}_{\beta}$ the components of 
\[ \mathbf{G}= \left(\hat{g}_{kj-\frac{1}{2}}, \hat{g}_{kj+\frac{1}{2}}, \hat{g}_{y kj-\frac{1}{2}}, 
\hat{g}_{y kj+\frac{1}{2}} \right)^T. \]

The numerical flux and its first-order derivative in $x$ direction,  $\hat{f}_{i+\frac{1}{2}l}$ and $\hat{f}_{x i+\frac{1}{2}l}$, are evaluated at the solution points along cell boundary $x_{i+\frac{1}{2}}$, while those in  $y$ direction, $\hat{g}_{kj+\frac{1}{2}}$ and $\hat{g}_{y kj+\frac{1}{2}}$, are evaluated at solution points along cell boundary $y_{j+\frac{1}{2}}$.  The 1D Riemann solver is used in each direction, and the 1D Lagrange interpolation  \eqref{lagintp_q3} is applied to find the left and right side values of the first-order derivatives in the normal directions respectively. 

For unstructured grids,  multi-dimensional formulations of MCV method can be implemented in a straightforward manner as well. We hereby present  the numerical formulations for  hexagonal and pentagonal  mesh elements which are used in the  icosahedral geodesic grid for the spherical geometry. We denote the grid cell indexed by $i$ as 
 ${\mathcal C}_{i}$, such that the computational domain ${\mathcal D}$ is divided by 
\begin{equation}\label{eq:domain-hp}
   {\mathcal D} = \bigcup_{i=1}^{I} {\mathcal C}_{i},
\end{equation}
where $i$ is the index of mesh cells and $I$ the total cell number.

\begin{figure}[htbp]
\begin{center}
\includegraphics[height=0.49\textwidth,width=0.5\textwidth]{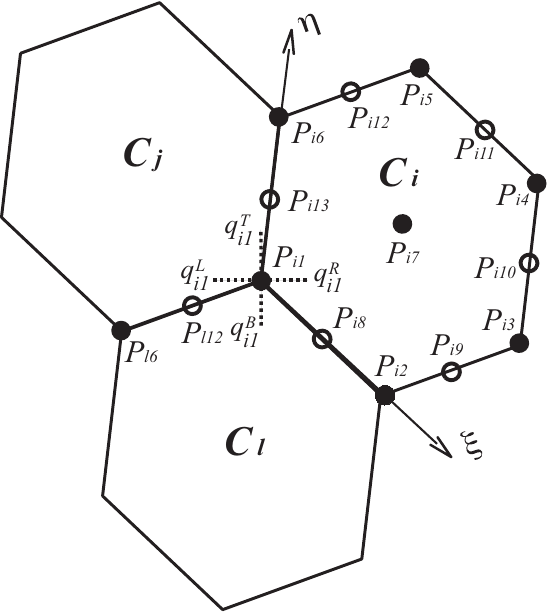}
\end{center}
\caption{Mesh configuration of the third-order MCV on a hexagonal element. The solution points are the 6 vertices and the center point denoted by solid circles where the unknown DOFs are updated. The values at the middle point of each boundary segment denoted by open circles are not predicted as the unknown DOFs but obtained from the cell-wise interpolation. }\label{DRP-1}
\end{figure}

Shown in Fig.\ref{DRP-1}, seven local DOFs (denoted by solid
circles over cell $i$), i.e. $q_{ik}\ \left(k=1,7\right)$ are defined as the point-value solutions for a
hexagonal element ${\mathcal C}_i$, which are located at the six vertices
from $P_{i1}$ to $P_{i6}$ (counter clockwise) and the cell center
$P_{i7}$.    

Single-cell based quadratic polynomial can be constructed based on
seven local DOFs within element ${\mathcal C}_i$ as
\begin{equation}\label{lagintp-h}
\mathcal{Q}_{i}(x,y)=c^{00}_{i}+c^{10}_{i}x+c^{01}_{i}y+c^{11}_{i}xy+c^{20}_{i}x^2+c^{02}_{i}y^2+c^{12}_{i}\left(x^2y+xy^2\right).
\end{equation}

The coefficients are uniquely determined by solving equation set
\[\mathcal{Q}_i(x_{ik},y_{ik})=q_{ik}, \ {\rm for} \ k=1,\cdots,7; \] 
where 
$\left(x_{ik},y_{ik}\right)$ is the location of point $P_{ik}$. 

Two kinds of moments are adopted as the constraint conditions, i.e. 
\begin{itemize}
 \item The PV moment
\begin{equation}
{q}_{ik}\left(t\right)=q\left(x_{ik},y_{ik},t\right),\
\left(k=1,6\right);
\end{equation}
\item The VIA moment
\begin{equation}
\overline{q}_i\left(t\right)=\frac{1}{A_i}\int_{{\mathcal C}_i} q(x,y,t)ds
\end{equation}
where $A_i$ is the area of hexagonal element ${\mathcal C}_i$.
\end{itemize}

From \eqref{lagintp-h}, the following relation holds between local
DOFs and constraint conditions
\begin{equation}
\left\{
\begin{array}{llll}
q_{ik}&=&q(x_{ik},y_{ik})&\left(k=1,6\right)\\
q_{i7}&=&a_7\overline{q}_i-\displaystyle \sum_{k=1}^6\left(a_k{q}_{ik}\right)
\end{array}
\right.,
\label{consre2d}
\end{equation}
where coefficients $a_1$ to $a_7$ are derived by integrating
\eqref{lagintp-h} over the hexagonal element ${\mathcal C}_i$. For example, a regular hexagon results in 
\begin{equation}
a_1= a_2=a_3=\frac{1}{40}; \ a_4= a_5= a_6=\frac{1}{15}; \ a_7=\frac{9}{40}.
\end{equation}

As shown later, the point values at the midpoints of boundary edges, from $P_{i8}$ to $ P_{i13}$, are also needed in computing 
the local reconstruction and the numerical fluxes. Since they are not the predicted variables at each time marching step, we evaluate them from the 
cell-wise interpolation \eqref{lagintp-h}. 

We update the point-value solutions,  $q_{ik}$, at the vertices using the  
 differential form of \eqref{eq-scalar-2d} as 
\begin{equation} \label{pv-update}
\frac{d q_{ik}}{d t}=-{\hat f}_{x ik}-{\hat g}_{y ik}, \  {\rm for} \  k=1,\cdots,6;
\end{equation}
where  ${\hat f}_{x ik}$ and ${\hat g}_{y ik}$ are the numerical approximations to the first-order derivatives of the flux functions ${\partial f}/{\partial x}$ and  ${\partial g}/{\partial y}$ respectively at vertex point $P_{ik}$, which is computed from a point-wise derivative Riemann solver.

Using Fig.\ref{DRP-1}, we discuss the procedure to compute the numerical fluxes at vertex point $P_{i1}$. 
The LLF Riemann solver in terms of the flux derivative in $x$ direction is written as
\begin{align} \label {llf-hx}
  \hat{f}_{x i1}=\frac{1}{2}\left( f^R_{xi1}+ f^L_{xi1}\right)
   -|\lambda^{(x)}_{\rm max}| \left(q^R_{xi1}-q^L_{xi1}\right),
\end{align}
and that in $y$ direction is
\begin{align} \label {llf-hy}
  \hat{g}_{y i1}=\frac{1}{2}\left( g^T_{yi1}+ g^B_{yi1}\right)
   -|\lambda^{(y)}_{\rm max}| \left(q^T_{yi1}-q^B_{yi1}\right),
\end{align}
where $\lambda^{(x)}_{\rm max}$ and  $\lambda^{(y)}_{\rm max}$ are the maximum values of $\partial f/\partial q$ and $\partial g/\partial q$ respectively in the surrounding cells. The left-, right-, top- and bottom-side values at point $P_{i1}$, $q^L_{xi1}$, $q^R_{xi1}$, $q^T_{yi1}$and $q^B_{yi1}$, are computed from the reconstructions over the neighboring cells.   
 Shown in Fig.\ref{DRP-1} for example, $q^R_{xi1}$ needs to be calculated within cell $\mathcal{C}_i$. We show the numerical steps next to find  $q^R_{xi1}$.

As aforementioned, the point values at the vertices and the cell center are the predicted variables, but the values of dependent variables ${q}$ at the center of the two connected edges, e.g. ${q}_{i8}$ and ${q}_{i13}$, must be found from the interpolations. They can be evaluated separately from two adjacent cells sharing the edge as
\begin{equation}
{q}_{i8}^i=\mathcal{Q}_i\left(x_{i8},y_{i8}\right),\ {q}_{i8}^l=\mathcal{Q}_l\left(x_{i8},y_{i8}\right),
\end{equation} 
and
\begin{equation}
{q}_{i13}^i=\mathcal{Q}_i\left(x_{i13},y_{i13}\right),\ {q}_{i13}^j=\mathcal{Q}_j\left(x_{i13},y_{i13}\right).
\end{equation} 

Generally, ${q}_{i8}^i\neq {q}_{i8}^l$ and $ {q}_{i13}^i\neq {q}_{i13}^j$. In the present model, the upstream value is adopted to evaluate the numerical flux across each edge using Simpson's rule. 

Given the three point values along each edge, the derivatives of the state variable $q$ with respect to $x$ and  $y$ can be obtained by using an isoparametric mapping  $(x,y) \rightarrow (\xi,\eta)$.

A local coordinate system $\left(\xi,\eta\right)$ is constructed along two edges, $\overline{P_{i1}P_{i2}} \rightarrow \xi$ and $\overline{P_{i1}P_{i6}} \rightarrow \eta$, which intersect at point $P_{i1}$. The local coordinate of points $P_{i1}$, $P_{i8}$, $P_{i13}$, $P_{i2}$ and $P_{i6}$ are $\left(0,0\right)$, $\left(0.5,0\right)$ $\left(0,0.5\right)$ $\left(1,0\right)$ and $\left(0,1\right)$, and points $P_{i8}$ and $P_{i13}$ are the centers of the arcs  $\overline{P_{i1}P_{i2}}$ and $\overline{P_{i1}P_{i6}}$. Given the point-wise values of $q$ at these five points, we can easily calculate the derivatives of $q$ with respect to $\xi$ and $\eta$ at point $P_{i1}$ by
\begin{equation}{
\left\{\begin{array}{l}
\left(\partial_{\xi}q\right)_{i1}= -3q_{i1}+4q_{i8}-q_{i2}\\
\left(\partial_{\eta}q\right)_{i1}= -3q_{i1}+4q_{i13}-q_{i6}
\end{array}\right..}\label{dlocal}
\end{equation}

Consequently, $q^R_{xi1}$ is computed by 
\begin{equation}
q^R_{xi1}=\left(\partial_{x}q\right)_{i1}=\frac{1}{|J_{i1}|}\left(\partial_{\xi}y \left(\partial_{\eta}q\right)_{i1}- \partial_{\eta}y \left(\partial_{\xi}q\right)_{i1} \right), 
\end{equation}
where 
\begin{equation}
J_{i1}=\left(\begin{array}{cc}\partial_{\xi}x&\partial_{\xi}y\\\partial_{\eta}x&\partial_{\eta}y\end{array}\right)_{i1}
\end{equation}
is the metric Jacobian at  point $P_{i1}$ which is determined by an isoparametric transformation using \eqref{dlocal}. All metric terms can be computed and stored at the grid generation stage for later use. 

Other values,  $q^L_{xi1}$, $q^T_{yi1}$and $q^B_{yi1}$, can be obtained via the same procedure over corresponding cells.  
The first-order derivatives of the numerical fluxes,  $f^L_{xi1}$, $f^R_{xi1}$, $g^T_{yi1}$ and $g^B_{yi1}$, are then immediately computed. 
With the numerical fluxes in \eqref{llf-hx} and \eqref{llf-hy} known, the prognostic solutions at the six vertices can be updated by  \eqref{pv-update}. 

The evolution equation for the point value at cell center $P_{i7}$ is derived from the constraint on the VIA moment.  
Integrating  \eqref{eq-scalar-2d} over control
volume $\mathcal{C}_i$, a finite volume formulation is obtained to
update VIA moment as
\begin{equation}
\displaystyle
\frac{ d \overline{q_i}}{d t}=-\frac{1}{A_i}\sum_{k=1}^6\int_{l_k}\mathbf{f}\cdot\mathbf{n}_{l_k}dl\label{flux-form}
\end{equation}
where $l_k\ \left(k=1,6\right)$ are the boundary edges compassing the
control volume $\mathcal{C}_i$. $\mathbf{f}=(f,g)$ is the vector of the flux function and
$\mathbf{n}_{l_k}=({n}_{xl_k},{n}_{yl_k})$ the outward normal unit of edge $l_k$.

Line integration in Eq.\eqref{flux-form} is computed by 3-point
Simpson's rule as, for example along edge $l_6\equiv \overline{P_{i6}P_{i1}}$,
\begin{equation}
\int_{l_6}\mathbf{f}\cdot\mathbf{n}_{l_6}dl=\frac{\left|l_6\right|}{6}\left(\mathbf{f}_{i6}+\mathbf{f}_{i1}+4\mathbf{f}_{i13}\right)\cdot\mathbf{n}_{l_6},
\end{equation}
where $\mathbf{f}_{i1}=\mathbf{f}(q_{i1})$, $\mathbf{f}_{i6}=\mathbf{f}(q_{i6})$ and $\mathbf{f}_{i13}=\mathbf{f}(q_{i13})$ since all the point values of the state variable, $q_{i1}$, $q_{i6}$ and $q_{i13}$, are readily known as shown above. 
The same procedure applies to the line integrations along other edges to get the numerical fluxes required in \eqref{flux-form}. 

Next, we derive the evolution equation to update the point value at cell center ($P_{i7}$) by differentiating \eqref{consre2d} with respect to time, and use  \eqref{pv-update} and \eqref{flux-form}, which yields  
\begin{equation}
\frac{d q_{i7}}{dt} =  -\frac{a_7}{A_i}\displaystyle  \sum_{k=1}^6\int_{l_k}\mathbf{f}\cdot\mathbf{n}_{l_k}dl-\sum_{k=1}^6a_k\left( 
{\hat f}_{x ik}+{\hat g}_{y ik}\right).
\label{dtconsre2d}
\end{equation}
The right-hand side of \eqref{dtconsre2d} are readily given as described above. \eqref{dtconsre2d} implies the finite volume formulation \eqref{flux-form} as a constraint condition on the VIA moment that ensures the numerical conservativeness.  

We finally get the semi-discretized equations, \eqref{pv-update} and \eqref{dtconsre2d}, to update the all point value solutions.  The third-order accuracy of the spatial discretization is shown in \citep{chen12} by a Taylor expansion analysis.  The  Runge-Kutta scheme \eqref{rk3} is used for time integration in the explicit model. 

It should be also noted that there are 12 pentagons in the icosahedral geodesic grids. In this case, six solution points are located in each mesh cell, five at the cell vertices and one at the cell center. We use the following polynomial for interpolation, 
\begin{equation}\label{lagintp-p}
\mathcal{Q}_{i}(x,y)=c^{00}_{i}+c^{10}_{i}x+c^{01}_{i}y+c^{11}_{i}xy+c^{20}_{i}x^2+c^{02}_{i}y^2, 
\end{equation} 
and all the numerical procedure discussed above for the hexagonal element applies to the pentagonal elements.

\subsection{Limiting projection}\label{sec:limiter}
High order schemes tend to generate spurious oscillations in the vicinity of discontinuities in numerical solutions, which might develop into unphysical disturbances and even cause computational failure in the worst case and thus must be suppressed in numerical solutions. Artificial viscosity, in form of either diffusion or filtering  operators, is widely used in atmospheric models. See \cite{ja-wi2011} for a thorough and comprehensive discussion on this sort methods.  However, 
the operators rely on some parameters which need to be optimized through steps in an ad hoc fashion. 

Another approach which is more popular in computational fluid dynamics (CFD), particularly  in computational aerodynamics where strong discontinuities (like shock waves) become dominant, is to use the nonlinear limiters to project the high-order reconstruction to
lower order. The typical works of this sort are found as the monotonic upstream-centered scheme for conservation laws (MUSCL) scheme \cite{vanLeer1979}, total variation diminishing (TVD) scheme \cite{harten1983}, essentially non-oscillatory (ENO) \cite{harten1987,shu1,shu2} and  weighted essentially non-oscillatory (WENO) \cite{jiang1996} schemes. The WENO limiter effectively reduces the numerical dissipation errors around smooth region and can retain the high-order convergence. However, the current implementations of the WENO limiting  in the high-order schemes with local reconstructions, such as DG and SE,  cannot make the full use of the local DOFs \cite{qiu2005}, and the solutions are heavily dependent on the TVB (total variation bounded) criterion that determines the ``troubled cells'' in an ad hoc fashion. 
How to suppress numerical oscillations is an unresolved problem for all spectral-like high order methods. 

In atmospheric modeling, even though the Mach number is too small (less than 0.2 in general) to generate shock wave and strong discontinuities in density, large gradients are usually observed in physical fields, such as temperature and moisture around front and cloud edge. So, it is one of the desired properties for advection schemes used to transport these quantities to be oscillation-less and positivity-preserving in some situations. 

We have devised two limiting projection schemes to MCV3 method to suppress numerical oscillation, which effectively eliminate the spurious oscillations while retain high-order accuracy for the smooth solutions. Both limiters are constructed over compact stencil including only the target cell and its immediate neighbors, where all the sub-grid information is used  in the spatial reconstructions. Thus, both scheme are better suited for
the local high-order reconstruction schemes where sub-grid information is available.

\subsubsection{The MCV-WENO4 limiter} \index{MCV limiter}

A WENO-type limiter to MCV scheme was proposed in \cite{stx2015} to solve 
hyperbolic conservation laws that may include  discontinuous solutions, like shock waves. The scheme, so-called MCV-WENO4 (multimoment
Constrained finite Volume with WENO limiter of 4th order) method, is an
extension of the MCV3 method by adding the 1st order derivative
(gradient or slope) at the cell center as an additional constraint for the cell-wise local
reconstruction in the spirit of  \cite{xiao01}. The gradient is computed from a limiting projection using the WENO
(weighted essentially non-oscillatory) reconstruction that is built from the nodal values
at 5 solution points within 3 neighboring cells. Different from other existing methods
where only the cell-average value is used in the WENO reconstruction, the present
method takes account of the solution structure within each mesh cell, and thus minimizes
the stencil for reconstruction. The resulting scheme has 4th-order accuracy and
is of significant advantage in algorithmic simplicity and computational efficiency.  

\subsubsection{The MCV3-BGS limiter}

A new limiting projection scheme for 3-point MCV method was devised in \cite{Deng2017}. Two fourth-order reconstruction function, MCV3-4L and MCV3-4R, are derived by employing the three point values in the target cell and an additional values  at the center of left and right neighboring cells respectively. The boundary gradient switching (BGS) algorithm, underlying the ENO concept, has been proposed  to design a non-oscillatory multi-moment scheme without degrading the fourth-order accuracy.  The basic idea of the BGS algorithm is to choose a spatial reconstruction between MCV3-4L and MCV3-4R schemes, which minimizes the difference in the derivatives of flux functions between the high-order profile  and the reconstruction with a slope limiter. This algorithm is easy to implement and free of and case-dependent ad hoc parameter. Thus, it is very promising for practical applications.

Compared to other existing methods, the present scheme has at least following advantages.
\begin{itemize}

\item The proposed scheme doesn't suffer from loss of accuracy, since both the candidate reconstructions, i.e.   MCV3-4L and MCV3-4R,   have 4th-order accuracy  without nonlinear weights. Our numerical tests verified that MCV3-BGS scheme has 4th-order convergence rate. 

\item MCV3-BGS scheme  does not  need the priori detector, such as the TVB criterion, to peak up the "troubled cells". With effective oscillation-suppressing mechanism and  well-controlled numerical dissipation, MCV3-BGS scheme resolves both smooth and discontinuous solutions.

\item Using the sub-grid DOFs, the spatial stencil used by MCV3-BGS is limited within three neighboring cells, which is very compact and suited for the grids with complex structures.
\end{itemize}

\begin{figure} [htbp]
\begin{center}
  \subfigure[ ]{
  \includegraphics[scale=0.5,angle=0,clip]{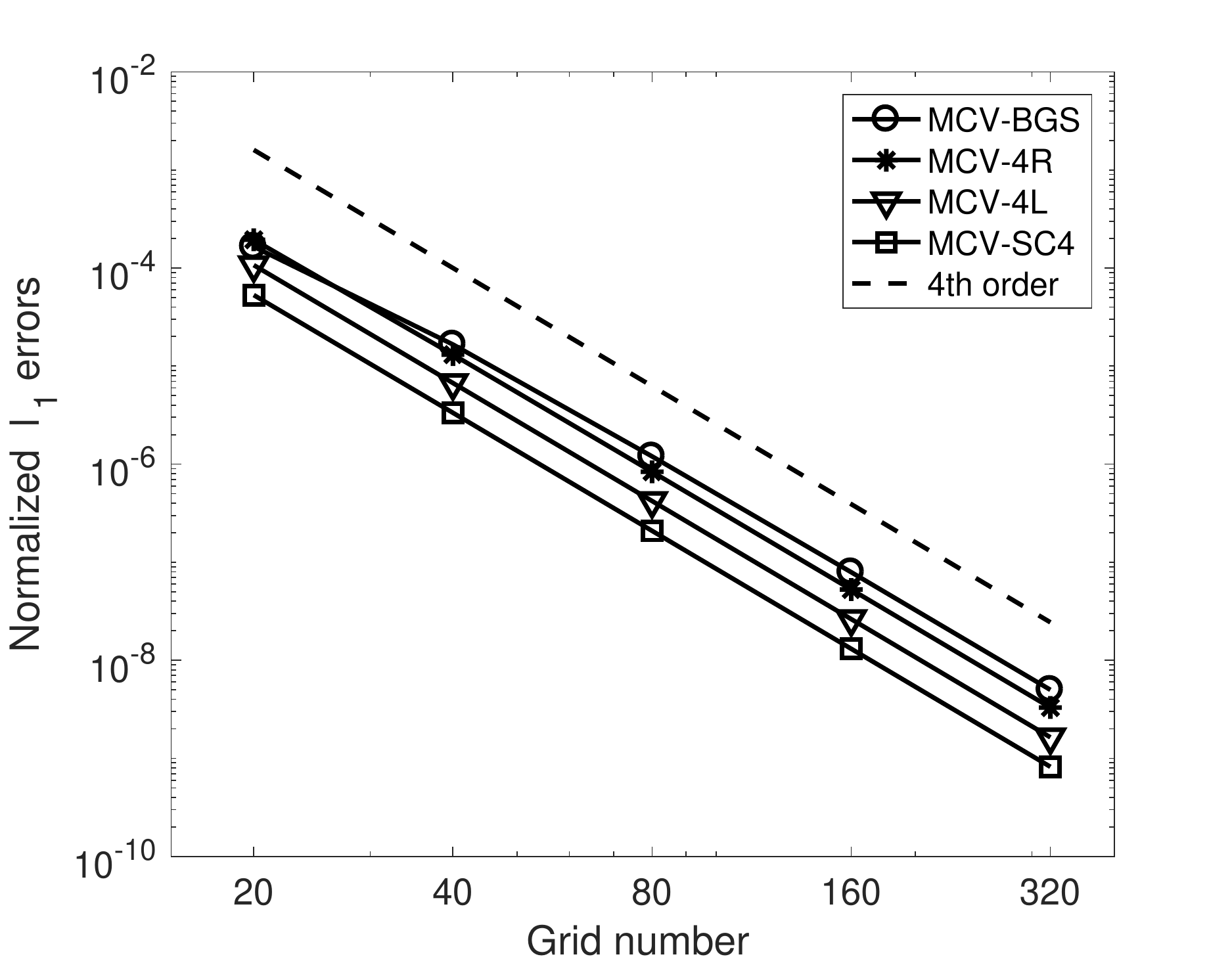}}
\end{center}
\vskip -\lastskip \vskip -3pt \caption{The normalized errors $\ell_1$ and convergence rate of the 1D Sine-wave advection using the MCV-BGS, MCV-4R, MCV-4L and MCV-SC4 schemes at CFL=0.2. }\label{fig:convergence_1D}
\end{figure}

The performance of the  MCV3-BGS scheme is verified by the widely used benchmark tests for both scalar and Euler conservation laws. A grid-refinement test of the sine-wave advection transport was carried out to verify the convergence rate of the MCV3-BGS scheme. It is observed from Fig.\ref{fig:convergence_1D}  that the MCV3-BGS scheme has a uniform 4th-order accuracy. 

\begin{figure*} [htbp]
\begin{center}
\subfigure[]
  {\includegraphics[width=0.4\textwidth]{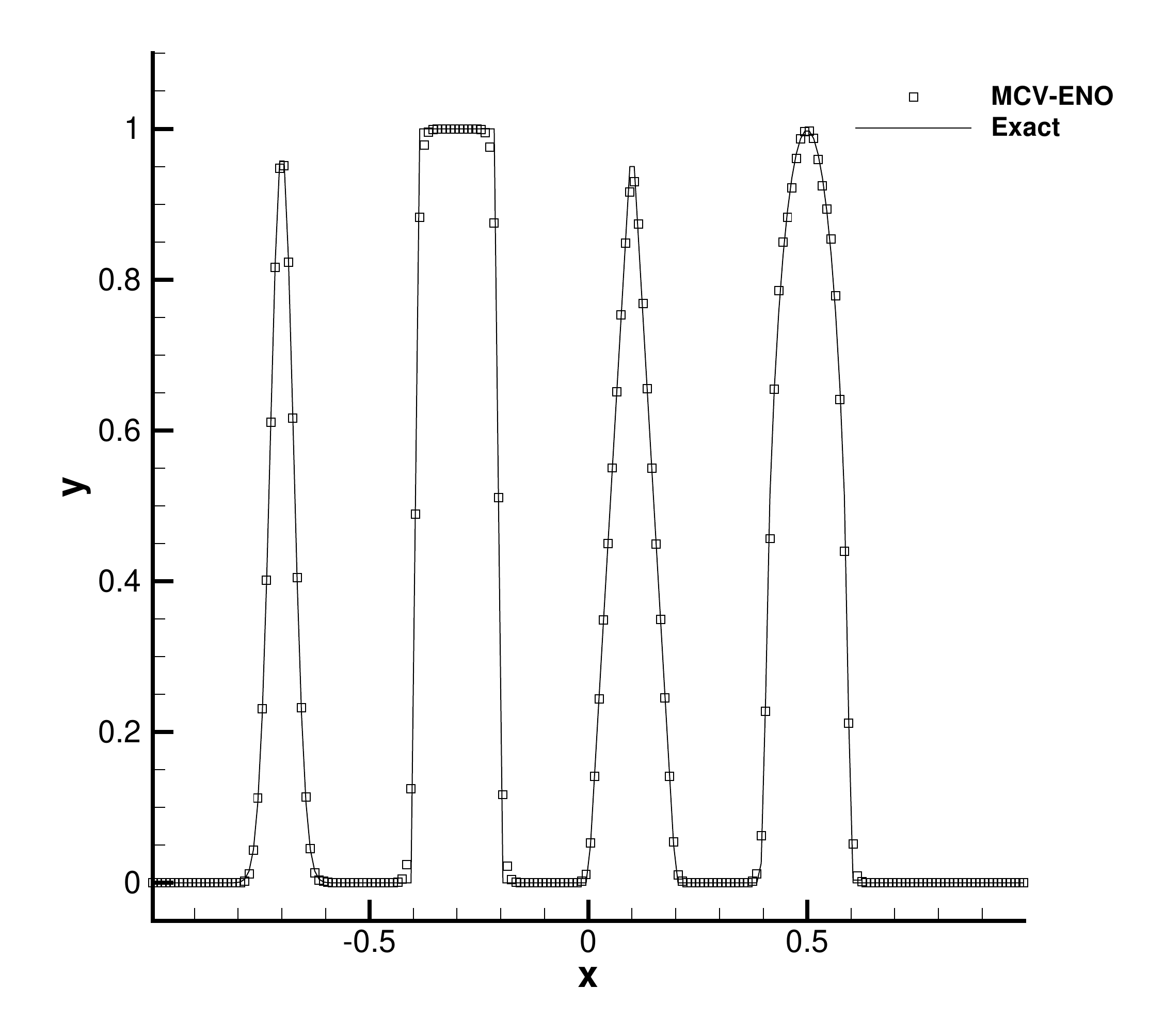}} \hspace{0.5cm}
 \subfigure[]
  {\includegraphics[width=0.4\textwidth]{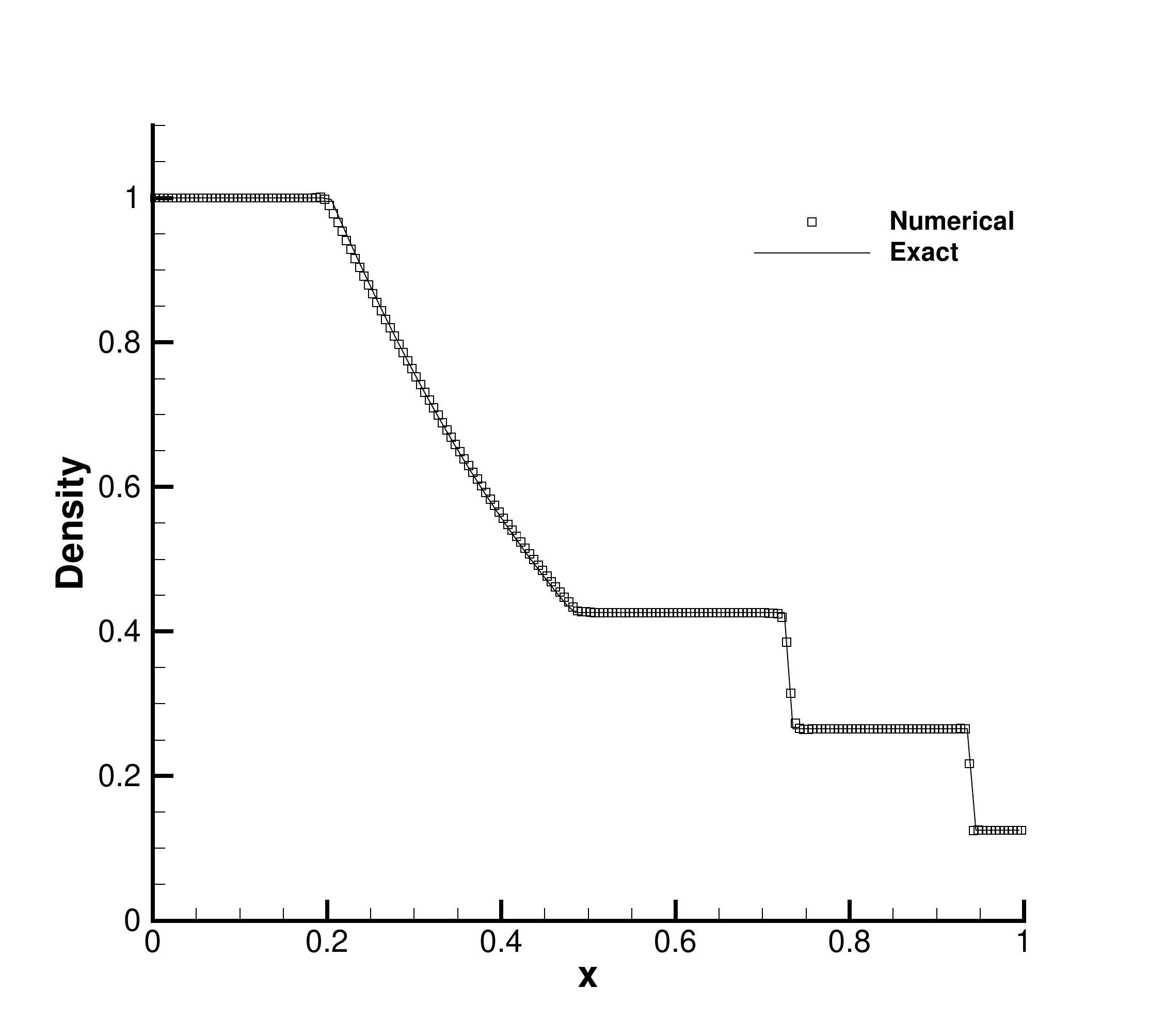}} 
\end{center}
\vskip -\lastskip \vskip -3pt 
\caption{Numerical results of  MCV3-BGS scheme. (a) advection test of a complex profile including smooth and discontinuous distributions \cite{jiang1996}. (b) Sod's test for Euler equations of compressible flow \cite{jiang1996}. }\label{fig:bgs1dtest}
\end{figure*}

Fig.\ref{fig:bgs1dtest} shows the numerical results of one dimensional benchmark tests for both advection and Euler equations. Both smooth and discontinuous solutions are accurately resolved, while the spurious oscillation is effectively removed. 

\begin{figure*} [htbp]
\begin{center}
\subfigure[]
  {\includegraphics[width=0.4\textwidth]{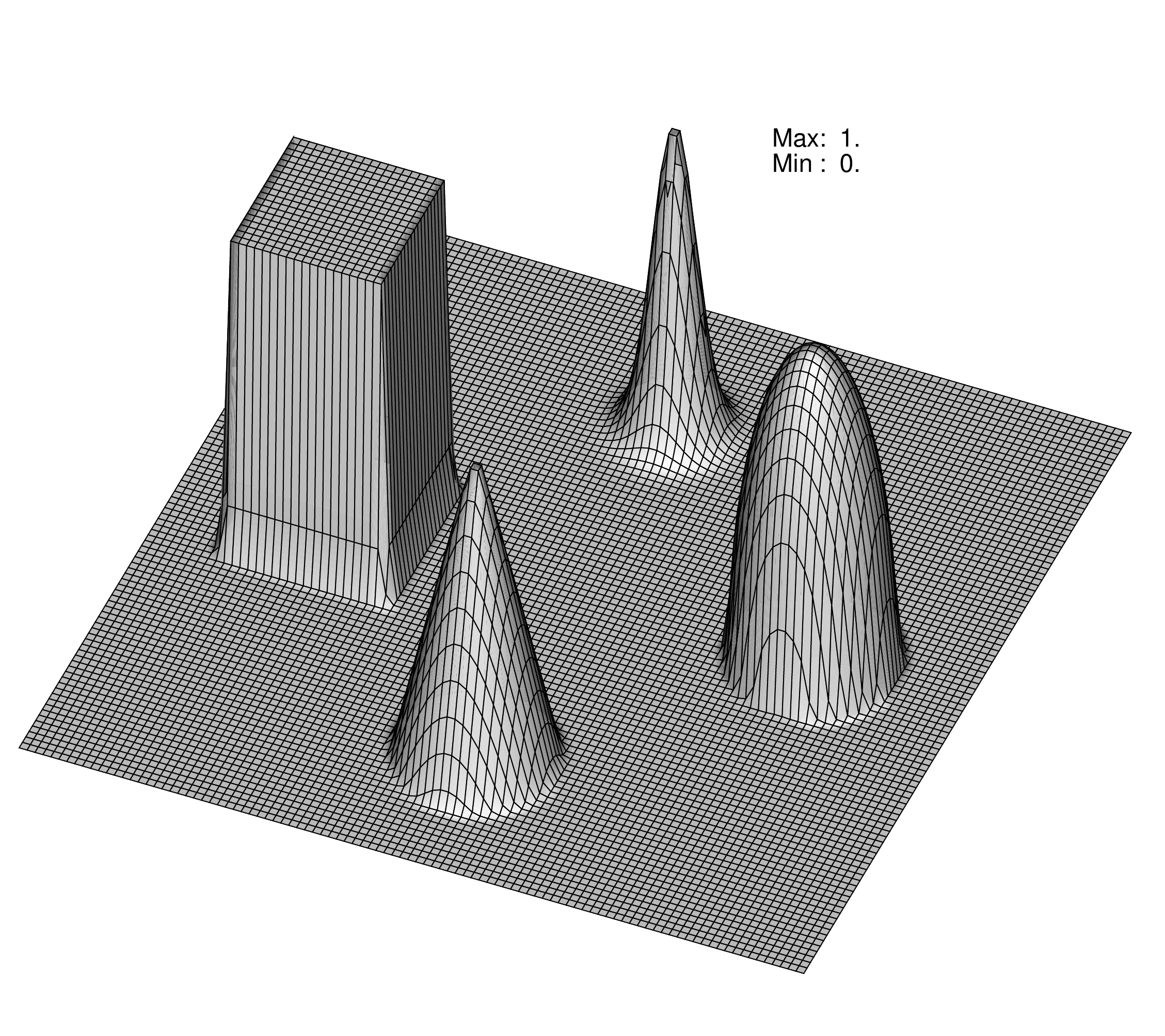}} \hspace{0.5cm}
 \subfigure[]
  {\includegraphics[width=0.4\textwidth]{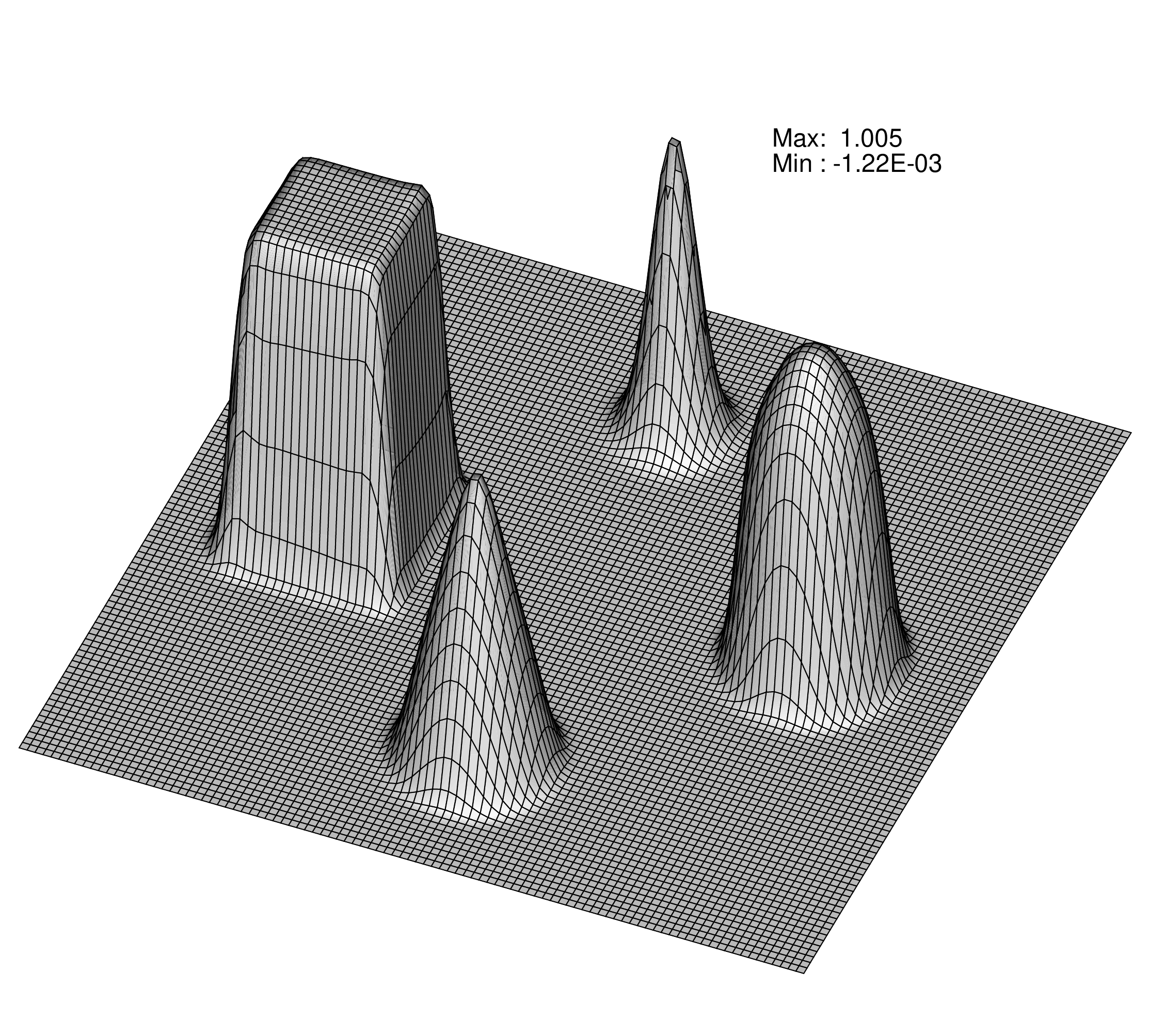}} 
\end{center}
\vskip -\lastskip \vskip -3pt 
\caption{Numerical result of 2D rotation of complex profile after one period with $100\times100$ cells. (a) the initial profile and (b) the numerical result.}\label{fig:bgs2dadvs}
\end{figure*}

We show in Fig. \ref{fig:bgs2dadvs} the results of a 2D advection test. It  is obvious that the shape of initial profile is faithfully preserved without visible numerical oscillations in 
vicinity of large gradients.   

\begin{figure*} [htbp]
\begin{center}
\subfigure[]
  {\includegraphics[width=0.9\textwidth]{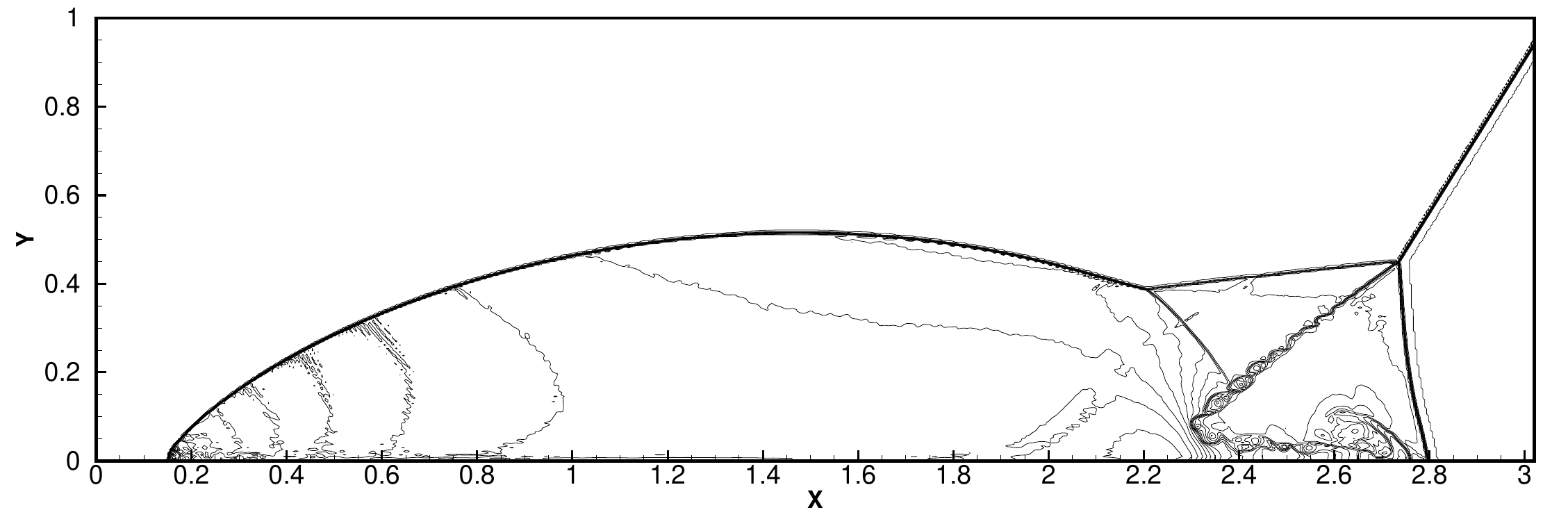}} \hspace{0.1cm}
 \subfigure[]
  {\includegraphics[width=0.45\textwidth]{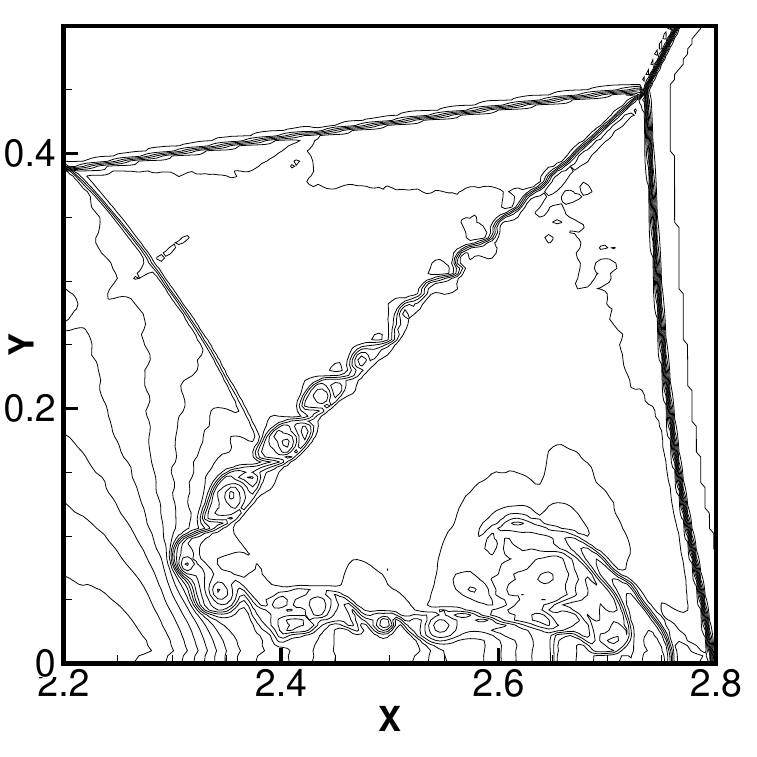}} 
\end{center}
\vskip -\lastskip \vskip -3pt 
\caption{Numerical result for density field of the double Mach reflection at $t=0.2$ with $250 \times800$ cells. (a) the density field in whole computational domain, (b) the enlarged view of vortex structures and instability along the slip lines.}\label{fig:bgs2deuler}
\end{figure*}

The numerical results of the double Mach reflection benchmark test for Euler equations \cite{woodward1984} are plotted in Fig.\ref{fig:bgs2deuler}. Both the strong discontinuities and the vortex structures are well resolved by MCV3-BGS scheme, which shows the well-controlled numerical dissipation in MCV3-BGS scheme.

Shown above, MCV3-BGS scheme is a high-fidelity scheme with local high order reconstruction to resolve both smooth and non-smooth solutions with appealing accuracy the robustness competitive to other existing schemes.

\section{Global shallow water models on quasi-uniform spherical grids }
\label{sec:research}

\begin{figure} [htbp]
\begin{center}
\subfigure[Yin-Yang grid]
{\includegraphics[width=0.3 \textwidth]{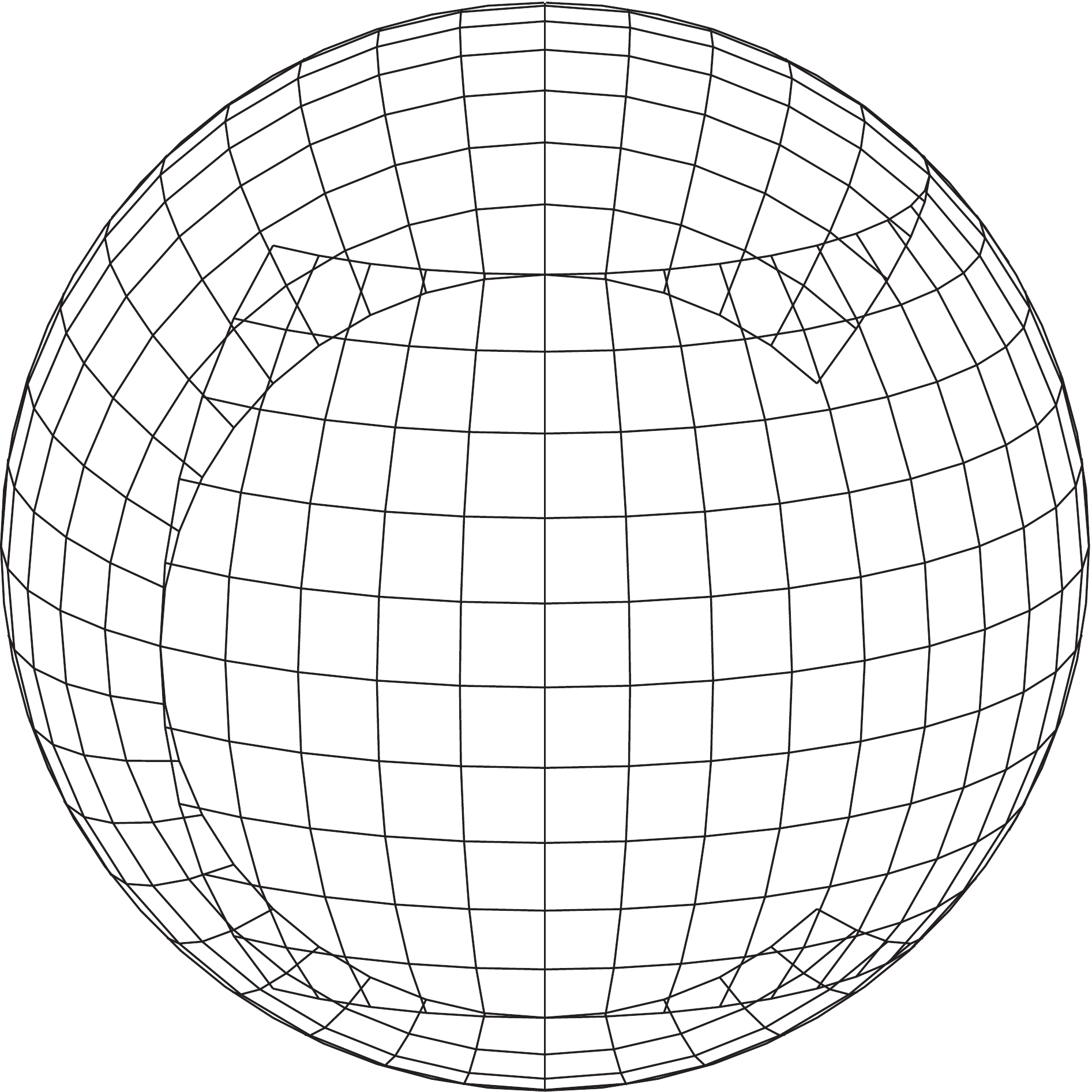}}
\subfigure[Cubed sphere grid]
{\includegraphics[width=0.3 \textwidth]{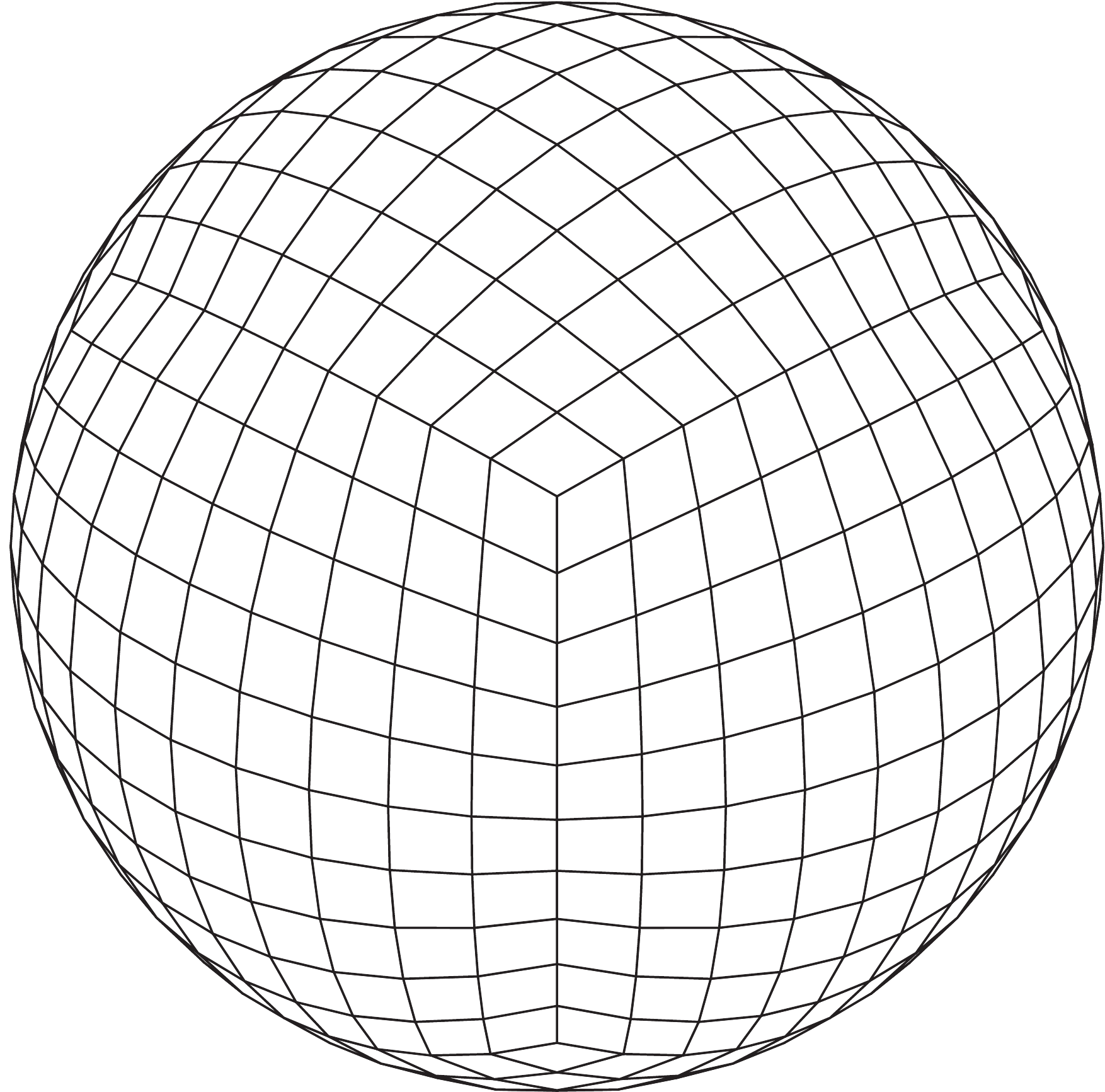}}
\subfigure[Icosahedral-hexagonal grid]
{\includegraphics[width=0.3\textwidth]{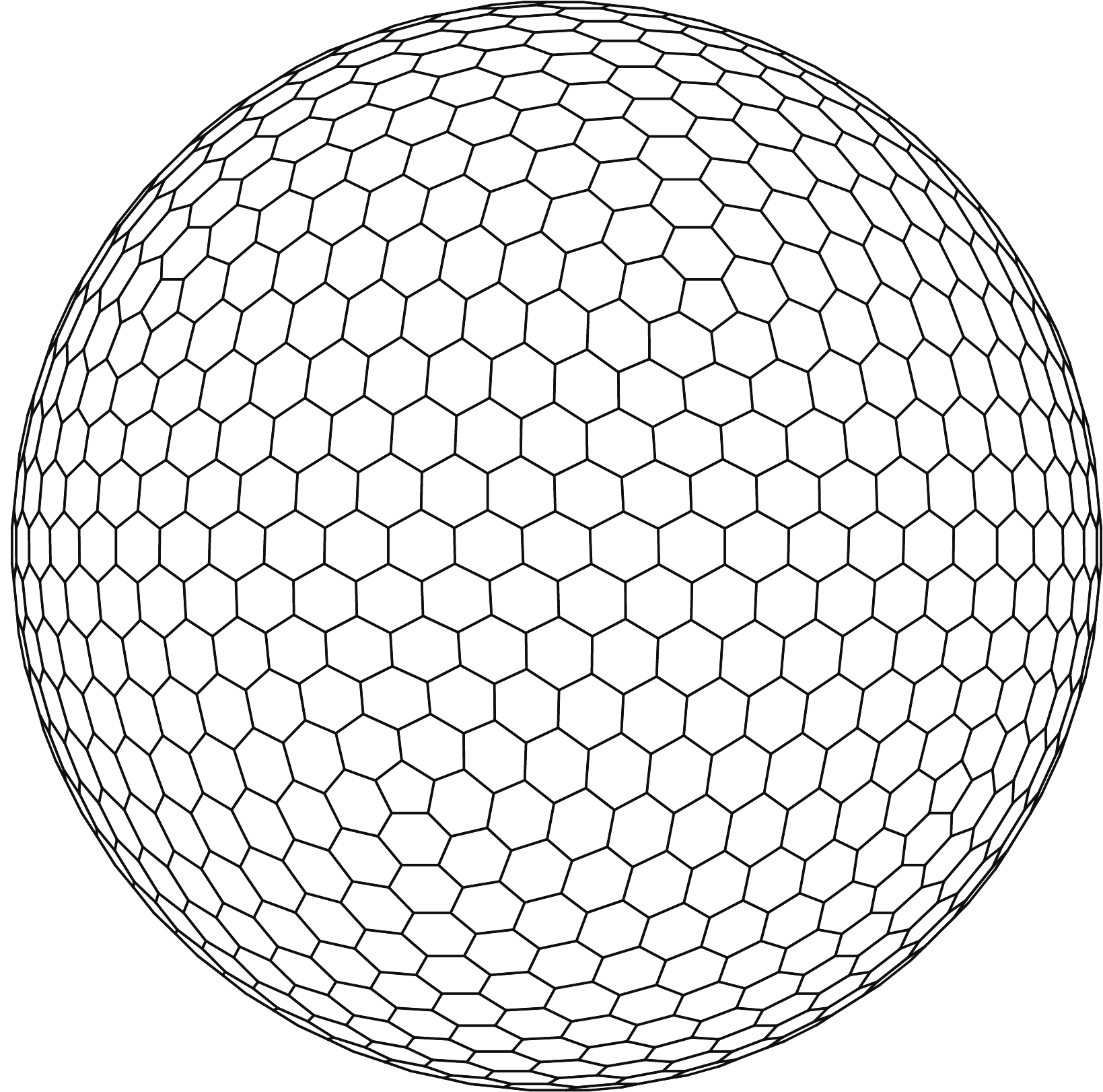}}
\end{center}
\caption{The three quasi-uniform spherical grids.}\label{3grids}
\end{figure}

We have implemented the MCV method to develop global shallow water models based on quasi-uniform grids, i.e.  Yin-Yang grid \citep{ks04}, cubed-sphere grid \citep{Sadourny1972} and icosahedron geodesic grid \citep{Sadourny1968,Williamson1968},  as shown in Fig.\ref{3grids}. 

These three quasi-uniform grids have gained an increasing popularity recently in developing  global atmospheric and oceanic models due to uniform area ratio of grid cells over the whole spherical surface. Despite the same desired uniformity in grid-cell area, they are constructed by different procedures.  Yin-Yang grid is constructed by overlapping two perpendicularly oriented LAT-LON grid components of low latitude part. With each of its component being part of the conventional LAT-LON grid, Yin-Yang grid provides a convenient platform readily to accommodate the numerical models originally developed for the LAT-LON grid. The cubed sphere grid is generated by mapping an inscribed cube onto the sphere, so it is composed of six identical patches connected each other by 12 patch boundaries. A gnomonic projection results in more uniform grid spacing, but the mesh on each patch is not orthogonal. Being another quasi-uniform grid, icosahedral geodesic grid projects an inscribed icosahedron, rather than the cube as in the cubed-sphere grid, onto the spherical surface. Both triangular and hexagonal elements can be generated from an icosahedron with quasi-uniform areas over the whole sphere. Compared to the triangular Delaunay tessellation, the hexagonal Voronoi tessellation is more attractive in global grid uniformity, arrangement of local degrees of freedom (DOFs) for reconstruction, as well as the accuracy and robustness in flux computation.

In order to facilitate a quantitative comparison among these grids, we define the grid resolution of the three grids by the number of partitions $N$ which are slightly different for the grids, i.e. the mesh element number along the short boundary edges in Yin-Yang grid, the mesh element number along the patch boundary edges in cubed sphere grid and the mesh element number along the boundary edges of the primary triangles in icosahedral grid. For a grid of $N$ level resolution, the total numbers of unknown DOFs of a scalar field are shown in Table \ref{tdofs}. 
\begin{table}[ht]
\caption{The total numbers of the unknown DOFs in the third-order MCV models for different spherical grids.} \label{tdofs}
\begin{tabularx}{\textwidth}{l@{\extracolsep{\fill}}ccccc}
\hline\hline
& Spherical grid   & Yin-Yang & Cubed sphere&Icosahedral hexagon\\ \hline
&    Total DOFs   & $24N^2+16N+2$         & $24N^2+2$ &  $30N^2+2$  \\ \hline
\end{tabularx}
\end{table}

The ratios of the  minimum and maximum areas of  mesh cells are given in Table \ref{area-ratio}. 
\begin{table}[h]
\caption{Minimum and maximum area ratio of different grids.} \label{area-ratio}
\begin{tabularx}{\textwidth}{l@{\extracolsep{\fill}}ccccc}
\hline\hline
  &Yin-Yang grid &Cubed-sphere grid&Icosahedral grid & Icosahedral grid \\
&                         &                               & with pentagonal cells & without pentagonal cells\\
\hline
&  0.7628 (N=10)    &   0.7666  (N=10) & 0.6263 (N=9)    &   0.8654 (N=9)      \\
& 0.7349 (N=20)    &   0.7359   (N=20)& 0.6216 (N=18)    &   0.8558 (N=18)      \\
& 0.7210 (N=40)    &   0.7213    (N=40)&  0.6197 (N=36)   &   0.8524 (N=36)      \\
&0.7140 (N=80)    &   0.7142   (N=80)&  0.6191 (N=72)   &    0.8513 (N=72)     \\
\hline
\end{tabularx}
\end{table}
 
Using the low latitudinal part of the LAT-LON grid ($-\pi/4 \le \varphi \le \pi/4$), Yin-Yang grid has much more uniform grid spacing compared to the conventional LAT-LON grid. The cubed sphere grid using gnomonic partitioning is similar to Yin-Yang grid. The icosahedral grid consists of 12 pentagonal mesh elements at the vertices of the regular icosahedron, which have smaller area than other hexagonal elements.  Shown in Table \ref{area-ratio}, the area difference between the pentagonal and hexagonal cells are much more significant than the difference among the groups of the same cell type. The uniformity becomes worse for all grids when the grid resolution is refined. To this point, icosahedral grid is more advantageous and has the least dependency on the grid resolution. 

Considering the grid structures mentioned above, different numerical algorithms are needed for different grids. Yin-Yang and cubed sphere grids are quadrilateral composite grids, and can be solved by schemes for structured grid in a more efficient way, but extra manipulations are required for data transfer across the patch boundaries.  The icosahedral is an non-quadrilateral grid which needs more complicated numerical procedure to solve the governing equations. The MCV formulation presented in this document is a well suited and efficient solver for computational fluid dynamics on icosahedral grid. Another issue distinguishing the Yin-Yang grid from other two is that Yin-Yang grid does not ensure the rigorous numerical conservativeness, which makes the cubed sphere grid and icosahedral grid more appealing for long range simulations. 

\subsection{Shallow water models on Yin-Yang grid and Cubed-sphere grid} \index{shallow water model}
As discussed above, both Yin-Yang and cubed-sphere grids can be mapped to a standard structured local grid. Thus, we can cast the shallow water equations for Yin-Yang and cubed-sphere grids into the same form. 

We consider the continuity and the momentum equations of shallow water equation in the 2D vector form, 
\begin{eqnarray}
&& \frac{\partial h}{\partial t} + \nabla \cdot (h \mathbf{v})=0, \label{eq:vector_swe_mass}     \\
&& \frac{\partial \mathbf{v} }{\partial t} + (\zeta +\mu)\mathbf{k}\times\mathbf{v} + \nabla (\Phi+K) =0, \label{eq:vector_swe_momentum}
\end{eqnarray}
where $h$ is the height of the fluid over the bottom mountain, $\mathbf{v}$ the velocity vector, 
 $\Phi=g(h+h_s)$ the geopotential height, $g$ gravitational
constant, $h_s$ the height of the bottom mountain, $K =
\frac{1}{2}\mathbf{v} \cdot \mathbf{v}$ the kinetic energy. $\zeta$
is relative vorticity defined as $\zeta \equiv \mathbf{k}\cdot
(\nabla\times\mathbf{v})$ and $\mathbf{k}$ is local unit  outward vector
normal to the surface of the sphere.  $\mu=2\Omega
\sin(\varphi)$ is the Coriolis parameter and $\Omega$ the rotation speed of the Earth.

The shallow water equations \eqref{eq:vector_swe_mass}  and \eqref{eq:vector_swe_momentum} on the curvilinear coordinate system $(\xi,\eta)$ are written as
\begin{align}
&\frac{\partial (\sqrt{\Gamma}h)}{\partial t} + \left( \frac{\partial
(\sqrt{\Gamma} h \tilde{u})}{\partial \xi} + \frac{\partial (\sqrt{\Gamma}
h \tilde{v})}{\partial \eta} \right)=0,
\label{eq:con_contra_variant_component1}
\\
&\frac{\partial u}{\partial t} + \frac{\partial (\Phi+K)}{\partial \xi}=-\sqrt{\Gamma}\tilde{v}(\mu+\zeta) \label{eq:con_contra_variant_component2},\\
&\frac{\partial v}{\partial t} + \frac{\partial (\Phi+K)}{\partial
\eta}
=-\sqrt{\Gamma}\tilde{u}(\mu+\zeta),\label{eq:con_contra_variant_component3}
\end{align}
where $\sqrt{\Gamma}$ is the Jacobian of the transformation. 
 $(u,v)$ and $(\tilde{u},\tilde{v})$ denote  
the covariant and contravariant components of velocity on the curvilinear coordinate,
respectively. $\zeta=\frac{1}{\sqrt{\Gamma}}( \frac{\partial v}{\partial
\xi}-\frac{\partial u}{\partial \eta} )$ and
$K=\frac{1}{2}(u\tilde{u}+v\tilde{v})$.

For Yin-Yang grid, the spherical LAT-LON coordinate is used, i.e. $(\xi,\eta)=(\lambda,\varphi)$.  For the cubed sphere grid, the curvilinear coordinates are defined by the length of the arcs, $(\xi,\eta)=(R\alpha,R\beta)$, using central angle projection $(\alpha,\beta)\in[-\pi/4,\pi/4]$.  The detailed covariant metric tensor $\Gamma_{ij}$, contravariant metric tensor $\Gamma^{ij}$ and Jacobian of the transformation $\sqrt{\Gamma}$ can be referred to \cite{chen14}.

\subsection{Shallow water model on icosahedral-hexagonal grid}
We write the shallow water equations in three-dimensional Cartesian coordinate form \citep{wi92,swarztrauber1998}.  The governing equations of flux-form are written as
\begin{equation}
\partial_t\mathbf{q}+\partial_x\mathbf{F}+\partial_y\mathbf{G}+\partial_z\mathbf{H}=\mathbf{S},\label{SWE}
\end{equation}
where dependent variables $\mathbf{q}=\left[h,\ hu,\ hv,\ hw\right]^T$ are water depth and momentum components in $x$, $y$ and $z$ directions, and the flux functions are
\begin{equation}
\mathbf{F}=\left(hu,\ hu^2+\frac{1}{2g}\Phi^2,\ huv,\ huw\right)^T,\label{fluxx}
\end{equation}
\begin{equation}
\mathbf{G}=\left(hv,\ huv,\ hv^2+\frac{1}{2g}\Phi^2,\ hvw\right)^T,\label{fluxy}
\end{equation}
\begin{equation}
\mathbf{H}=\left(hw,\ huw,\ hvw,\ hw^2+\frac{1}{2g}\Phi^2\right)^T.\label{fluxz}
\end{equation}
The source terms include Coriolis force and topographic effects as
\begin{equation}
\mathbf{S}=\left(\begin{array}{c}0\\{\tilde \mu}\left(zhv-yhw\right)+gh_s\partial_x\Phi\\{\tilde \mu}\left(xhw-zhu\right)+gh_s\partial_y\Phi\\{\tilde \mu}\left(yhu-xhv\right)+gh_s\partial_z\Phi\end{array}\right),
\end{equation}
where $\left(x,y,z\right)$ indicates the position in Cartesian coordinate and ${\tilde \mu}={\mu}/{R}$. 
Same as in the previous cases, total height is adopted in the flux terms of \eqref{SWE} to ensure the C-property. 

The three dimensional governing equations \eqref{SWE} are then projected to the plane tangential to the sphere, which restricts the velocity (or momentum) vector in the tangential direction of the spherical surface. We map the predicted momentum vector $\left[hu,\ hv,\ hw\right]^T$ at each time step by the direction correction matrix \citep{wi92,swarztrauber1998},
\begin{equation}
\mathcal{P}=\frac{1}{R^2}\left(\begin{array}{ccc}R^2-x^2 & -xy & -xz \\ -xy & R^2-y^2 & -yz \\ -xz & -yz & R^2-z^2\end{array}\right).
\end{equation} 

The numerical formulations of  MCV method presented in section 2 are implemented to solve shallow water equations on the  three quasi-uniform spherical grids, see \cite{chen14} for more technical details.

\subsection{Numerical tests of global shallow water models on spherical grids}

We have extensively verified the global shallow water models by widely used benchmark tests, including those in  \cite{wi92},  \cite{lauter2005} and \cite{gal04}.

\subsubsection{The convergence rates of the shollow water models}

We have examined the convergence rates (accuracy) of the shollow water models with both steady-state geostrophic flow test \cite{wi92} and time-dependent flow test \cite{lauter2005}.  

The steady-flow test is the case 2 of Williamson's test set\cite{wi92}. The true solution to this test is a steady flow with all the physical fields remaining unchanged identical to the initial condition. The numerical errors, $(\phi_n-\phi_e)$, of the tests with different flow directions, $\gamma=0$, $\frac{\pi}{4}$ and $\frac{\pi}{2}$, are given in Table \ref{f-direction-c2}. The numerical errors don't show significant dependency on the flow direction, which reveals the robustness of the MCV models on different spherical grids. We may owe this partly to the locality of the MCV scheme by which quadratic reconstruction can be built over single cell. It is particularly beneficial for the data communication needed across the patch boundaries in the Yin-Yang and cubed sphere grids.

\begin{table}[htbp]
\caption{Normalized errors of case 2 at day 5 for different flow directions.} \label{f-direction-c2}
\begin{tabularx}{\textwidth}{l@{\extracolsep{\fill}}ccccc}
\hline\hline
 &\multicolumn{4}{c}{Yin-Yang grid  ($N = 20$)}\\ 
\hline

& Flow direction &{$l_1$} error &{$l_2$} error &{$l_\infty$} error\\
\hline
& $\gamma=0$                &  $2.13\times10^{-4}$  & $2.36\times10^{-4}$ & $4.14\times10^{-4}$ \\
& $\gamma=\frac{\pi}{4}$    &  $1.84\times10^{-4}$  & $2.11\times10^{-4}$ & $4.11\times10^{-4}$ \\
& $\gamma=\frac{\pi}{2}$    &  $2.19\times10^{-4}$  & $2.44\times10^{-4}$ & $4.10\times10^{-4}$ \\

\hline\hline
 &\multicolumn{4}{c}{Cubed-sphere grid ($N = 20$)}\\ 
\hline

&Flow direction            & $l_1$ error  &$l_2$ error  &$l_\infty$ error\\ \hline
& $\gamma=0$             & $1.59\times10^{-4}$& $1.91\times10^{-4} $& $3.67\times10^{-4}$\\
& $\gamma=\frac{\pi}{4}$ & $1.76\times10^{-4}$ &$ 1.98\times10^{-4}$ & $4.04\times10^{-4}$ \\
& $\gamma=\frac{\pi}{2}$ & $1.59\times10^{-4}$ & $1.91\times10^{-4} $& $3.67\times10^{-4} $\\ 
\hline\hline
 &\multicolumn{4}{c}{Icosahedral grid ($N = 18$)}\\ 
\hline

&Flow direction      & $l_1$ error  &$l_2$ error  &$l_\infty$ error\\ \hline
& $\gamma=0$                  & $8.27\times10^{-5}$      & $9.00\times10^{-5}$    & $1.27\times10^{-4}$    \\
& $\gamma=\frac{\pi}{4}$  & $7.15\times10^{-5}$     & $8.22\times10^{-5}$     & $1.77\times10^{-4}$   \\
& $\gamma=\frac{\pi}{2}$  & $7.98\times10^{-5} $    & $9.05\times10^{-5}  $  & $1.66\times10^{-4}  $ \\ \hline

\end{tabularx}
\end{table}

Grid refinement tests were conducted to check the convergence rate of the proposed shallow water models.
We simulated the westerly steady geostrophic flow for 5 days on gradually refined grids with  three models.  The numerical errors of $l_1$, $l_2$ and $l_\infty$ norms in the  height field are shown in Table \ref{convergence}. Uniform third order accuracy are verified for all models on different spherical grids. 

The numerical results obtained by the present models are competitive to other existing models, for example those reported in  \cite{tom01} and \cite{ringler10} on icosahedral-hexagonal grid, particularly when the grid is refined due to the higher order convergence rate of the present models.  
Compared to the Yin-Yang and cubed sphere grids, the  icosahedral-hexagonal grid generates smaller numerical errors as shown in Table \ref{convergence} where similar numbers of the unknown DOFs are used. It reveals the major advantage of the icosahedral-hexagonal grid. Namely, the icosahedral-hexagonal grid has more homogeneous topological connections among the mesh cells, and  the special treatment for data transfer at the  patch boundaries is minimized. It is also observed that the icosahedra-grid model has more uniform convergence rate than the Yin-Yang and  cubed-sphere grids, since the extra errors generated by patch boundaries usually converge more slowly compared with the errors within each patch, which becomes more significant with refined grid resolution.  Overall, the cubed-sphere grid model appears to be more accurate than the Yin-Yang grid model though both of them need data communication across the patch boundaries. One major reason is that the cell edges of the cubed-sphere grid are exactly matched along the patch boundary which does not only benefit for the numerical conservation but also for the interpolation accuracy, whereas the Yin-Yang grid has irregularly overlapped regions between the two component grids. As mentioned before the Yin-Yang grid does not ensures the numerical conservation.

\begin{table}[htbp] 
\caption{Normalized errors and convergence rate of case 2 at day 5 for different grids ($\gamma=0$).}\label{convergence}
\begin{tabularx}{\textwidth}{l@{\extracolsep{\fill}}cccccccc}
\hline\hline
 &\multicolumn{7}{c}{Yin-Yang grid}\\ 
\hline

& Resolution &\multicolumn{2}{c}{$l_1$}&\multicolumn{2}{c}{$l_2$}&\multicolumn{2}{c}{$l_\infty$}\\
\cline{3-4}\cline{5-6}\cline{7-8}
& & error & order & error &order & error & order \\ \hline
& $N=10$    &  $1.68\times10^{-3}$ &   -  & $1.87\times10^{-3}$ &   -  & $3.36\times10^{-3}$ &  -   \\
& $N=20$    &  $2.13\times10^{-4}$ & 2.98 & $2.36\times10^{-4}$ & 2.99 & $4.14\times10^{-4}$ & 3.02 \\
& $N=40$    &  $2.71\times10^{-5}$ & 2.97 & $2.99\times10^{-5}$ & 2.98 & $5.21\times10^{-5}$ & 2.99 \\
& $N=80$    &  $3.42\times10^{-6}$ & 2.99 & $3.80\times10^{-6}$ & 2.98 & $6.49\times10^{-6}$ & 3.00 \\

\hline\hline
 &\multicolumn{7}{c}{Cubed-sphere grid}\\ 
\hline

& Resolution   &\multicolumn{2}{c}{$l_1$}&\multicolumn{2}{c}{$l_2$}&\multicolumn{2}{c}{$l_\infty$}\\
\cline{3-4}\cline{5-6}\cline{7-8}
&                 & error         & order  & error         &order  & error    & order \\ \hline
& $N=10$ &$ 1.29\times10^{-3}$ & -    & $1.53\times10^{-3}$ & -    & $3.01\times10^{-3} $& -    \\
& $N=20$ & $1.59\times10^{-4} $& 3.01 & $1.91\times10^{-4}$ & 3.00 & $3.67\times10^{-4}$ & 3.04 \\
& $N=40$ & $1.99\times10^{-5} $& 3.00 & $2.39\times10^{-5} $& 3.00 & $4.54\times10^{-5} $& 3.02 \\
& $N=80$ & $2.50\times10^{-6} $& 3.00 & $2.99\times10^{-6}$& 3.00 & $5.66\times10^{-6} $& 3.00\\ 

\hline\hline
 &\multicolumn{7}{c}{Icosahedral grid}\\ 
\hline

& Resolution   &\multicolumn{2}{c}{$l_1$}&\multicolumn{2}{c}{$l_2$}&\multicolumn{2}{c}{$l_\infty$}\\
\cline{3-4}\cline{5-6}\cline{7-8}
&                 & error         & order  & error         &order  & error    & order \\ \hline
& $N=9$    & $6.63\times10^{-4} $ & -          & $7.17\times10^{-4} $ & -        & $1.02\times10^{-3} $ & -     \\
& $N=18$  & $8.27\times10^{-5}  $ & 3.00   & $9.00\times10^{-5}  $ & 2.99  & $1.27\times10^{-4}  $&  3.00 \\
& $N=36$  & $1.03\times10^{-5} $  & 3.00   & $1.12\times10^{-5} $  & 3.00  & $1.59\times10^{-5}  $ &  2.99 \\
& $N=72$  & $1.29\times10^{-6}$ & 3.00   & $1.41\times10^{-6} $ & 3.00  & $1.99\times10^{-6}  $&  3.00 \\ \hline
\end{tabularx}
\end{table}
A time-dependent zonal flow with analytical solution was introduced in \cite{lauter2005}. This new test is a more  challenging problem supplementary to Williamson's test case 2 to evaluate the convergence rate of shallow water models \cite{pudykiewicz2011}. 

\begin{figure}[htbp]
\begin{center}
\subfigure[The height field on cubed-sphere grid ($N=20$).]
{\includegraphics[width=0.48\textwidth]{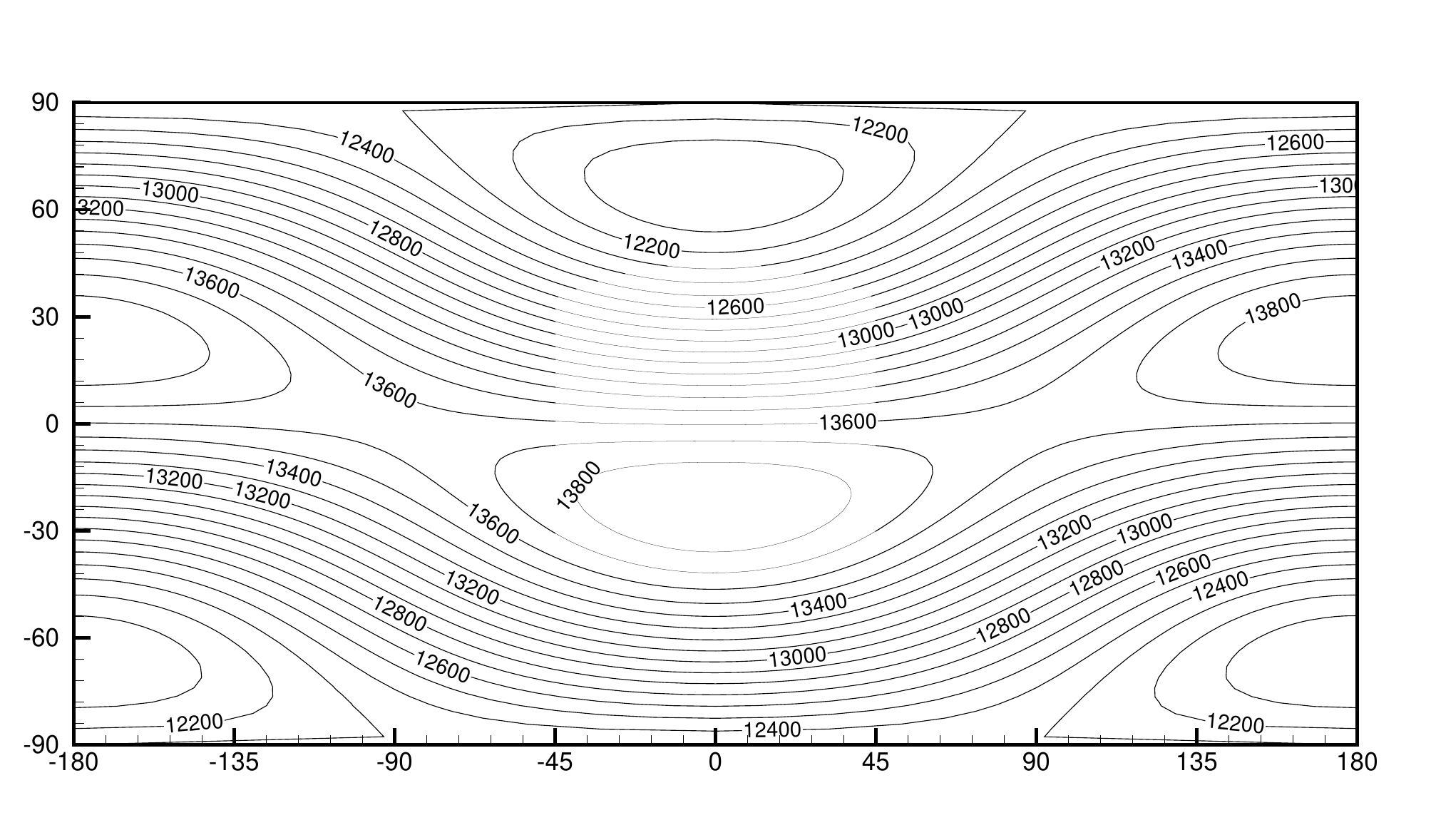}}
\subfigure[The absolute error on Yin-Yang grid ($N=20$).]
{\includegraphics[width=0.48\textwidth]{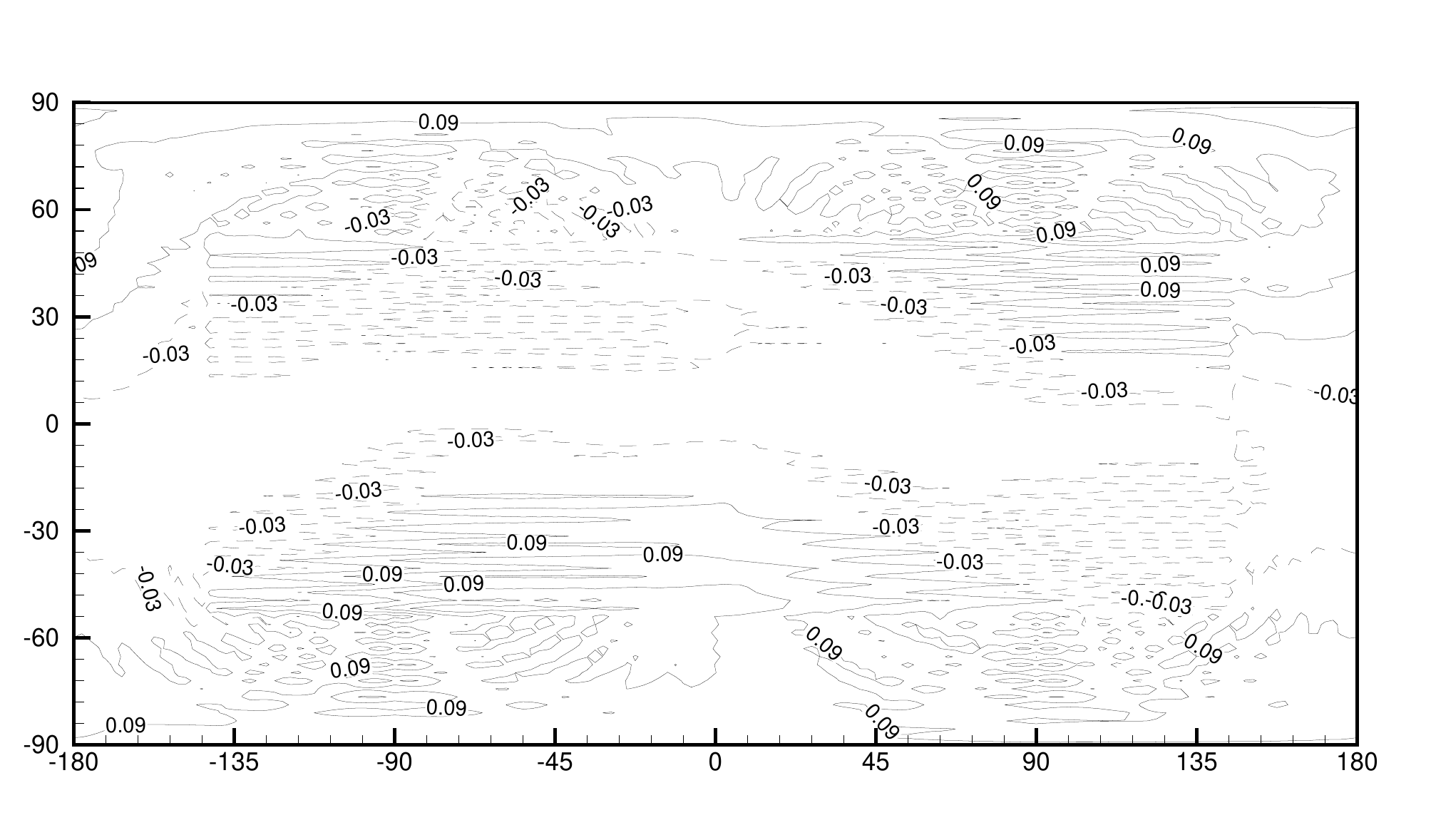}}\\
\subfigure[The absolute error on cubed-sphere grid ($N=20$).]
{\includegraphics[width=0.48\textwidth]{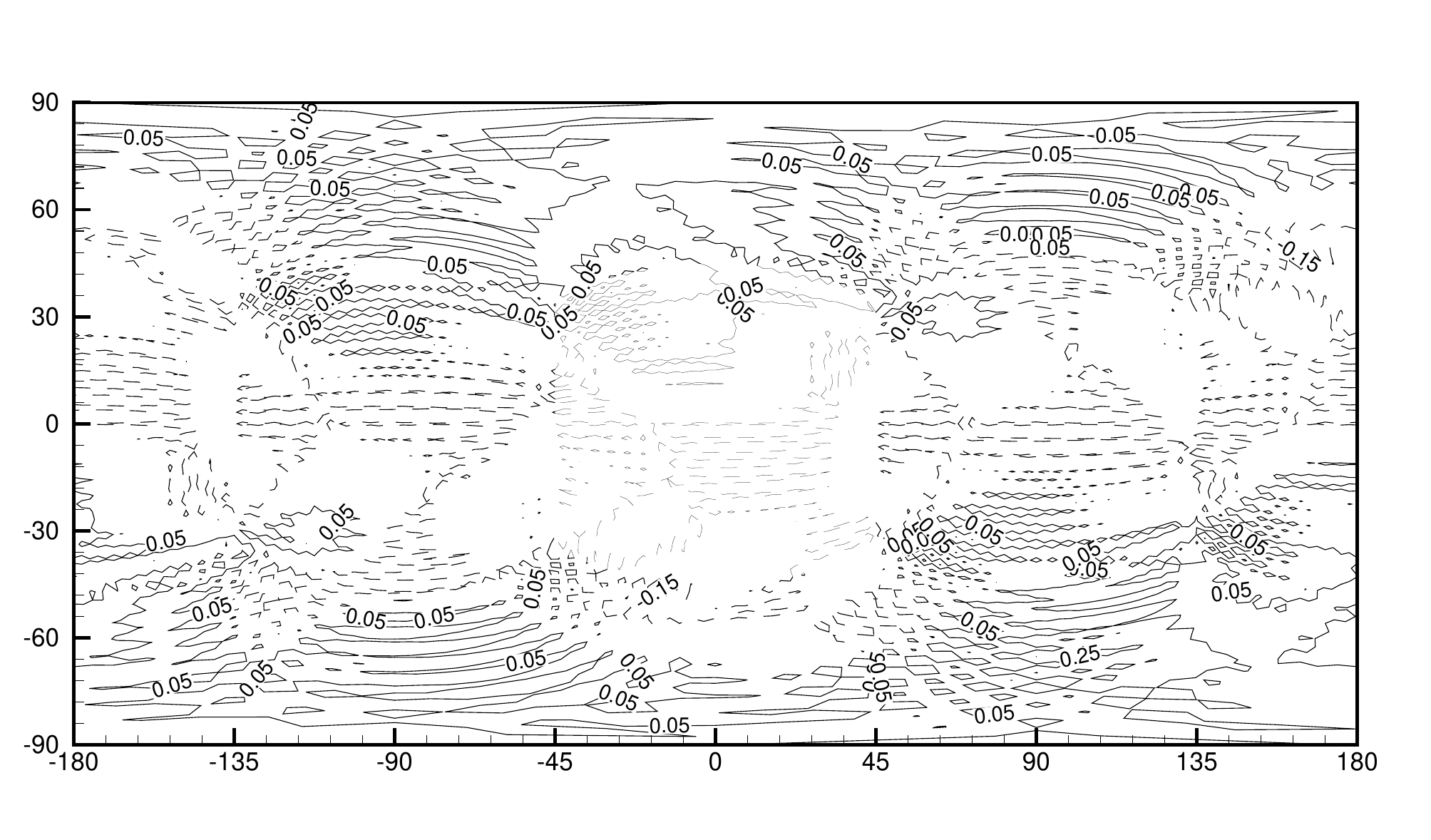}}
\subfigure[The absolute error on icosahedral grid ($N=18$).]
{\includegraphics[width=0.48\textwidth]{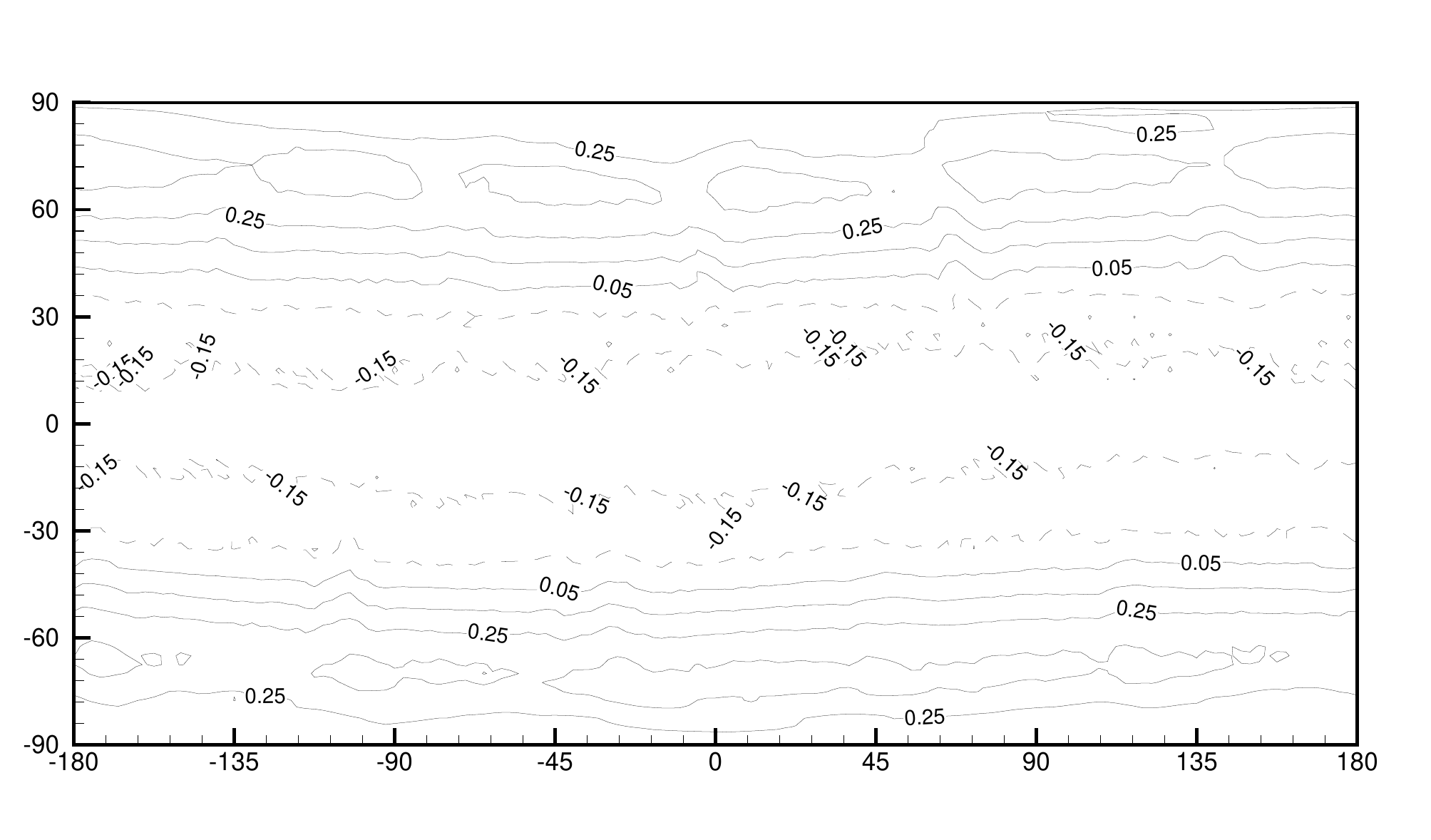}}
\end{center}
\caption{Numerical results of time dependent zonal flow problem after 12 hours. Contour lines vary from 12000m to 13800m with an interval of 100m for height plot, from -0.15m to 0.45m with an interval of 0.1 for error plots.}\label{tdzfh}
\end{figure}

\begin{figure}[htbp]
\begin{center}
\subfigure[Yin-Yang grid]
{\includegraphics[width=0.33\textwidth]{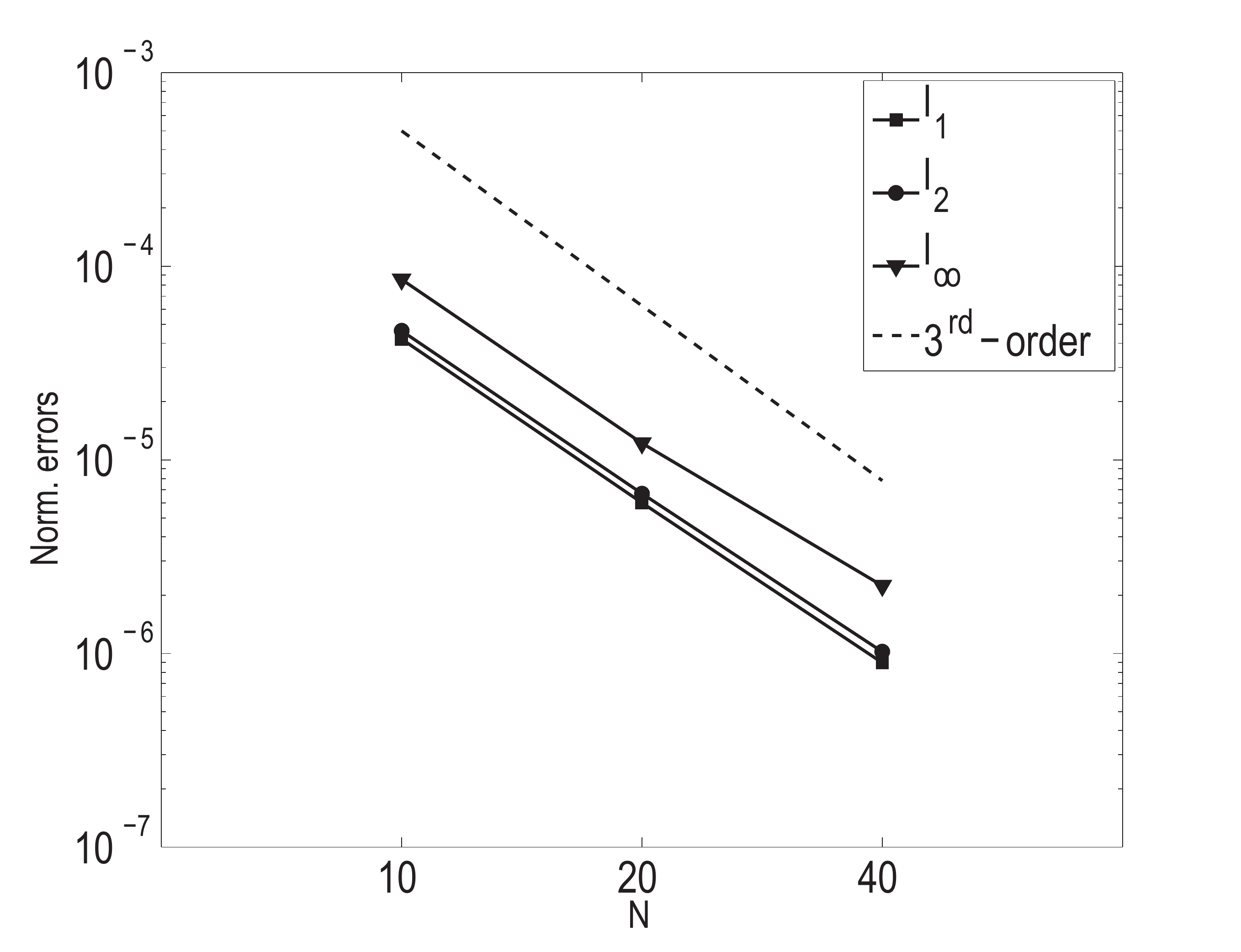}}
\subfigure[Cubed-sphere grid]
{\includegraphics[width=0.33\textwidth]{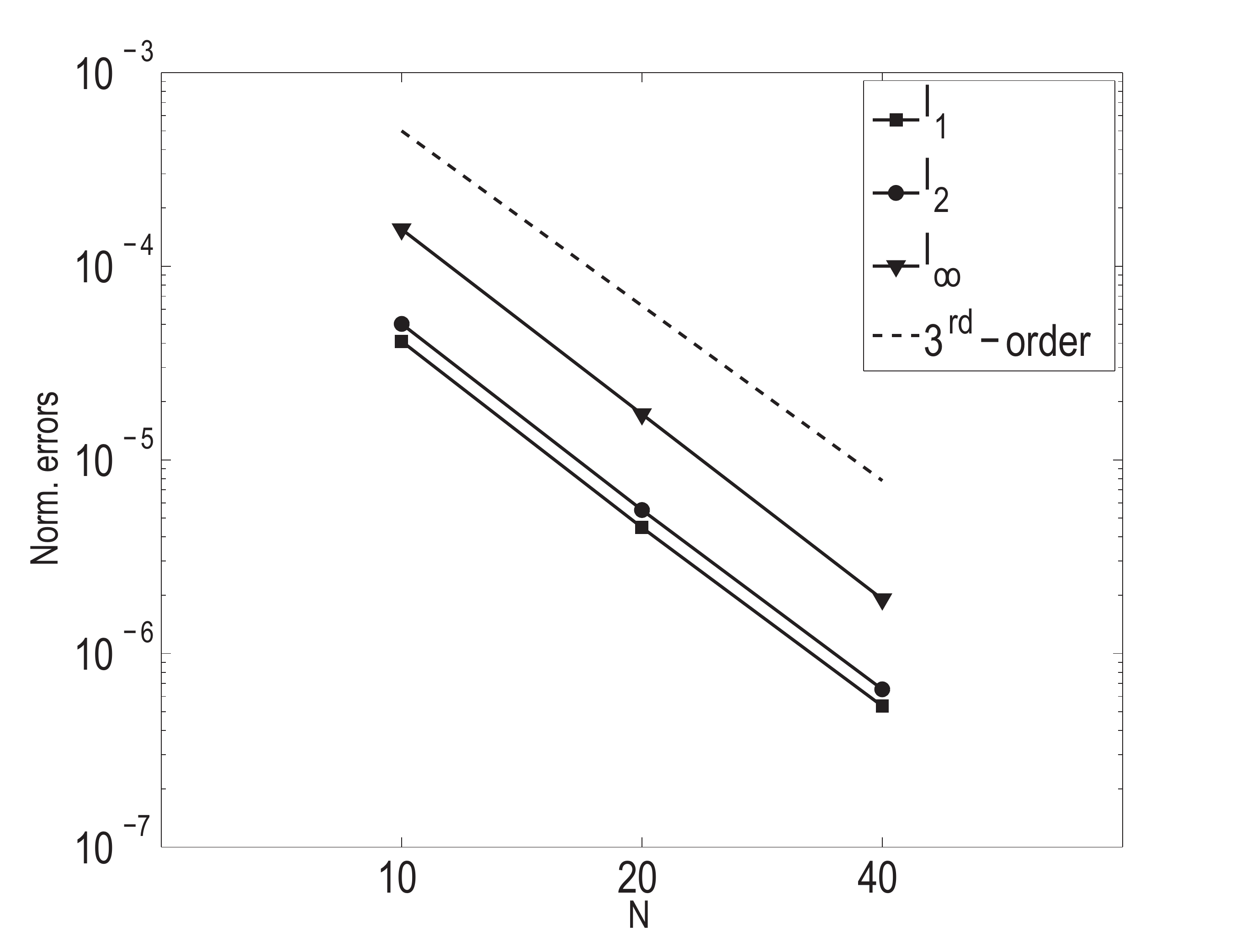}}
\subfigure[Icosahedral grid]
{\includegraphics[width=0.33\textwidth]{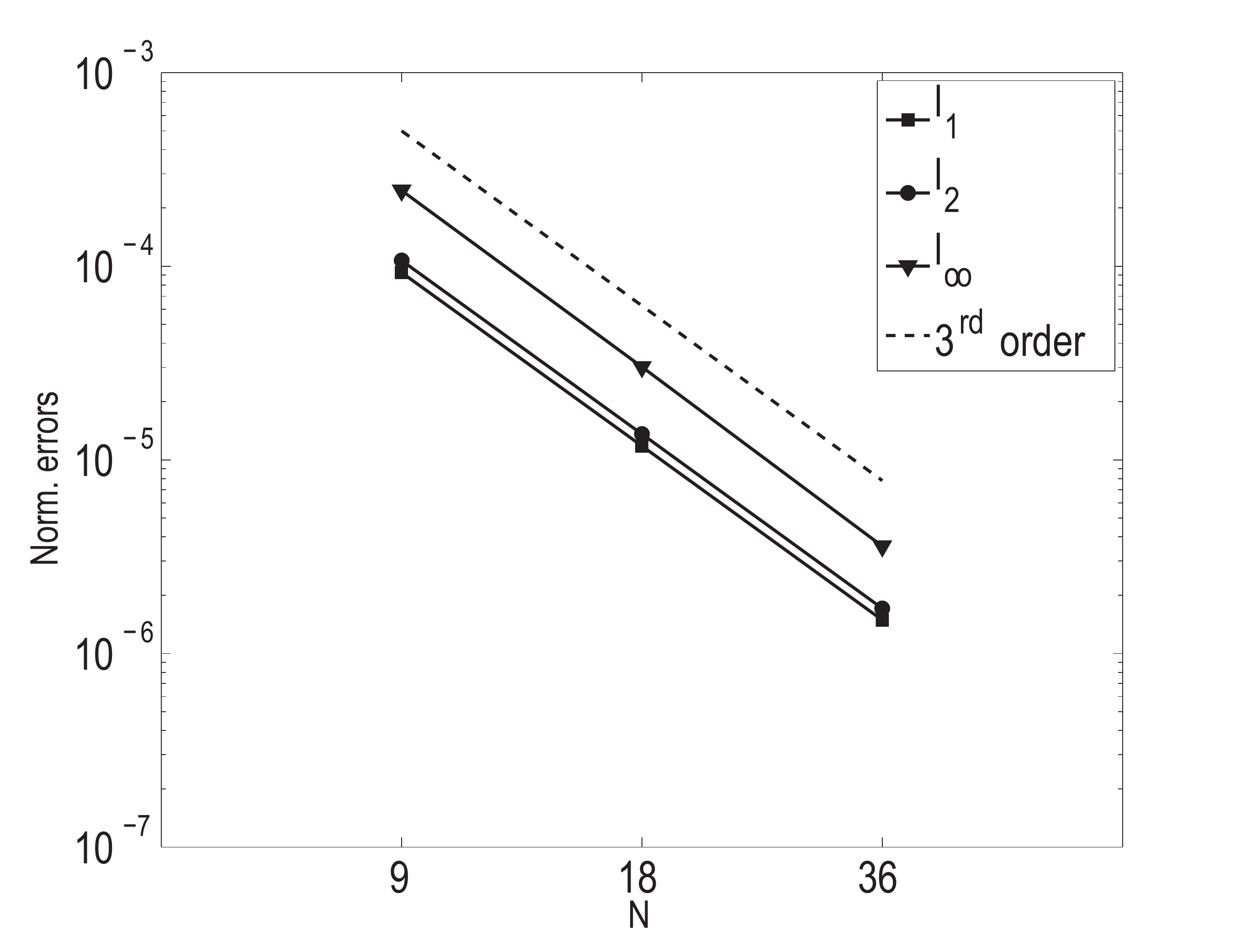}}
\end{center}
\caption{Convergence of normalized errors on refined grids of time dependent zonal flow. Dashed lines denote the third-order slope.}\label{tdzfe}
\end{figure}

We check this test with refined resolutions on different grids, i.e. $N=10,\ 20,\ 40$ on cubed-sphere and Yin-Yang grids and $N=9,\ 18,\ 36$ on icosahedral grid. The height field after integrating the MCV models for 12 hours are shown Fig.\ref{tdzfh}(a) for numerical results on cubed-sphere grid with $N=20$. We don't depict the exact solution and the numerical results of height field on other two grids here since they are visually identical. The other three panels of Fig.\ref{tdzfh} show the absolute errors for height fields on different grids. Obviously, we can observe the relatively larger errors generated around the inner boundaries between different patches of each global mesh. The convergence of the normalized errors of the MCV models are given in the three panels of Fig.\ref{tdzfe} for  different grids. All results show the third-order convergence rate as expected. We see a slightly slower convergence rate for $l_\infty$ error on Yin-Yang grid owing to the data communications across the overlapping zones of two components that usually degrade the local accuracy in numerical output.

\subsubsection{Williamson's test case 5: Zonal flow over an isolated mountain}

\begin{figure} [htbp]
\begin{center}
\subfigure[Yin-Yang grid (N=40).]
{\includegraphics[width=0.5\textwidth]{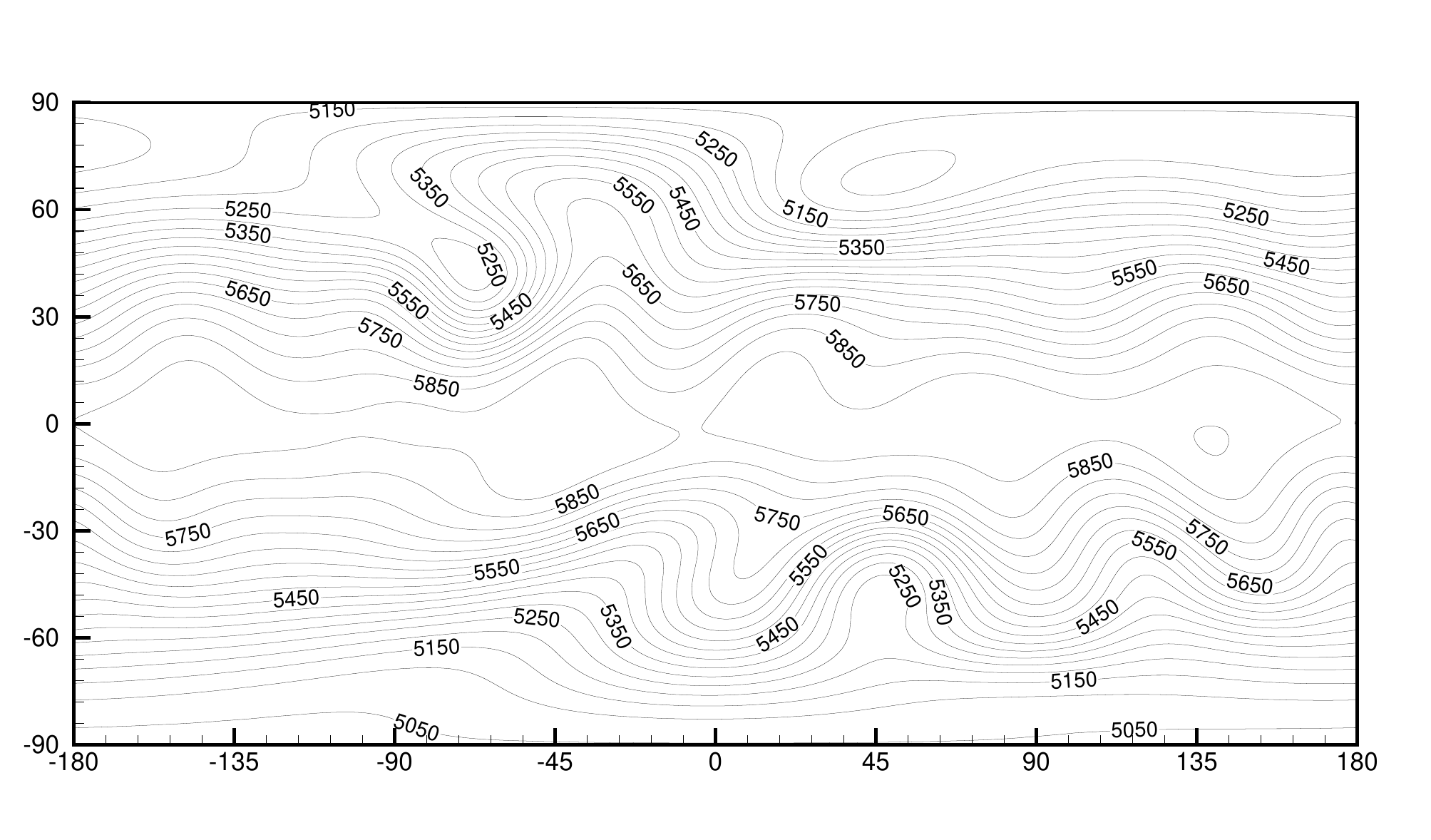}}
\subfigure[Cubed-sphere grid (N=40)]
{\includegraphics[width=0.5\textwidth]{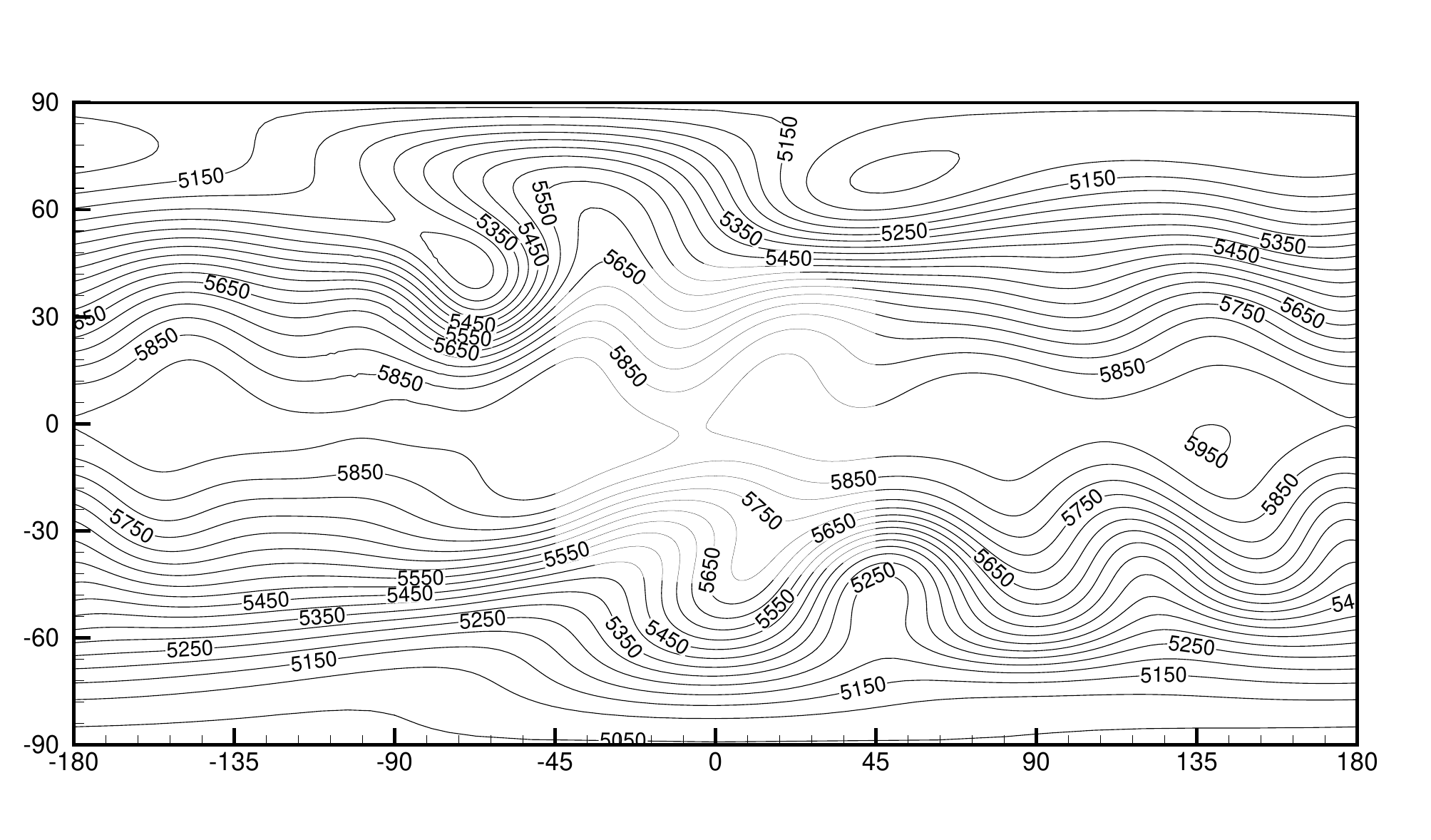}}\\
\subfigure[Icosahedral grid (N=36)]
{\includegraphics[width=0.5\textwidth]{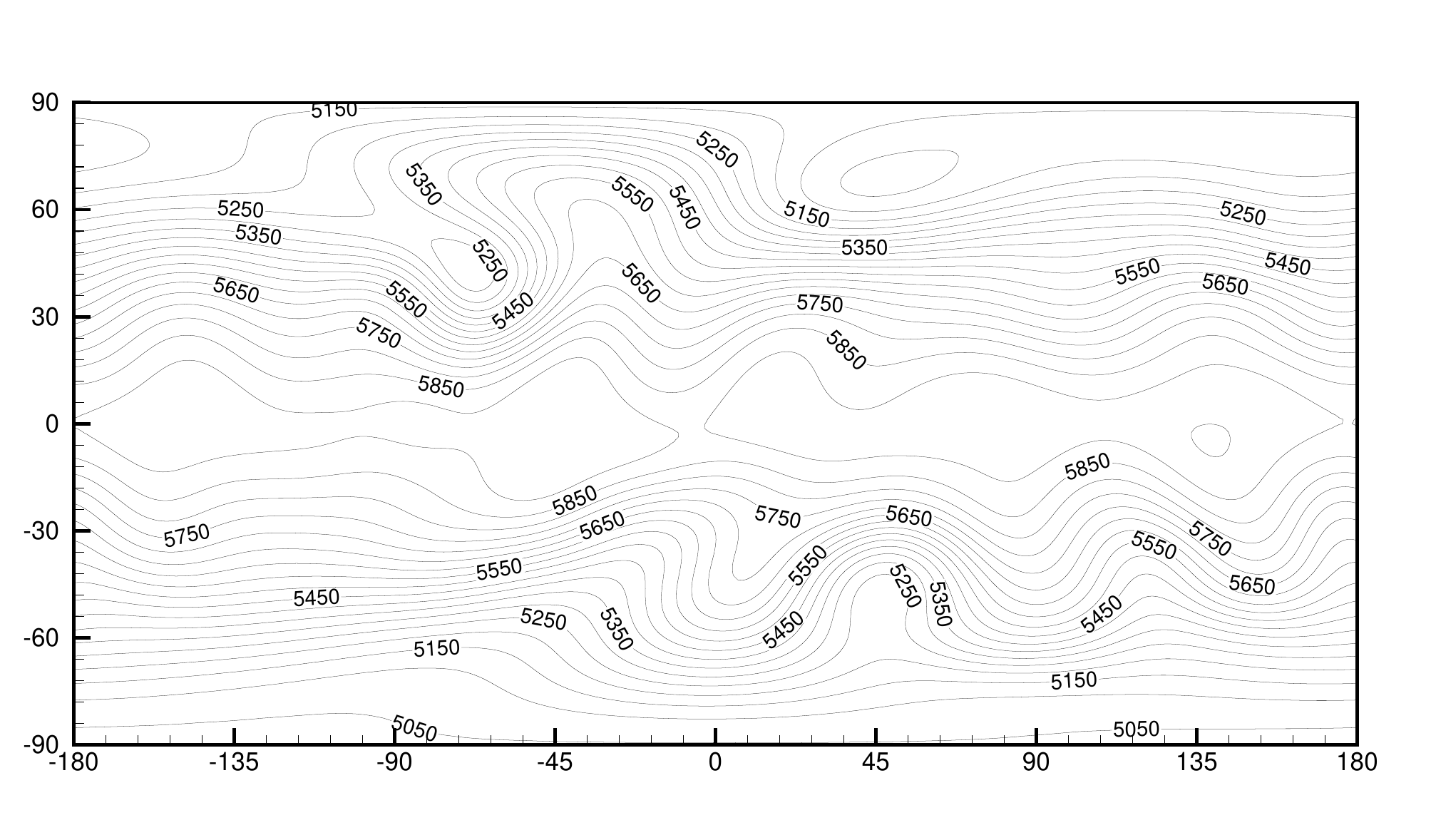}}\\
\end{center}
\caption{Numerical solutions of the total height field of mountain wave test (Williamson's test case 5).
}\label{case5_height}
\end{figure}

Numerical results of the total height field after integrating the MCV models for 15 days are shown in Fig.\ref{case5_height}. The spectral transform solution to this test on T213 grid is usually considered as the reference solution for verification\cite{chien95}. Our results agree well with the reference solution, except that present results are free of the nonphysical oscillations around boundary of bottom mountain due to the satisfaction of the C-property.

The models using cubed-sphere and icosahedral grids rigorously conserve the total mass. The conservation errors of total energy and potential enstrophy are also controlled to a low level. The cubed sphere grid shows a significant improvement in high resolution simulation. The icosahedral hexagonal grid demonstrates  overall more accurate results and a  smoother decline in the conservation errors when the computational grid is refined. It is observed that the result of icosahedral grid with a resolution of $N=36$ is comparable to the spectral transform solution on grid T42 (see Fig.5.3 in \cite{chien95}).

\subsubsection{Williamson's test case 6: Rossby-Haurwitz wave}

The Rossby-Haurwitz wave provides a good test bed for global shallow water models to simulate the middle-range dynamic processes up to two weeks. Being case 6 in Williamson test set, the details of initial conditions of this test were given in \cite{wi92}. The flow field of Rossby-Haurwitz wave test is complicated and consists of phenomena of many different scales. High-order numerical models with less numerical dissipation are usually expected to give better results to this test. Since there is no analytic solution available for this test, we consider the spectral transform solution on fine T213 grid \citep{chien95}  as the reference solution. 

\begin{figure} [htbp]
\begin{center}
\subfigure[Yin-Yang grid ($N=54$).]
{\includegraphics[width=0.48\textwidth]{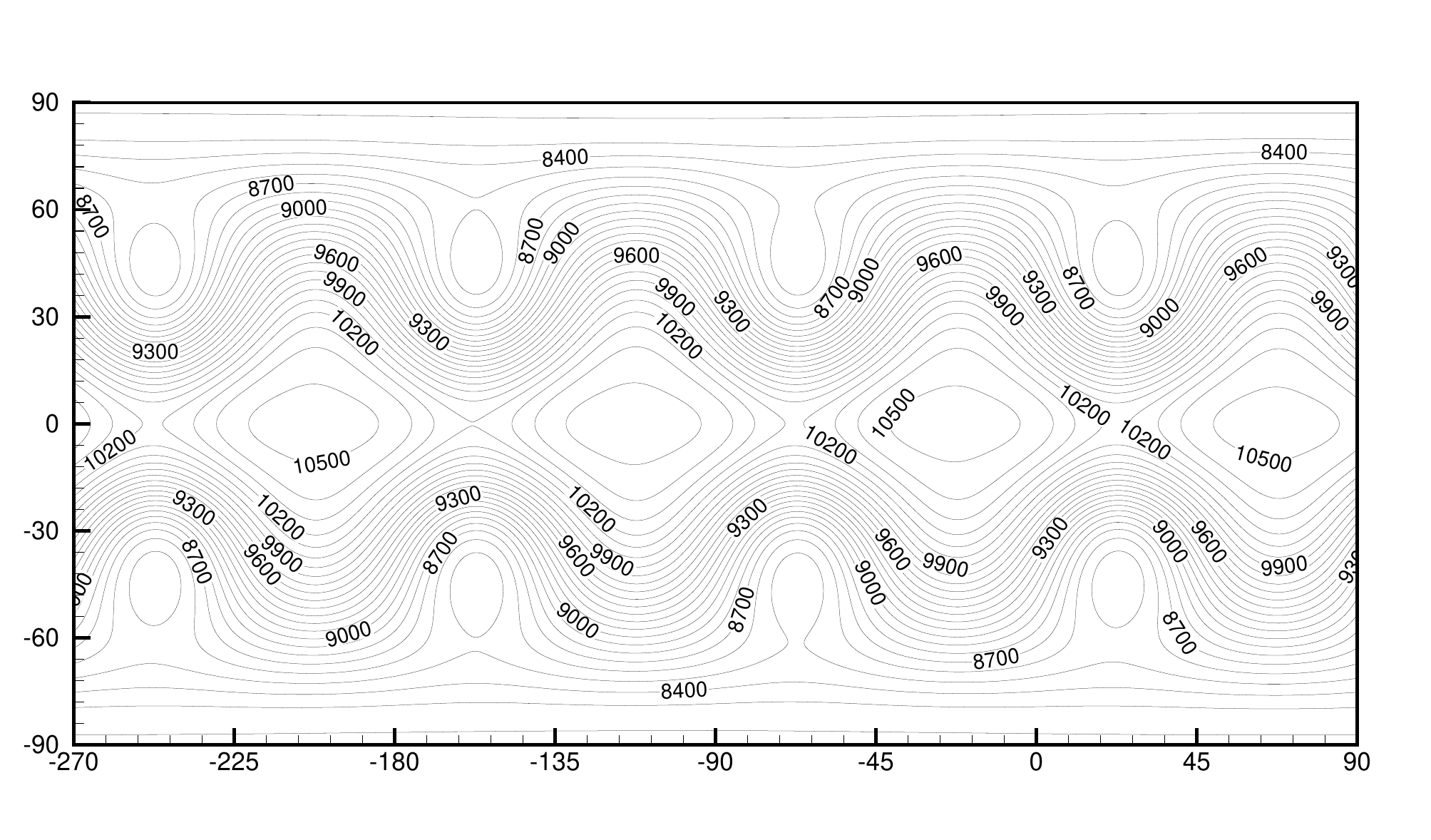}}
\subfigure[Yin-Yang grid ($N=108$)]
{\includegraphics[width=0.48\textwidth]{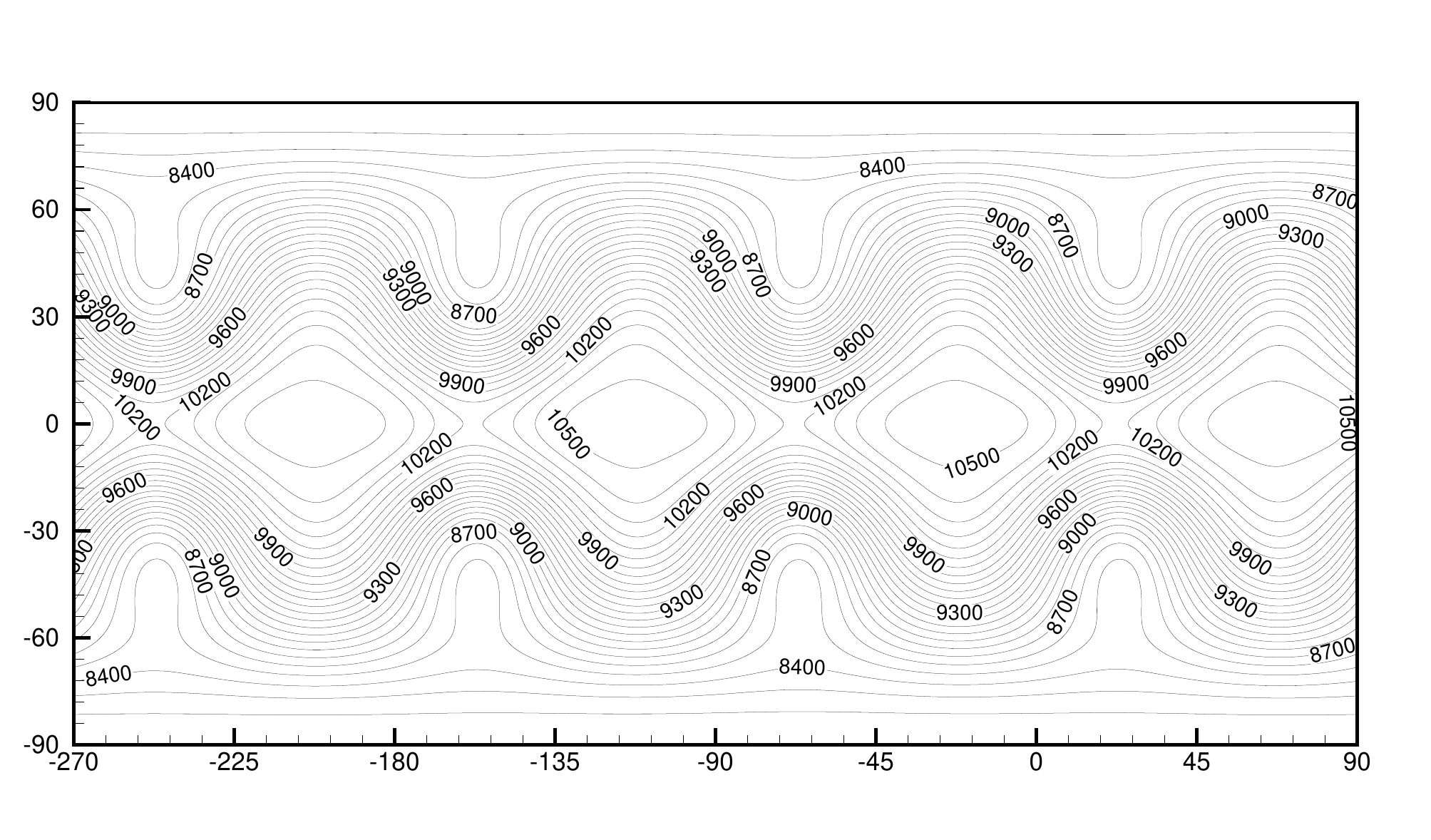}}
\subfigure[Cubed-sphere grid ($N=54$).]
{\includegraphics[width=0.48\textwidth]{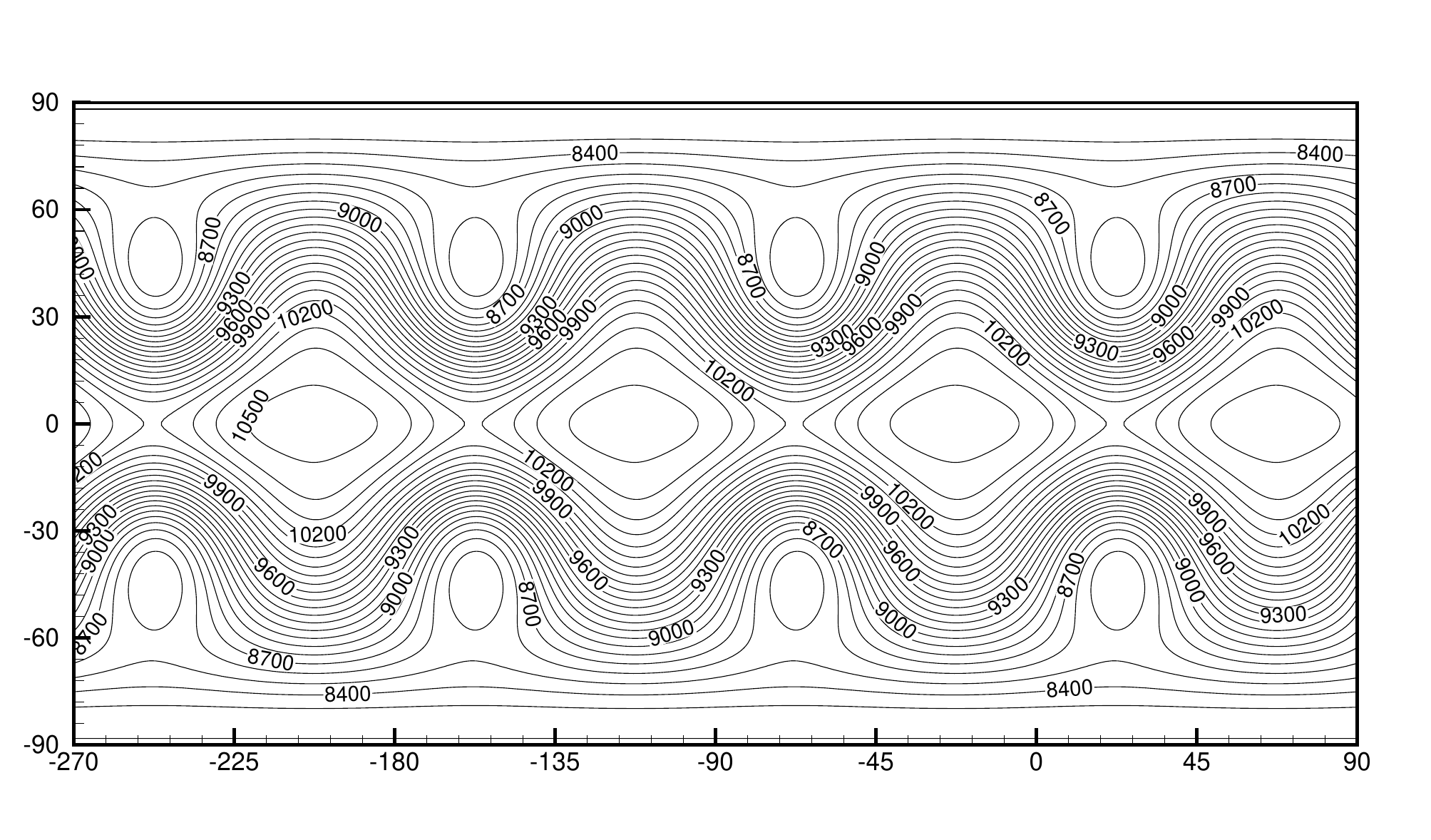}}
\subfigure[Cubed-sphere grid ($N=108$)]
{\includegraphics[width=0.48\textwidth]{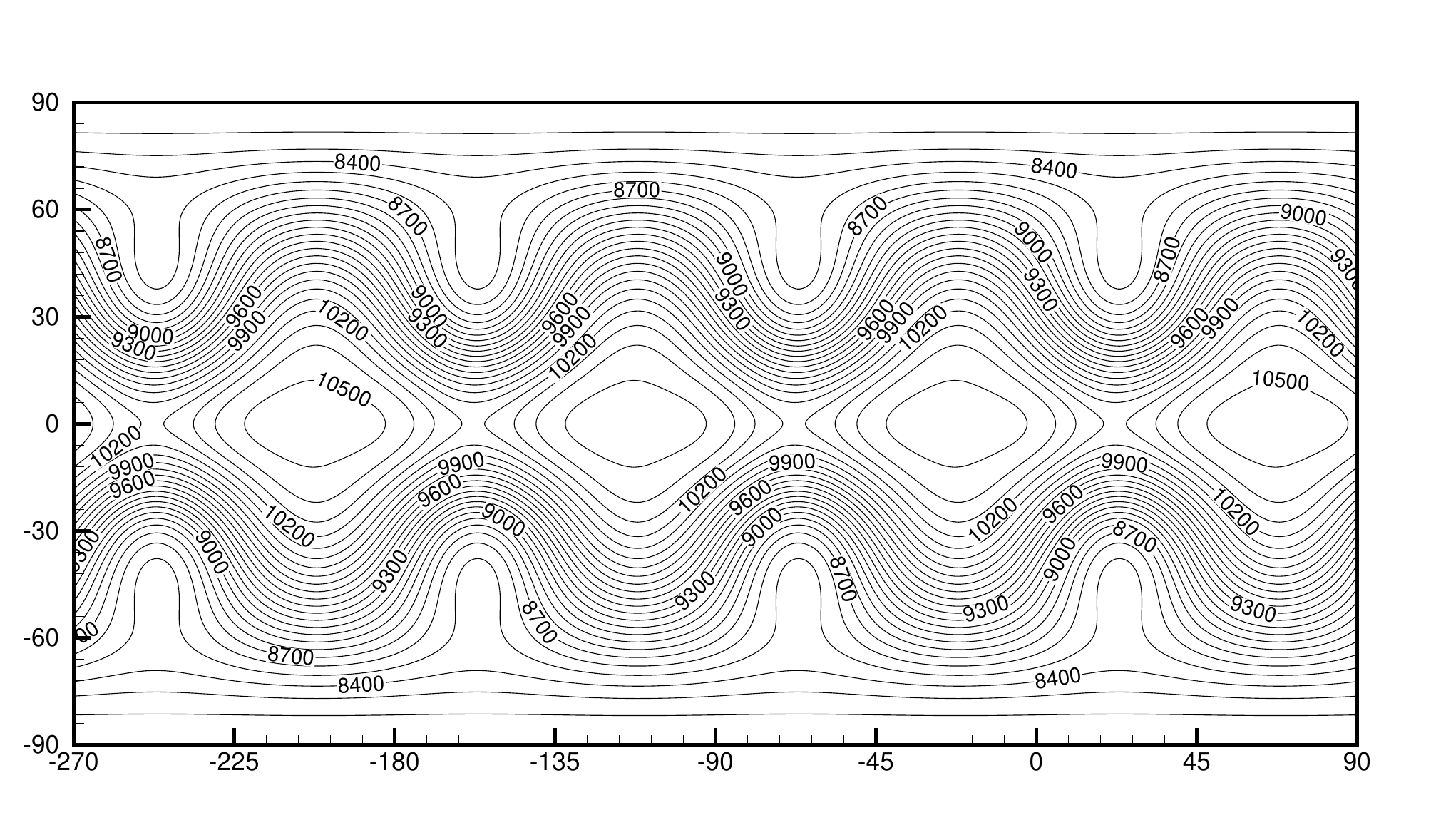}}
\subfigure[Icosahedral grid ($N=48$)]
{\includegraphics[width=0.48\textwidth]{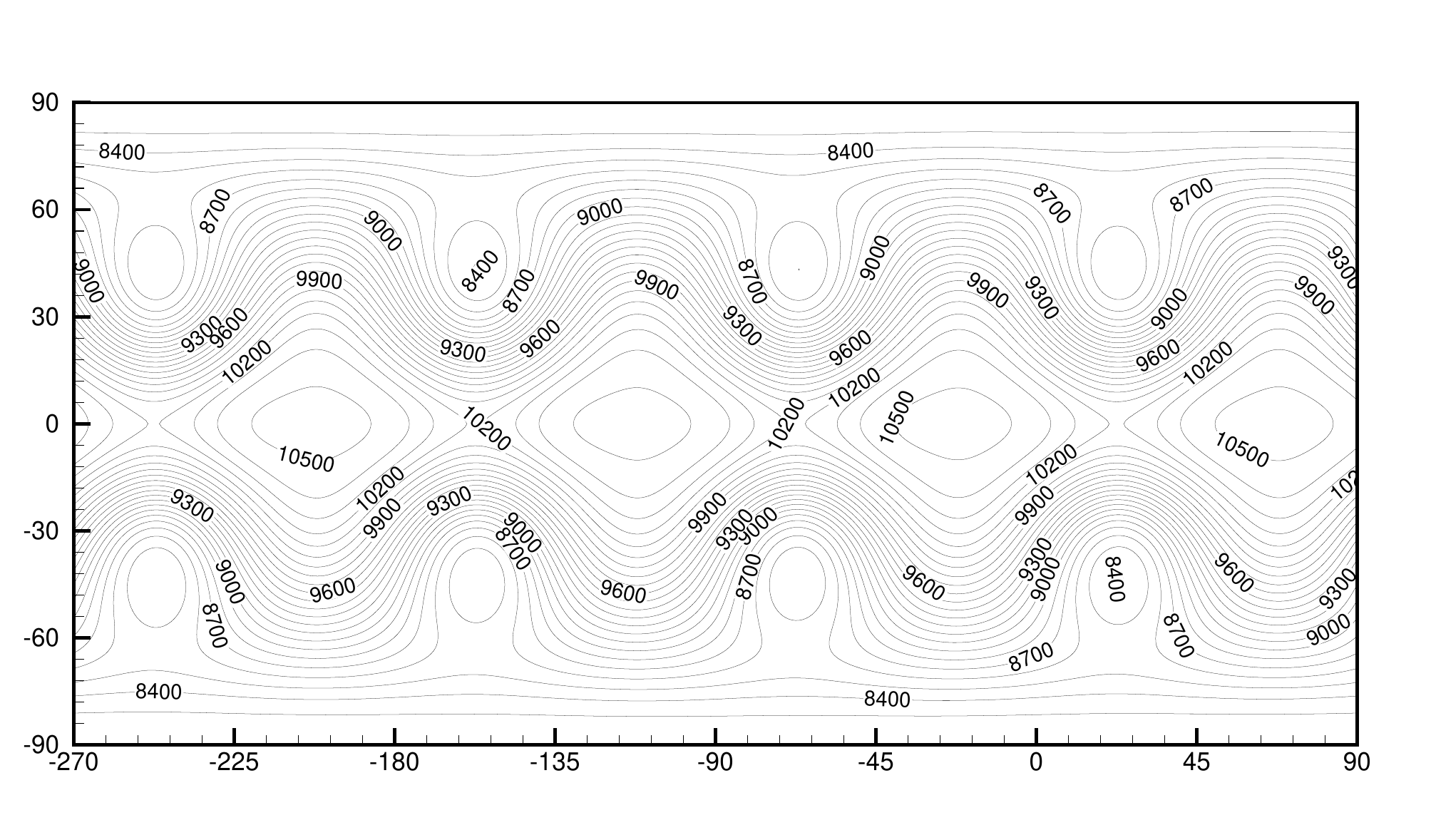}}
\subfigure[Icosahedral grid ($N=96$)]
{\includegraphics[width=0.48\textwidth]{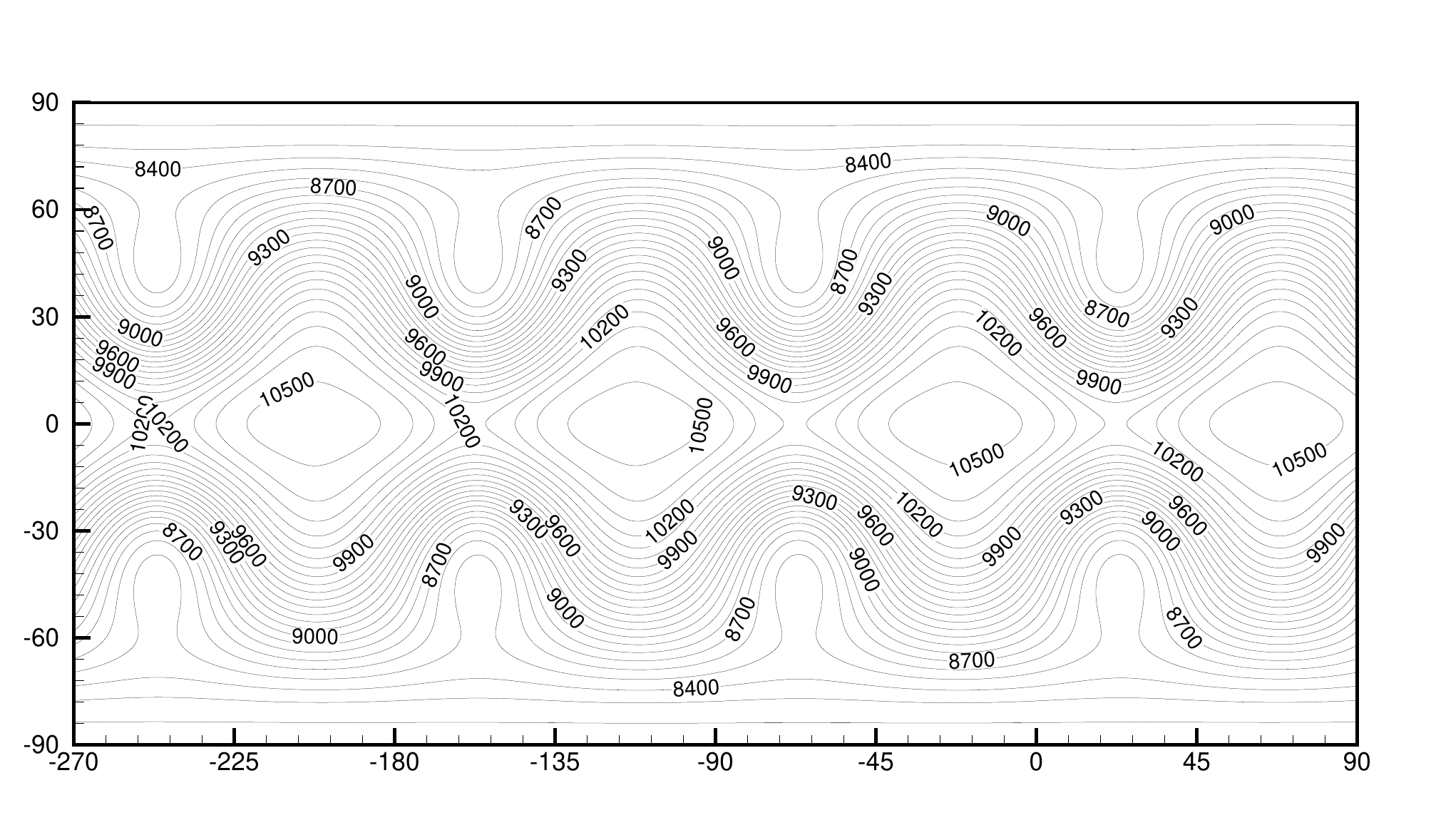}}
\end{center}
\caption{Numerical results of height fields at  day 14 of case 6.}\label{case6-height}
\end{figure}

Numerical solutions of the third order MCV model on three different grids with equivalent DOFs are shown in Fig.\ref{case6-height} for height fields at day 14. Numerical experiments were carried out with two sets of grid resolutions, i.e. $N=54$ and $108$ for Yin-Yang and cubed sphere grids, and  $N=48$ and $96$  for icosahedral hexagonal grid, equivalent to T42 and T63 respectively for the  spectral transform solutions.  From Fig.\ref{case6-height}, it is observed that the numerical results of the presented models look very similar to the spectral transform solutions at different levels of grid resolution (see Fig.5.7 in \cite{chien95}). It is found that with a refined grid, the numerical solutions of all three grids converge to the reference solution. 

We have examined the conservation of total energy and potential enstrophy. The total energy errors in the results of Yin-Yang and cubed sphere grids with $N=40$ and icosahedral grid with  $N=36$ are comparable to the spectral transform solution on grid T42.
  
\subsubsection{Barotropic jet flow}

A barotropic jet flow test was designed in \cite{gal04} to evaluate the effect of the spherical grids on numerical solutions. 

\begin{figure} [htbp]
\begin{center}
\subfigure[Yin-Yang grid ($N=45$).]
{\includegraphics[width=0.48\textwidth]{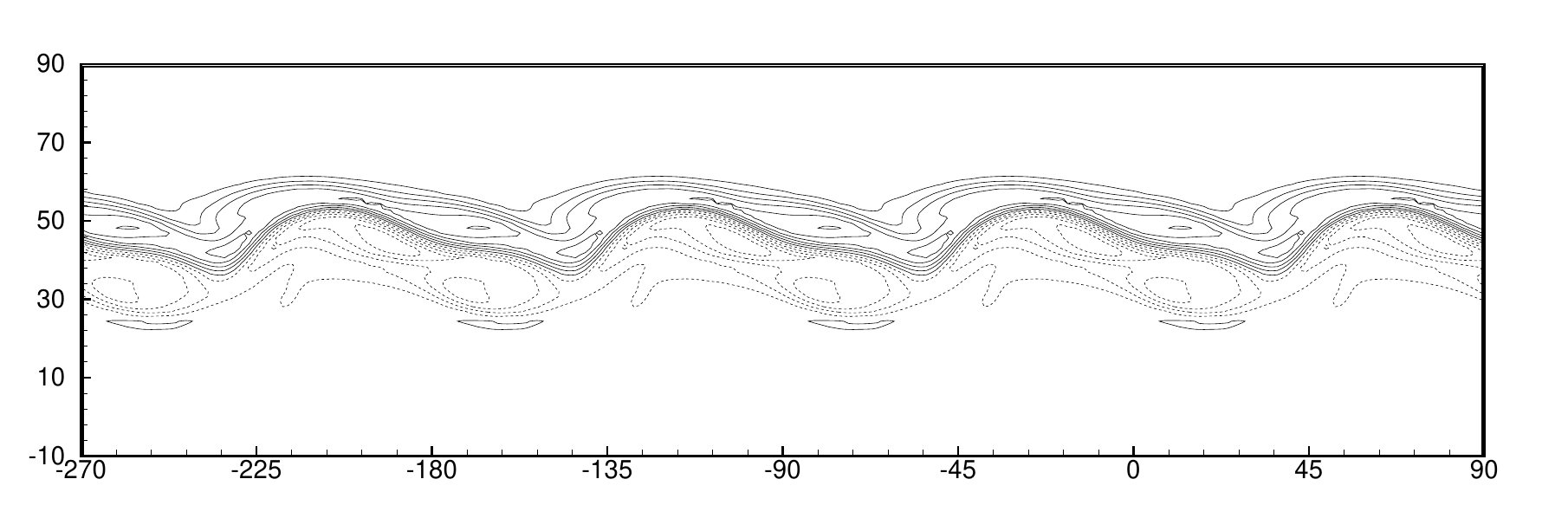}}
\subfigure[Yin-Yang grid ($N=180$)]
{\includegraphics[width=0.48\textwidth]{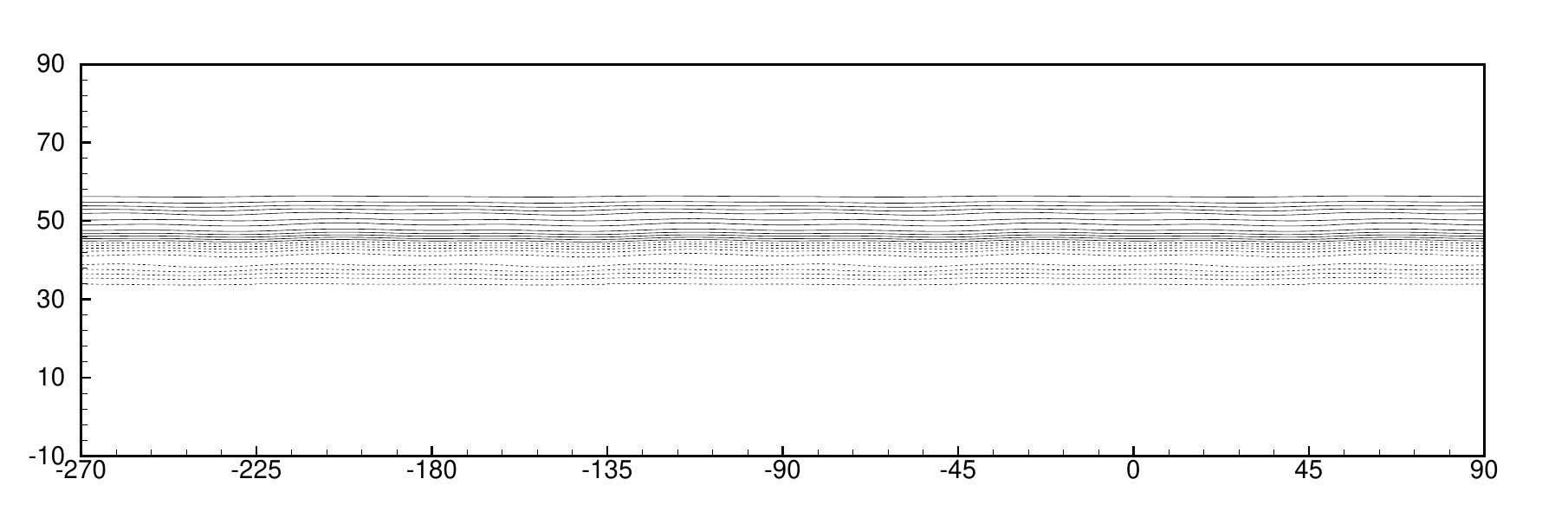}}
\subfigure[Cubed-sphere grid ($N=45$).]
{\includegraphics[width=0.48\textwidth]{./Figures/balanced_jet_vor_N45.pdf}}
\subfigure[Cubed-sphere grid ($N=180$)]
{\includegraphics[width=0.48\textwidth]{./Figures/balanced_jet_vor_N180.pdf}}
\subfigure[Icosahedral grid ($N=40$)]
{\includegraphics[width=0.48\textwidth]{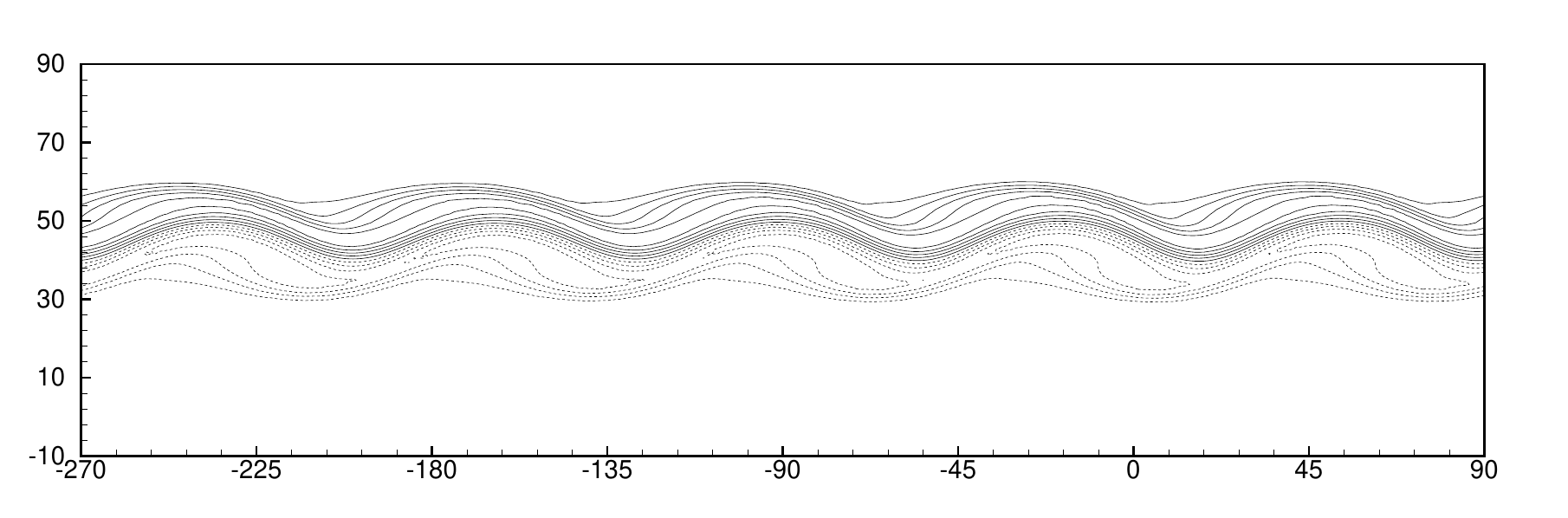}}
\subfigure[Icosahedral grid ($N=160$)]
{\includegraphics[width=0.48\textwidth]{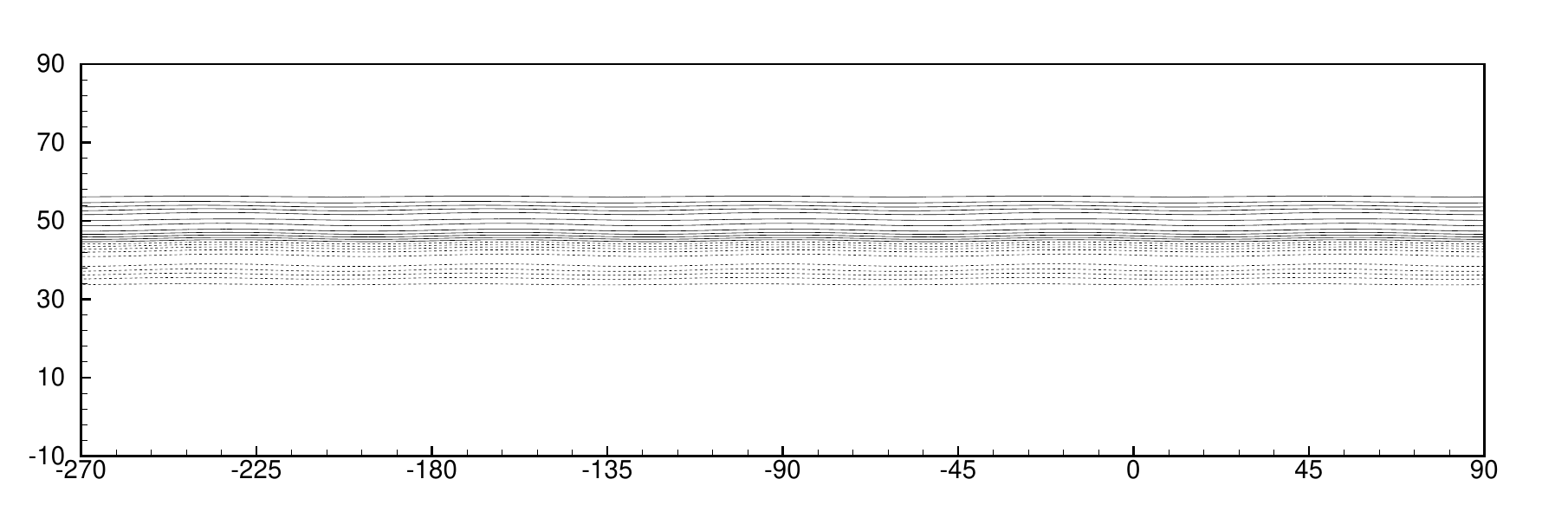}}
\end{center}
\caption{Numerical results at day five of vorticity field for the balanced case without initial perturbations.}\label{case7-height}
\end{figure}

We simulated the balanced  setup without any initial perturbation in the physical fields. Numerical results of relative vorticity field at day five are shown in Fig.\ref{case7-height} for three grids with different resolutions. On coarse grids,  $N=45$ is for Yin-Yang and cubed sphere grids and $N=40$ for icosahedral hexagonal grid,  instability develops significantly due to numerical errors. Yin-Yang grid appears to be the most erroneous probably because of the data transfer schemes for the overset boundaries between Yin and Yang components. Cubed sphere and icosahedral hexagonal grids produce much better results. A noticeable 4-wave pattern is observed for cubed sphere grid, which may corresponds to the 4 patch boundaries along the flow path. Similarly,  a 5-wave pattern is visible for the icosahedral grid mainly due to the extra numerical errors caused by five internal boundaries between different primary triangles of the spherical icosahedron. When the grids are refined,  i.e. $N=180$ for Yin-Yang and cubed sphere grids and $N=160$ for icosahedral hexagonal grid, the dynamic balance is preserved satisfactorily for all spherical grids. 
 
\begin{figure} [htbp]
\begin{center}
\subfigure[Relative vorticity field on grid $N=45$]
{\includegraphics[width=0.6\textwidth]{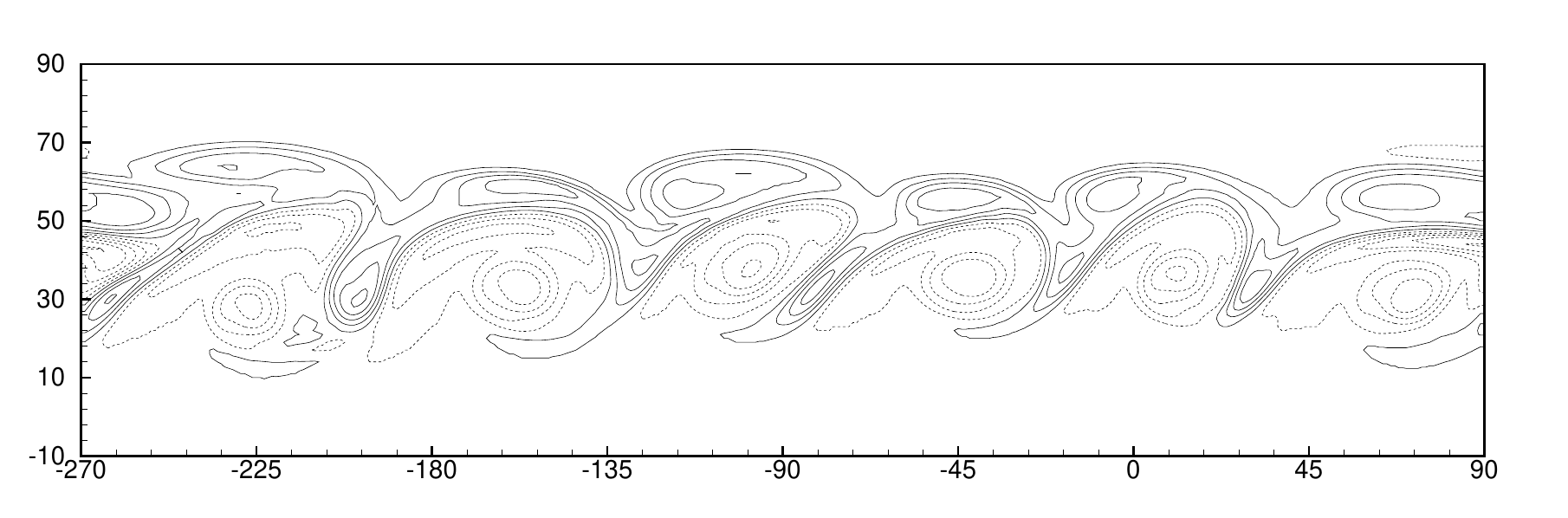}}\\
\subfigure[Relative vorticity field on grid $N=90$]
{\includegraphics[width=0.6\textwidth]{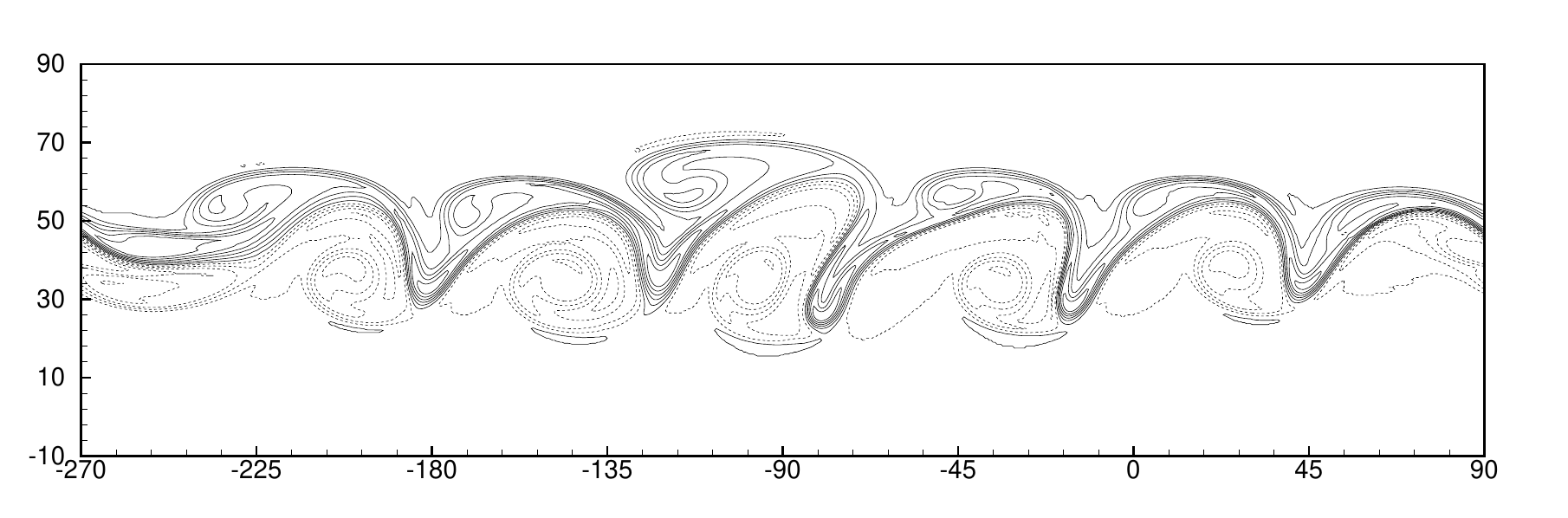}}\\
\subfigure[Relative vorticity field on grid $N=135$]
{\includegraphics[width=0.6\textwidth]{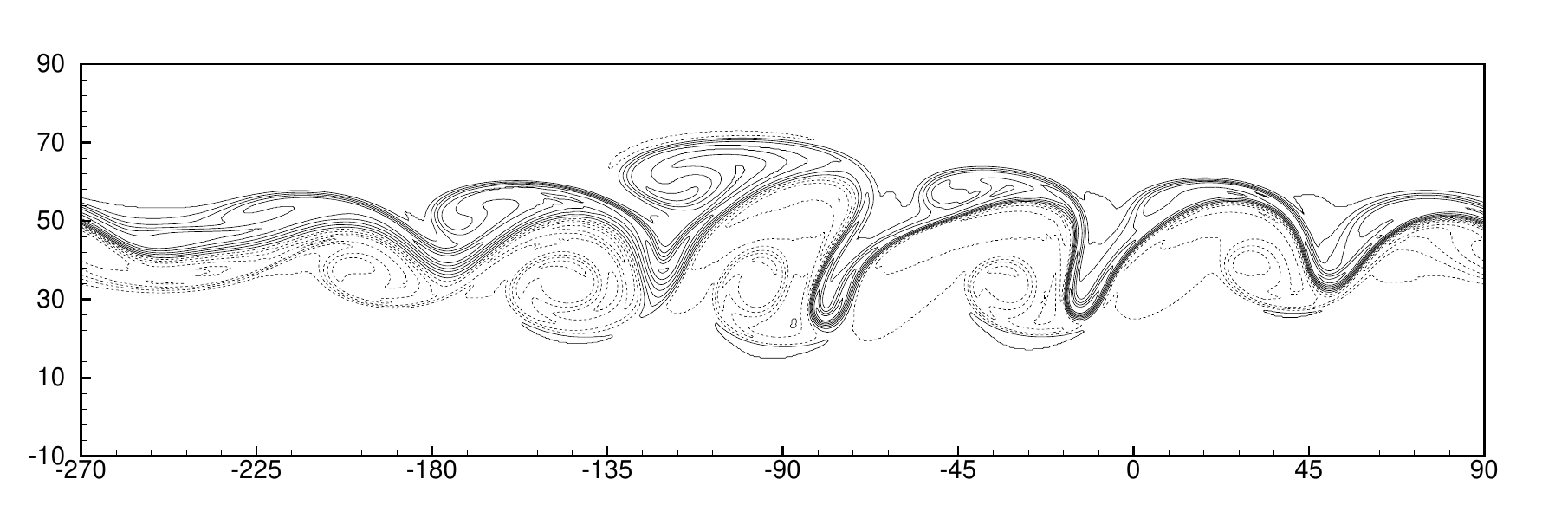}}\\
\subfigure[Relative vorticity field on grid $N=180$]
{\includegraphics[width=0.6\textwidth]{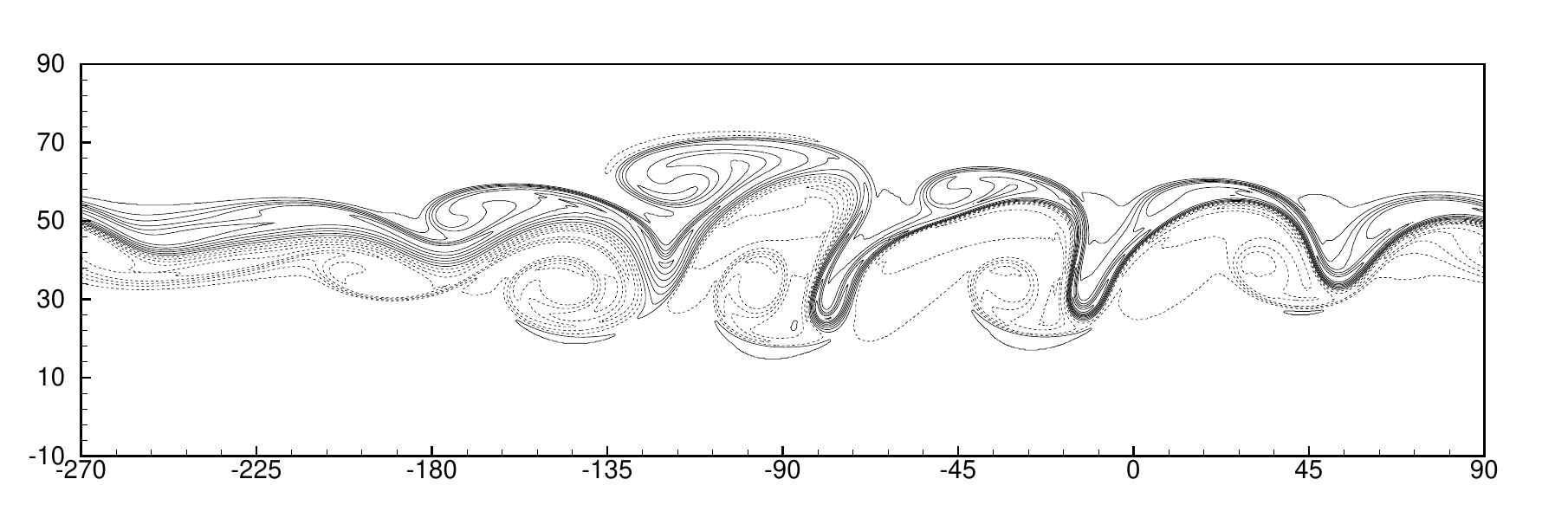}}\\
\end{center}
\caption{Numerical results at day six of vorticity field of perturbed case on Yin-Yang grid.}\label{case7-vor-yy}
\end{figure}

\begin{figure} [htbp]
\begin{center}
\subfigure[Relative vorticity field on grid $N=45$]
{\includegraphics[width=0.6\textwidth]{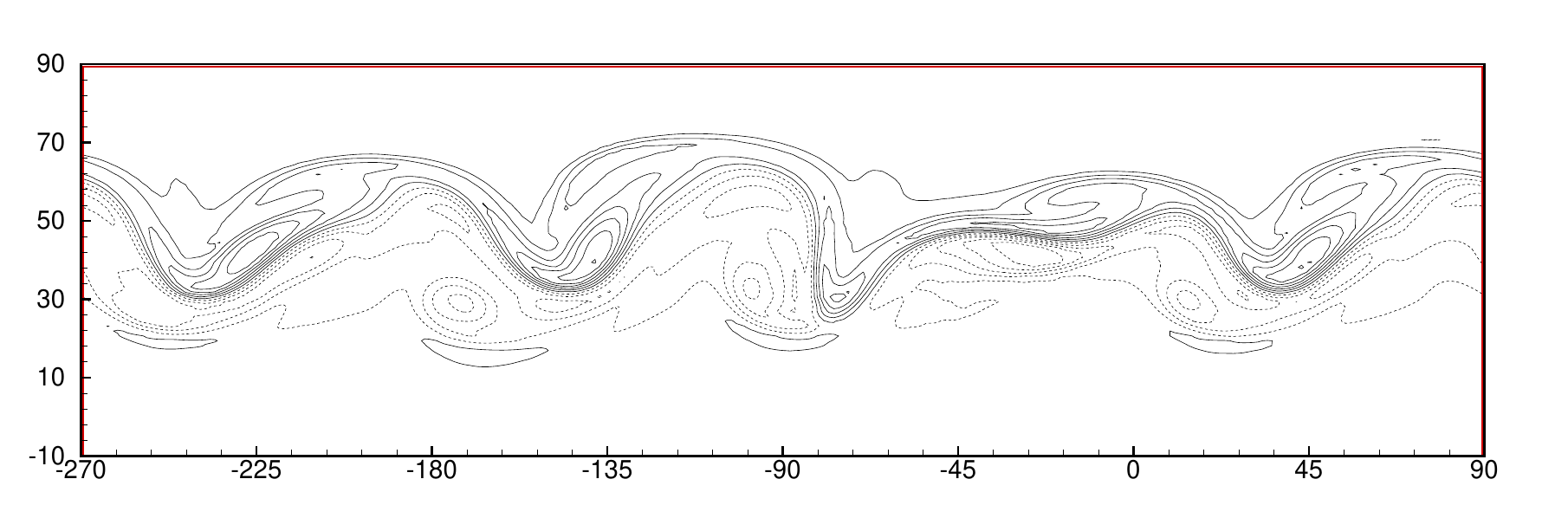}}\\
\subfigure[Relative vorticity field on grid $N=90$]
{\includegraphics[width=0.6\textwidth]{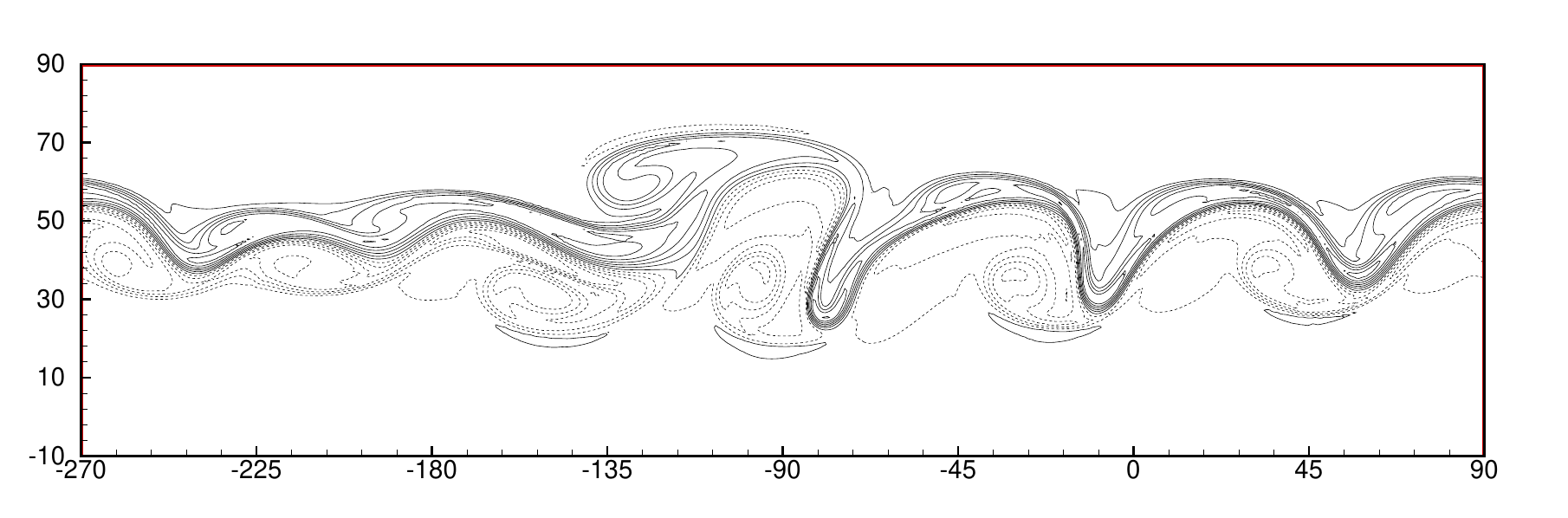}}\\
\subfigure[Relative vorticity field on grid $N=135$]
{\includegraphics[width=0.6\textwidth]{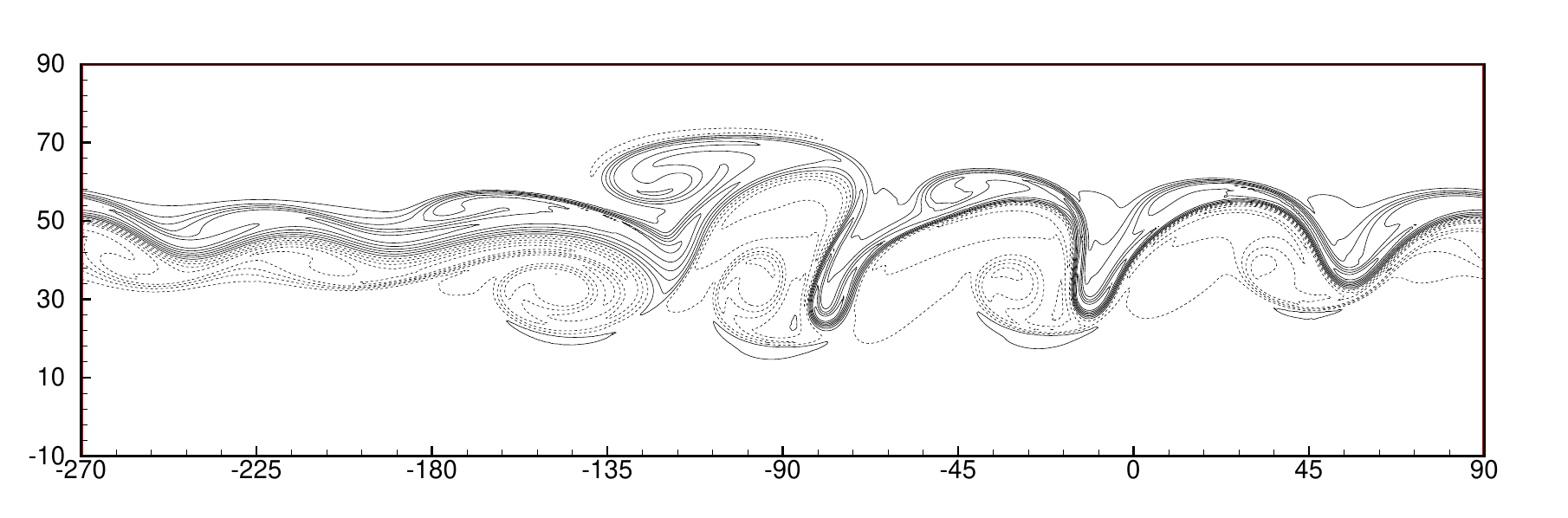}}\\
\subfigure[Relative vorticity field on grid $N=180$]
{\includegraphics[width=0.6\textwidth]{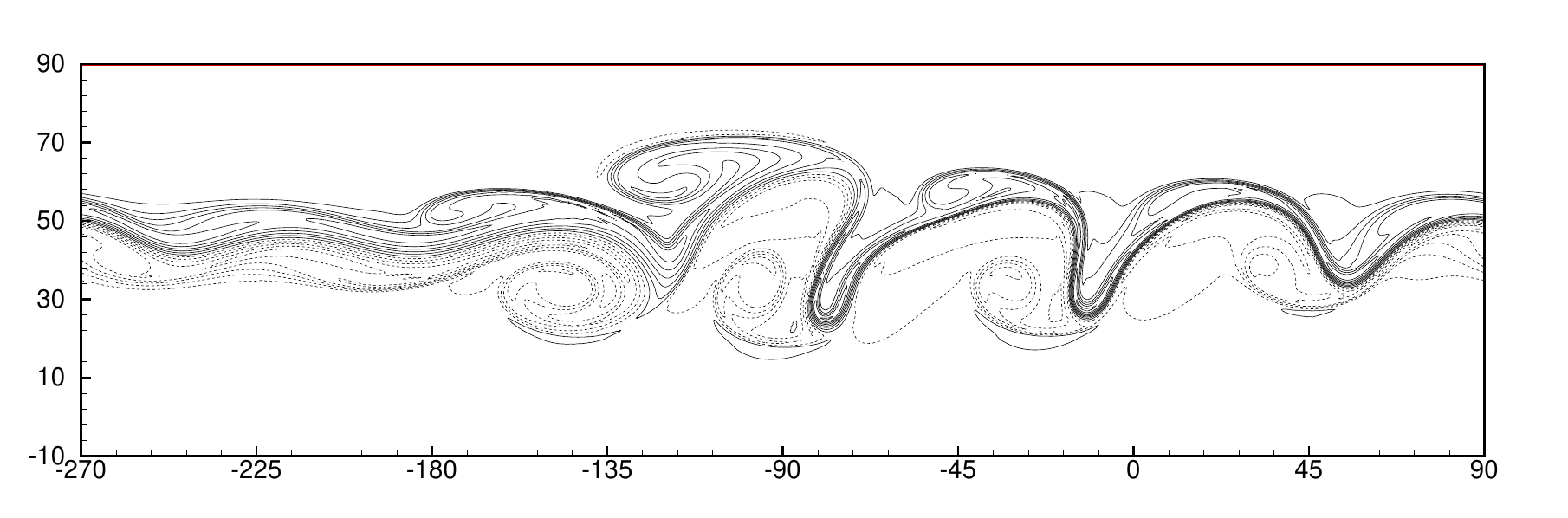}}\\
\end{center}
\caption{Numerical results at day six of vorticity filed of perturbed case on cubed-sphere grid.}\label{case7-vor-cs}
\end{figure}

\begin{figure} [htbp]
\begin{center}
\subfigure[Relative vorticity field on grid $N=40$]
{\includegraphics[width=0.6\textwidth]{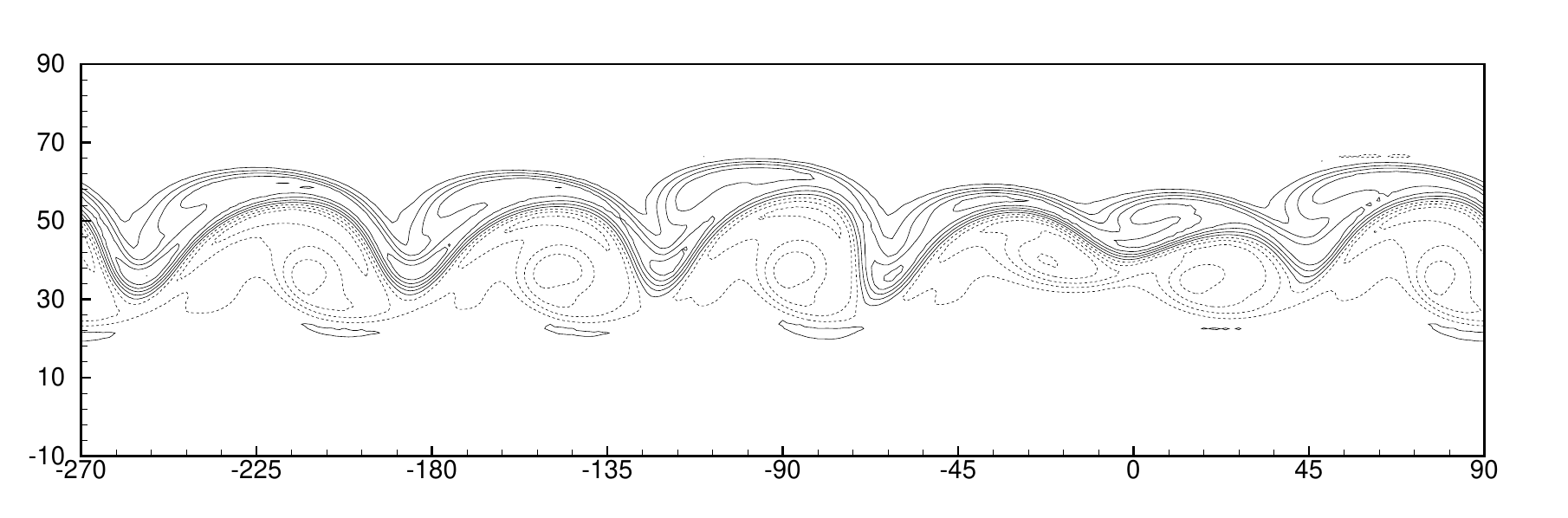}}\\
\subfigure[Relative vorticity field on grid $N=80$]
{\includegraphics[width=0.6\textwidth]{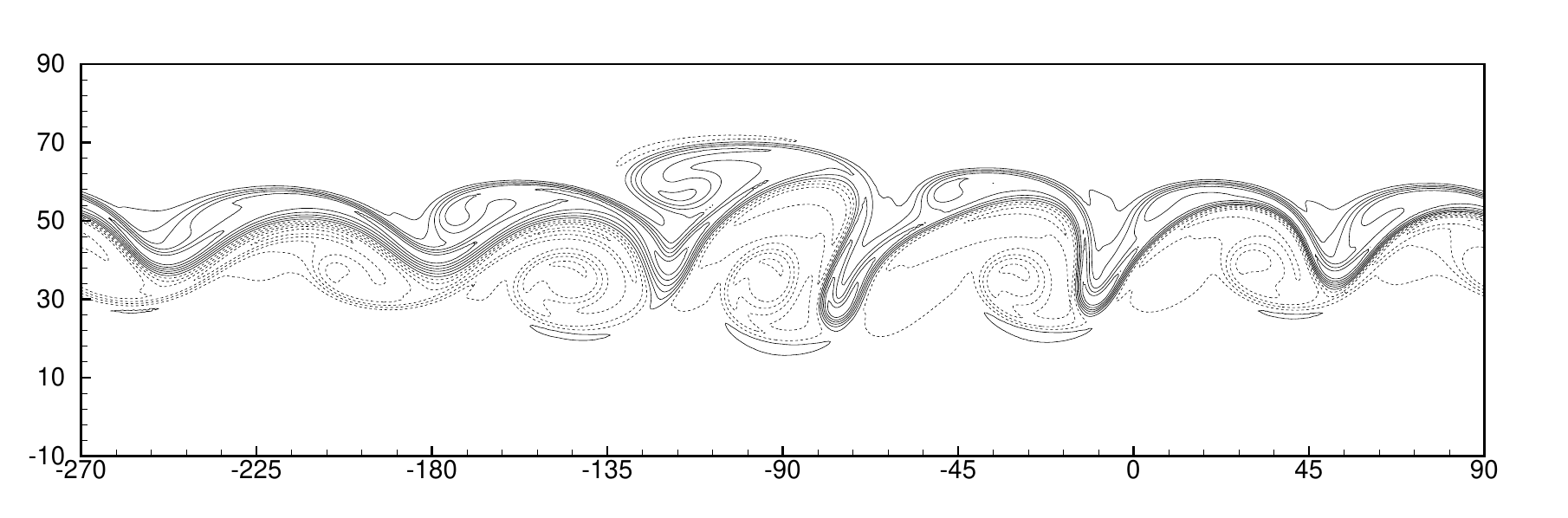}}\\
\subfigure[Relative vorticity field on grid $N=120$]
{\includegraphics[width=0.6\textwidth]{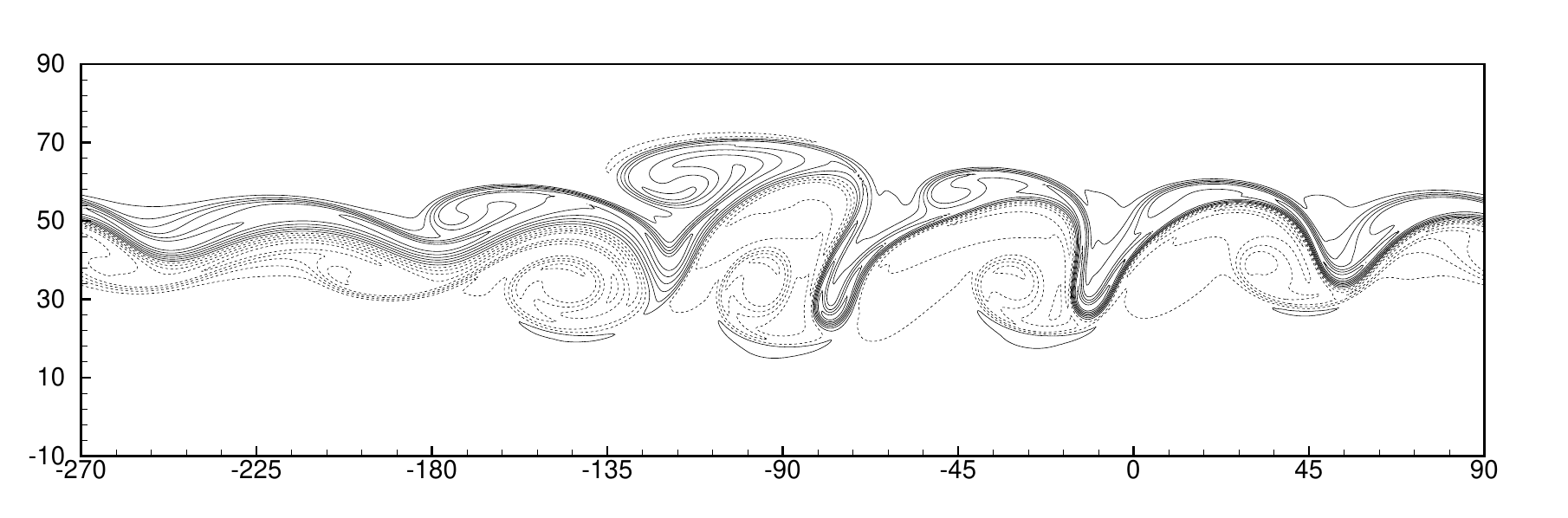}}\\
\subfigure[Relative vorticity field on grid $N=160$]
{\includegraphics[width=0.6\textwidth]{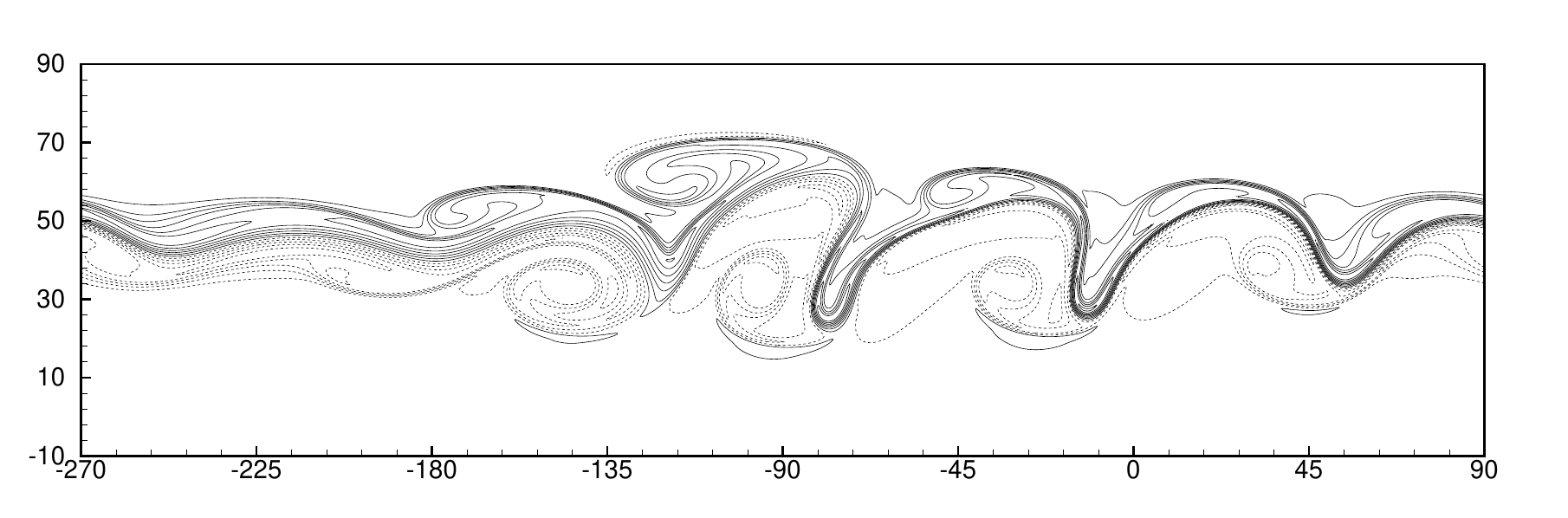}}\\
\end{center}
\caption{Numerical results at day six of vorticity filed of perturbed case on icosahedral-hexagonal grid.}\label{case7-vor-ih}
\end{figure}

We also experimented with initial condition which has a perturbation to height field to see how the barotropic instability develops from the jet flow. We run on three spherical grids with gradually refined  grids to check the convergence of relative vorticity fields at day 6. The reference solution is given in \cite{gal04} using spectral transform model on fine grid (T341). Numerical results of relative vorticity field at the 6th day are shown in Figs.\ref{case7-vor-yy}, \ref{case7-vor-cs} and \ref{case7-vor-ih} separately. All models can get the converged solution when the grids are refined. Cubed sphere and icosahedral grids converge more rapidly compared to Yin-Yang grid.  

\subsubsection{Williamson's test case 7: a ``real case" simulation}

This case starts from an analyzed height and wind fields on 500mb  of Dec. 21, 1978 as the initial conditions.  It  provides an example of actual flow to evaluate the performance of SWE models on different grids \cite{qaddouri2011}. Since the analytical solution doesn't exist in this test, the spectral transform solution on T213 grid is adopted in this study as the reference solution to check the performance of MCV models. 

\begin{figure}[tbhp]
\begin{center}
\subfigure[Reference solution]
{\includegraphics[width=0.49\textwidth]{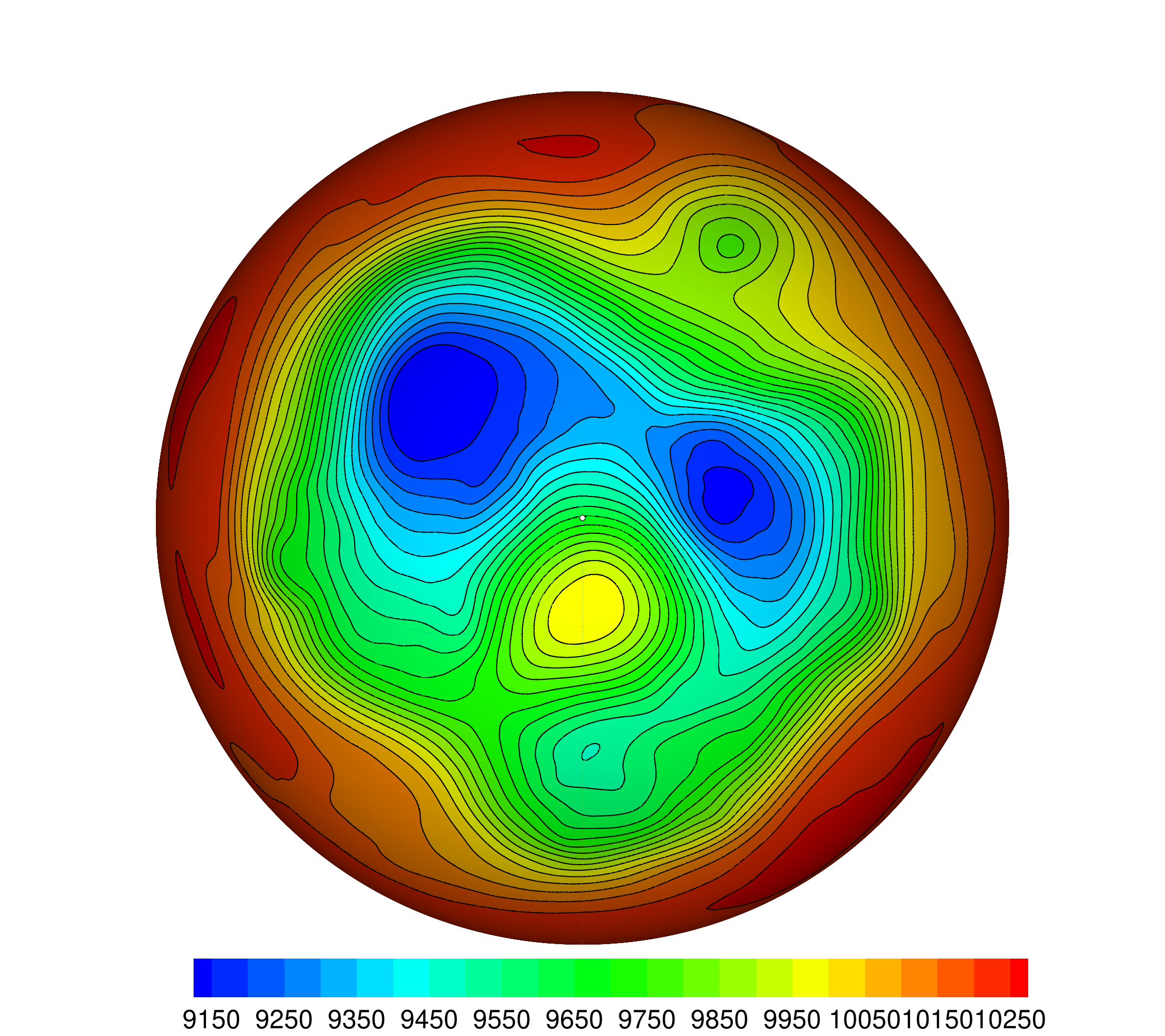} }
\subfigure[Yin-Yang grid]
{\includegraphics[width=0.49\textwidth]{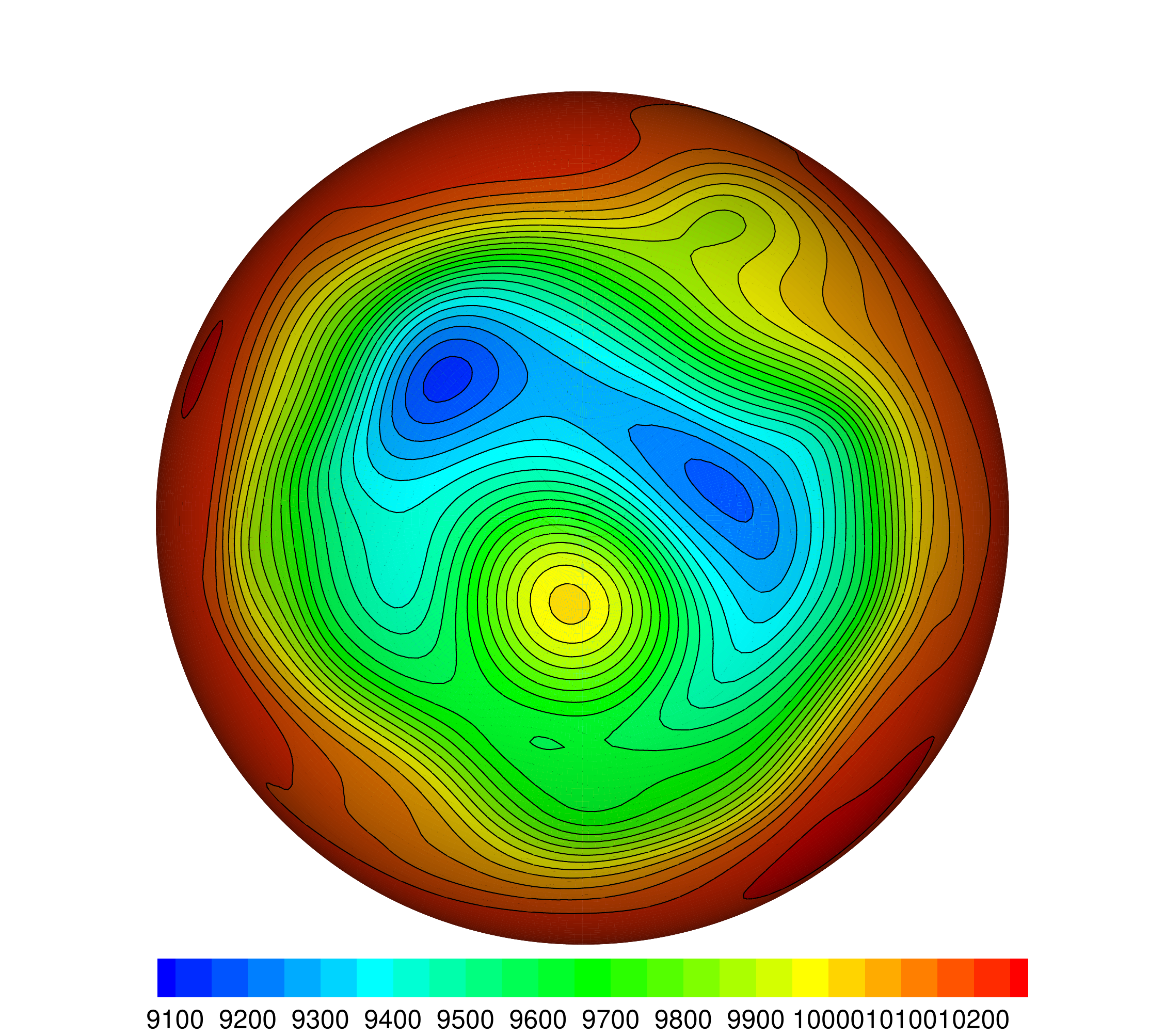} } \\
\subfigure[Cubed-sphere grid]
{\includegraphics[width=0.49\textwidth]{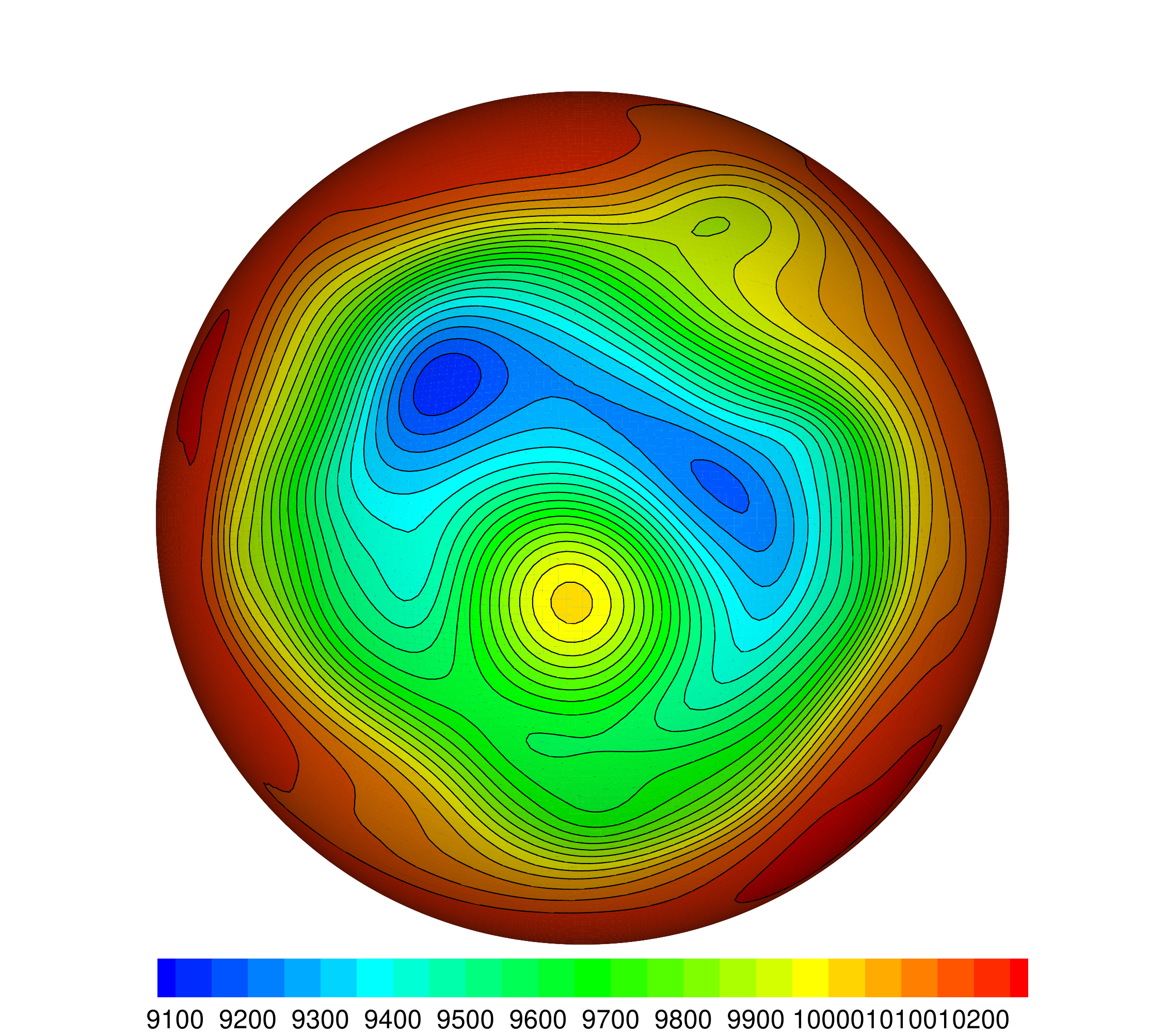} }
\subfigure[Icosahedral grid]
{\includegraphics[width=0.49\textwidth]{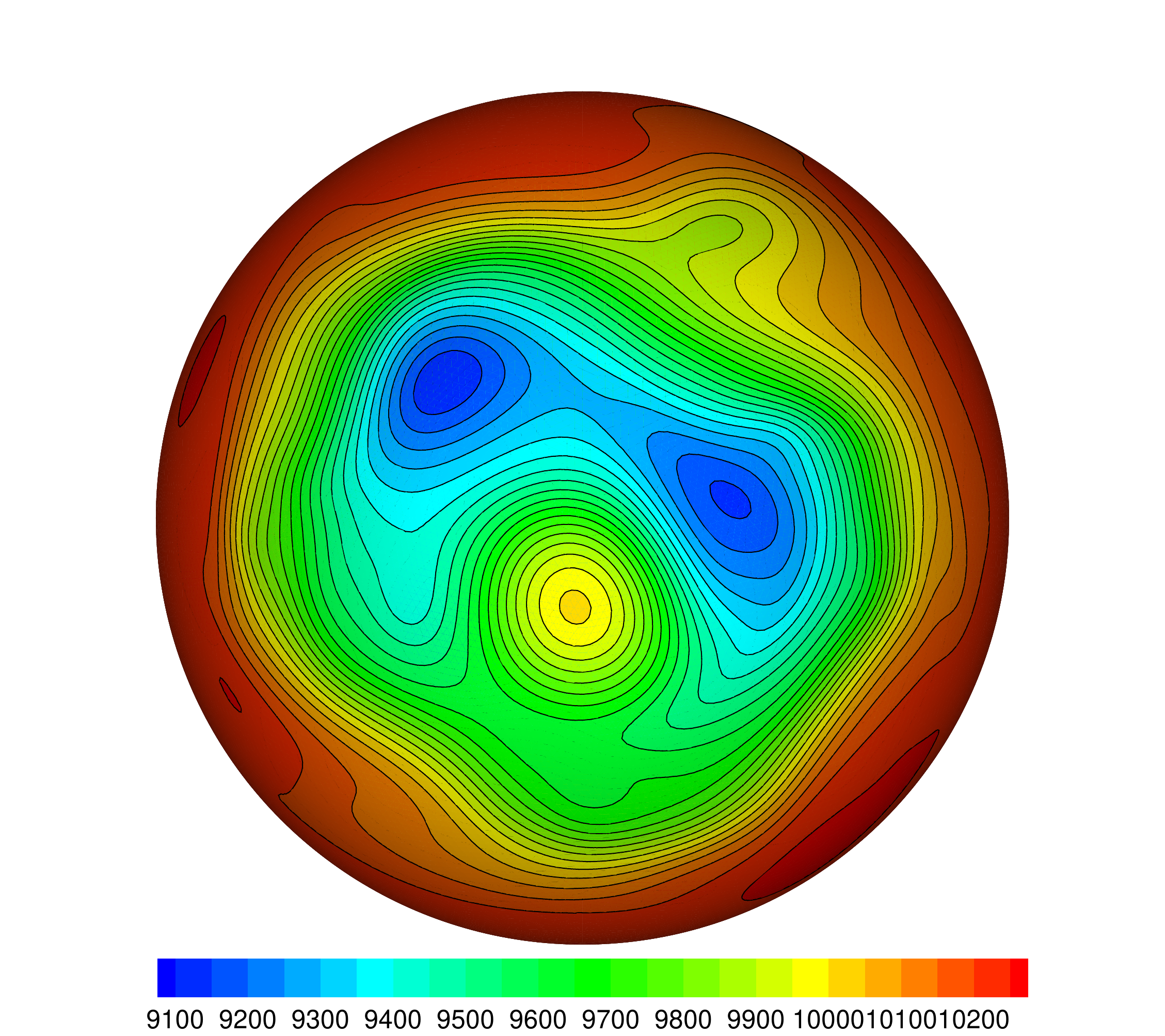} }
\end{center}
\caption{Numerical results of height field of Williamson's test case 7 at day 5. }\label{case7h}
\end{figure}

\begin{figure}[tbhp]
\begin{center}
\subfigure[Yin-Yang grid]
{\includegraphics[width=0.32\textwidth]{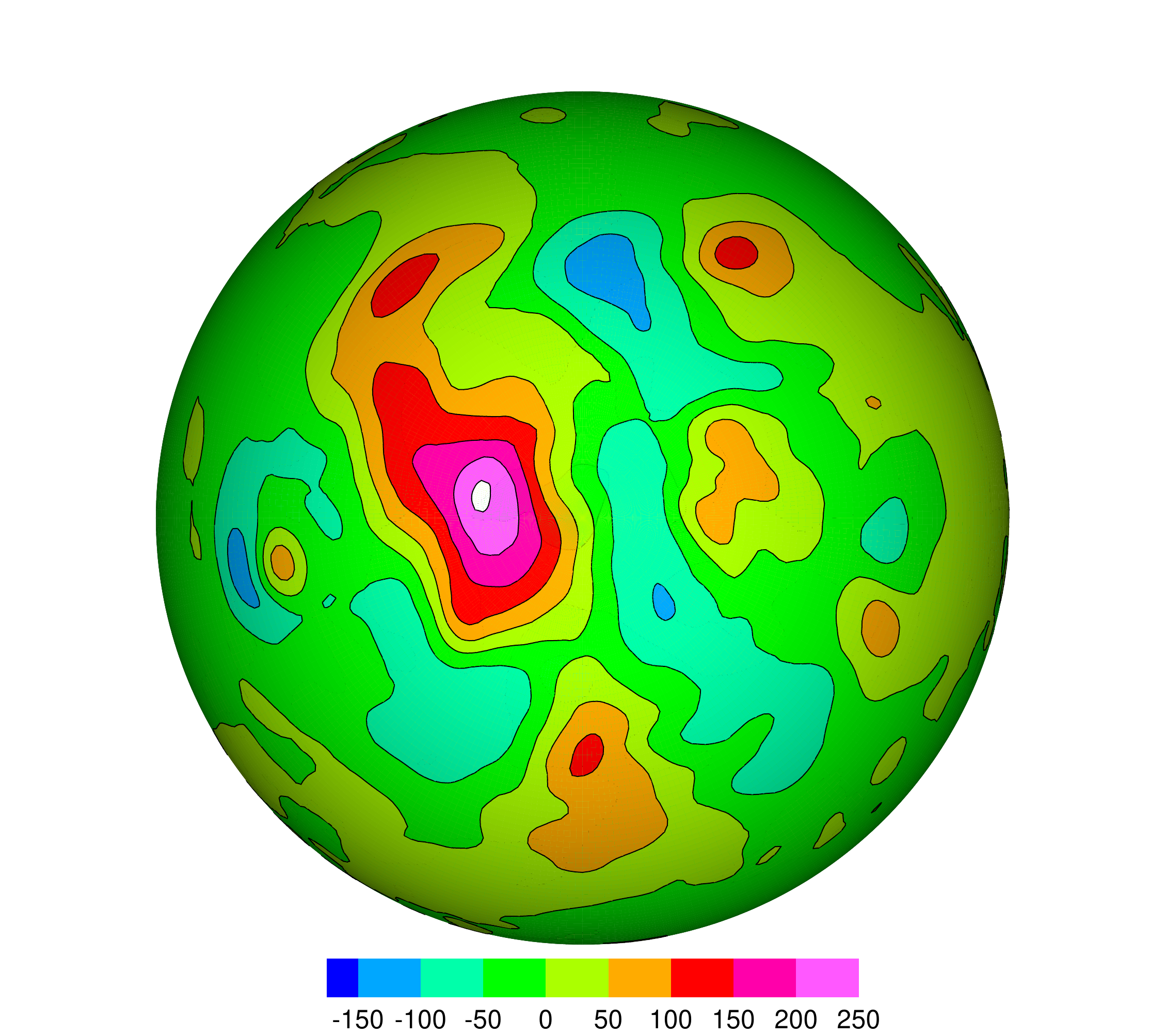} }
\subfigure[Cubed-sphere grid]
{\includegraphics[width=0.32\textwidth]{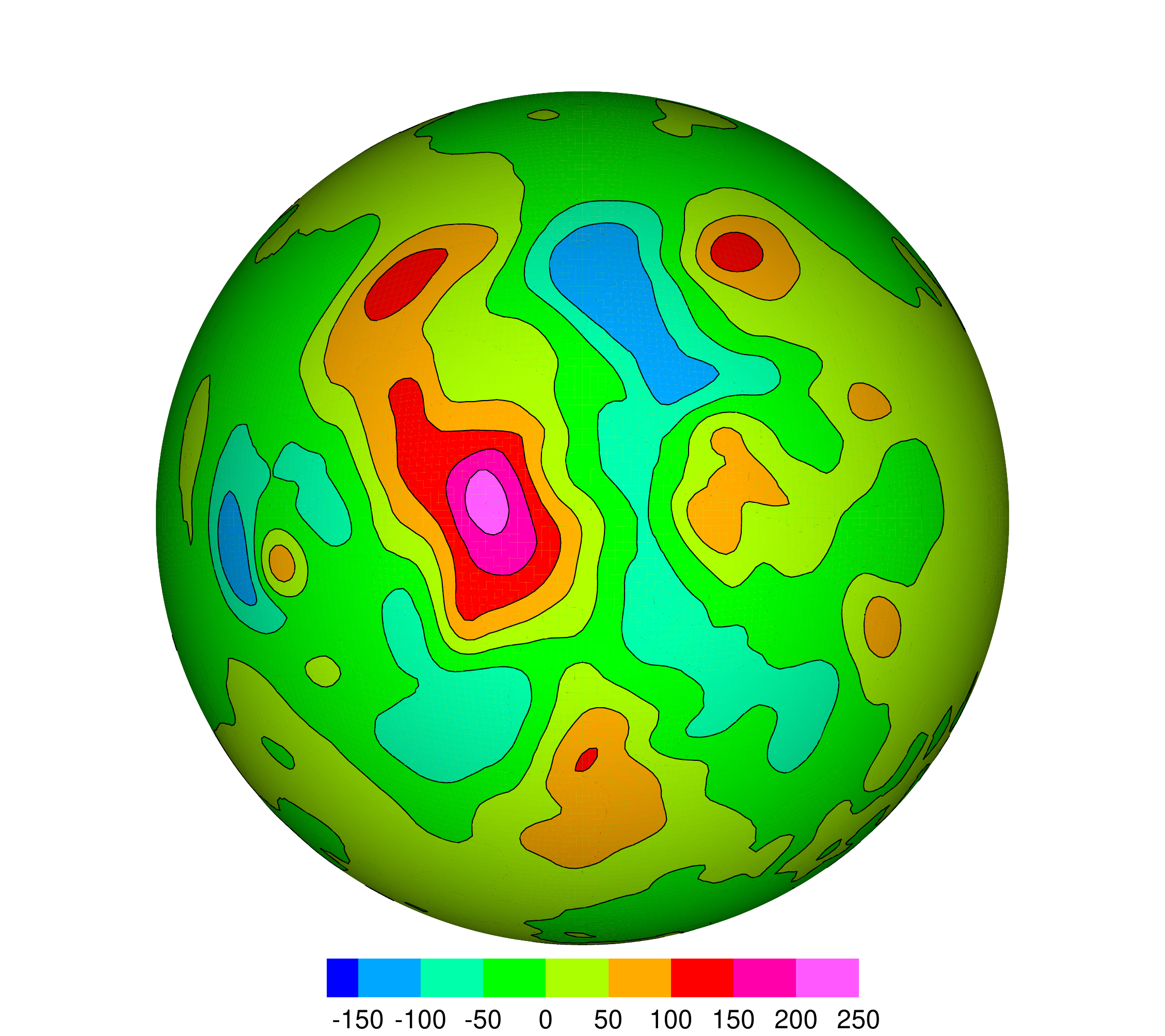} }
\subfigure[Icosahedral grid]
{\includegraphics[width=0.32\textwidth]{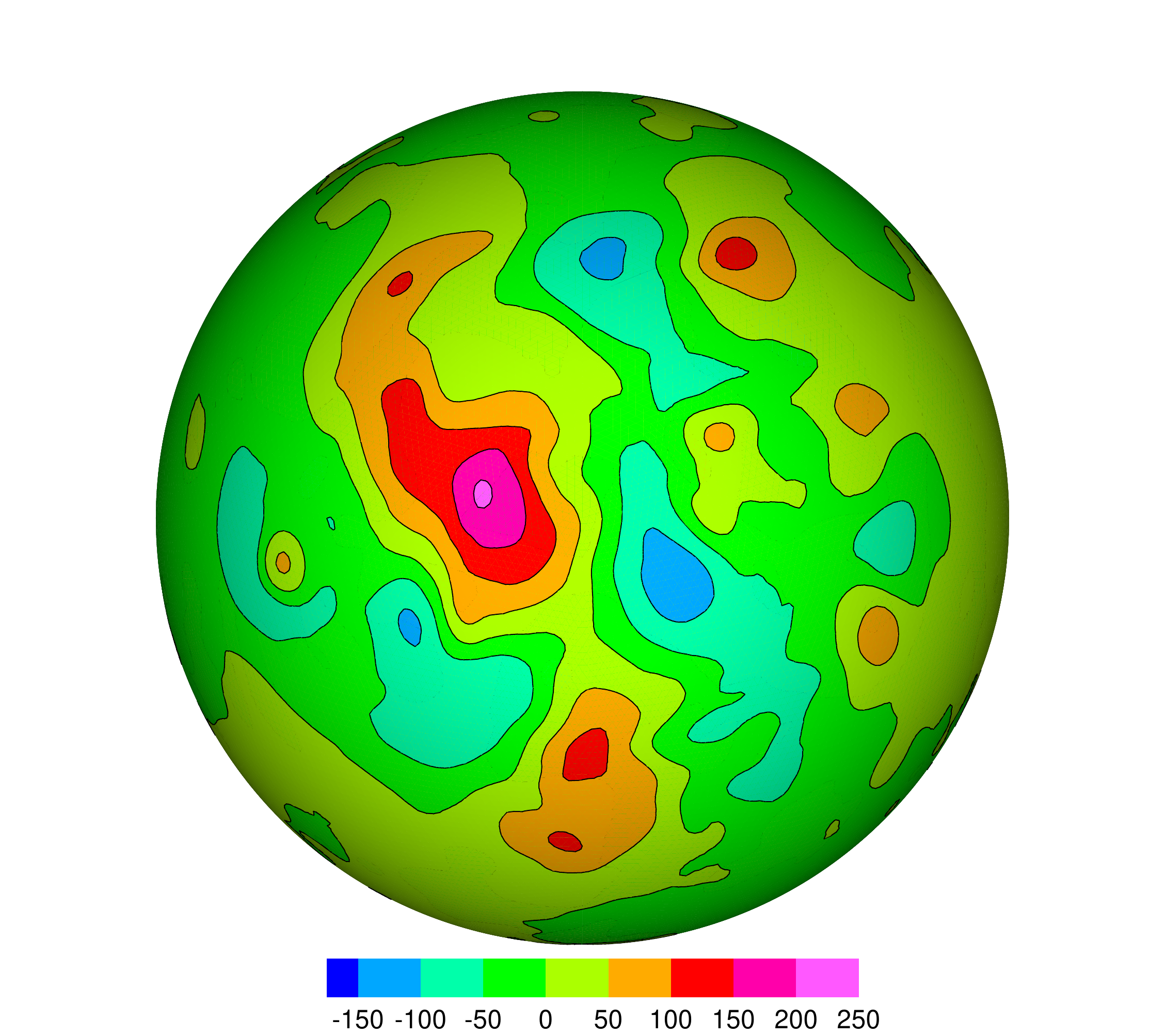}}
\end{center}
\caption{ Numerical errors in height fields shown in Fig.\ref{case7h}.  }\label{case7h-e}
\end{figure}

\begin{figure}[htbp]
\begin{center}
\subfigure[$l_1$ error]
{\includegraphics[width=0.33\textwidth]{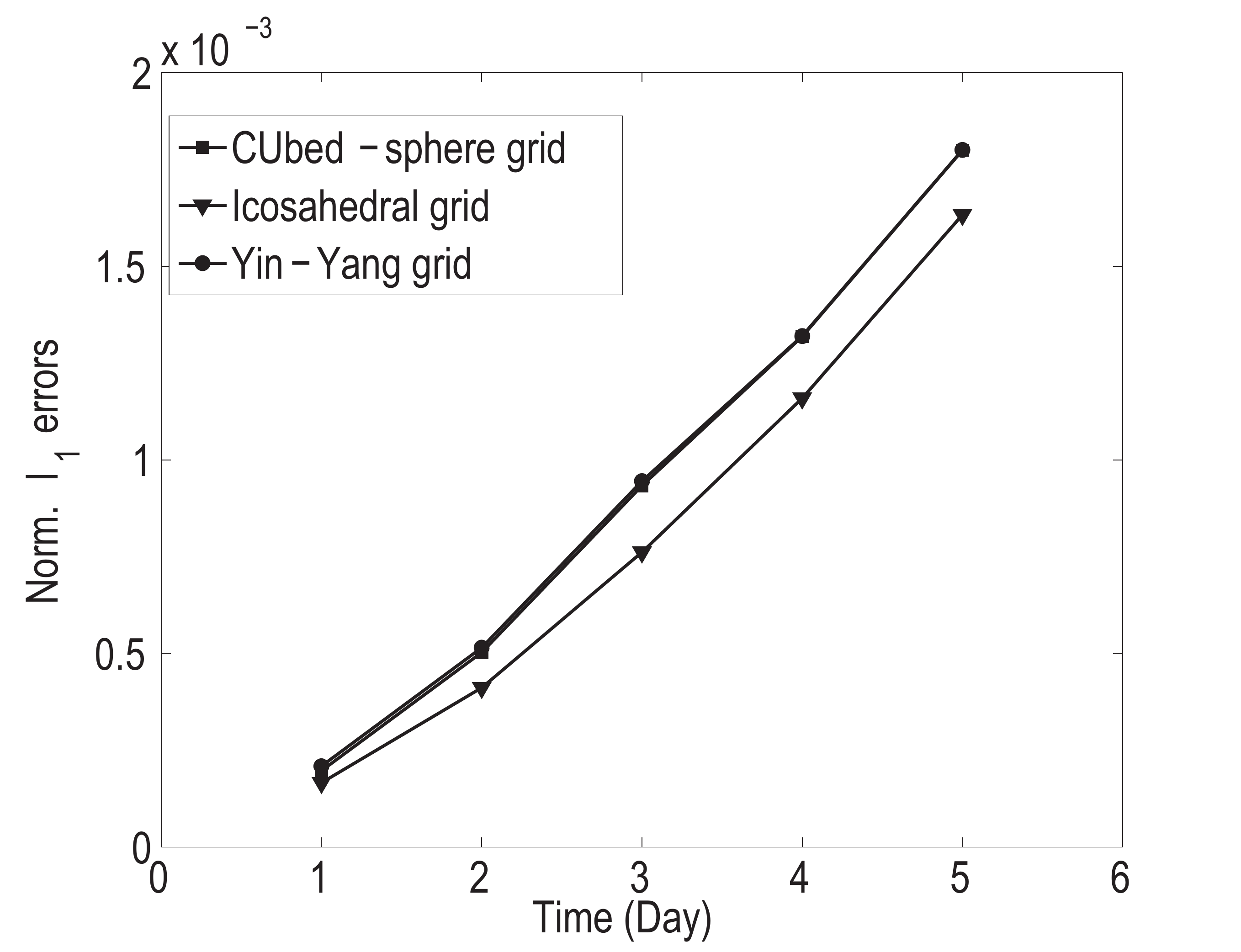}}
\subfigure[$l_2$ error]
{\includegraphics[width=0.33\textwidth]{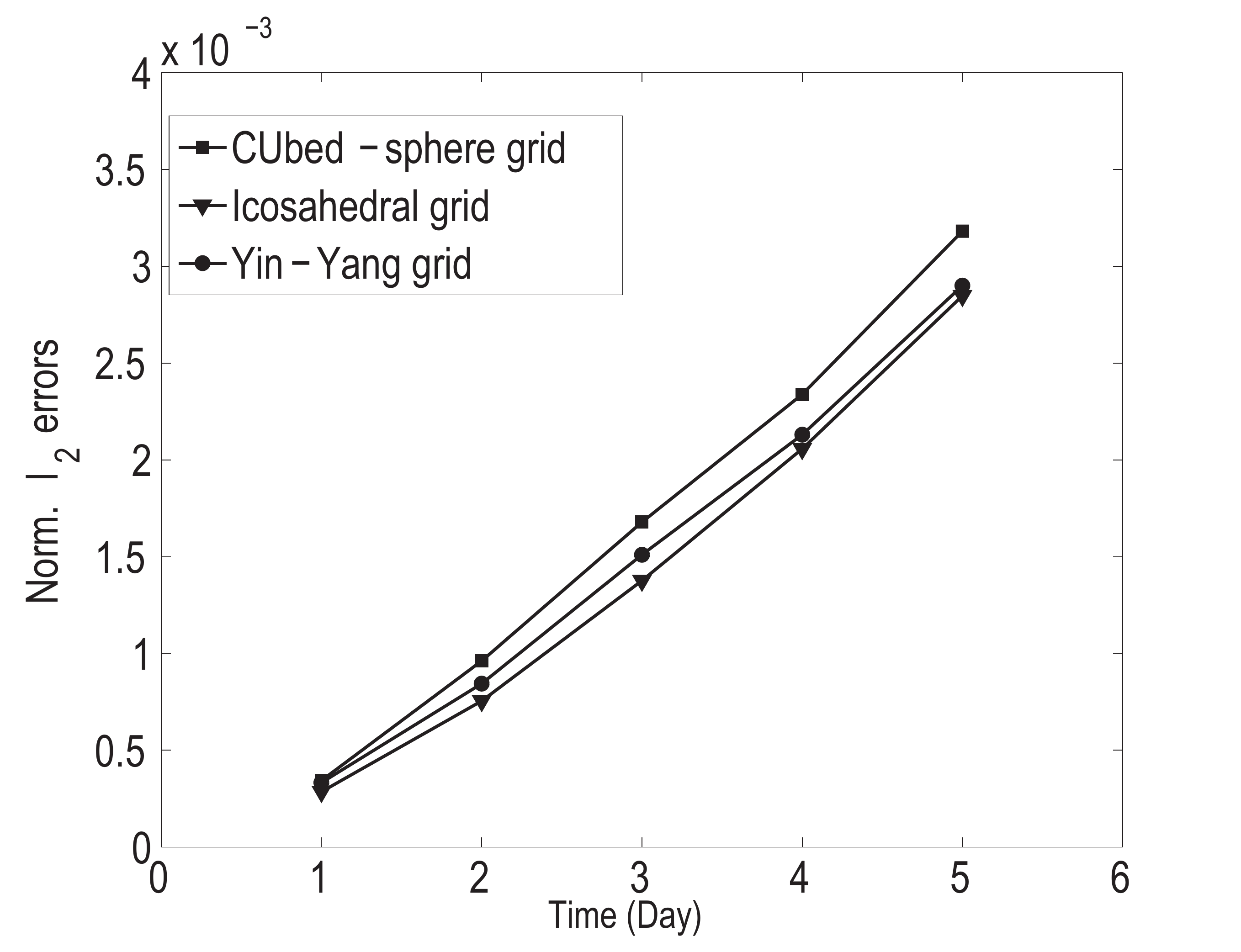}}
\subfigure[$l_\infty$ error]
{\includegraphics[width=0.33\textwidth]{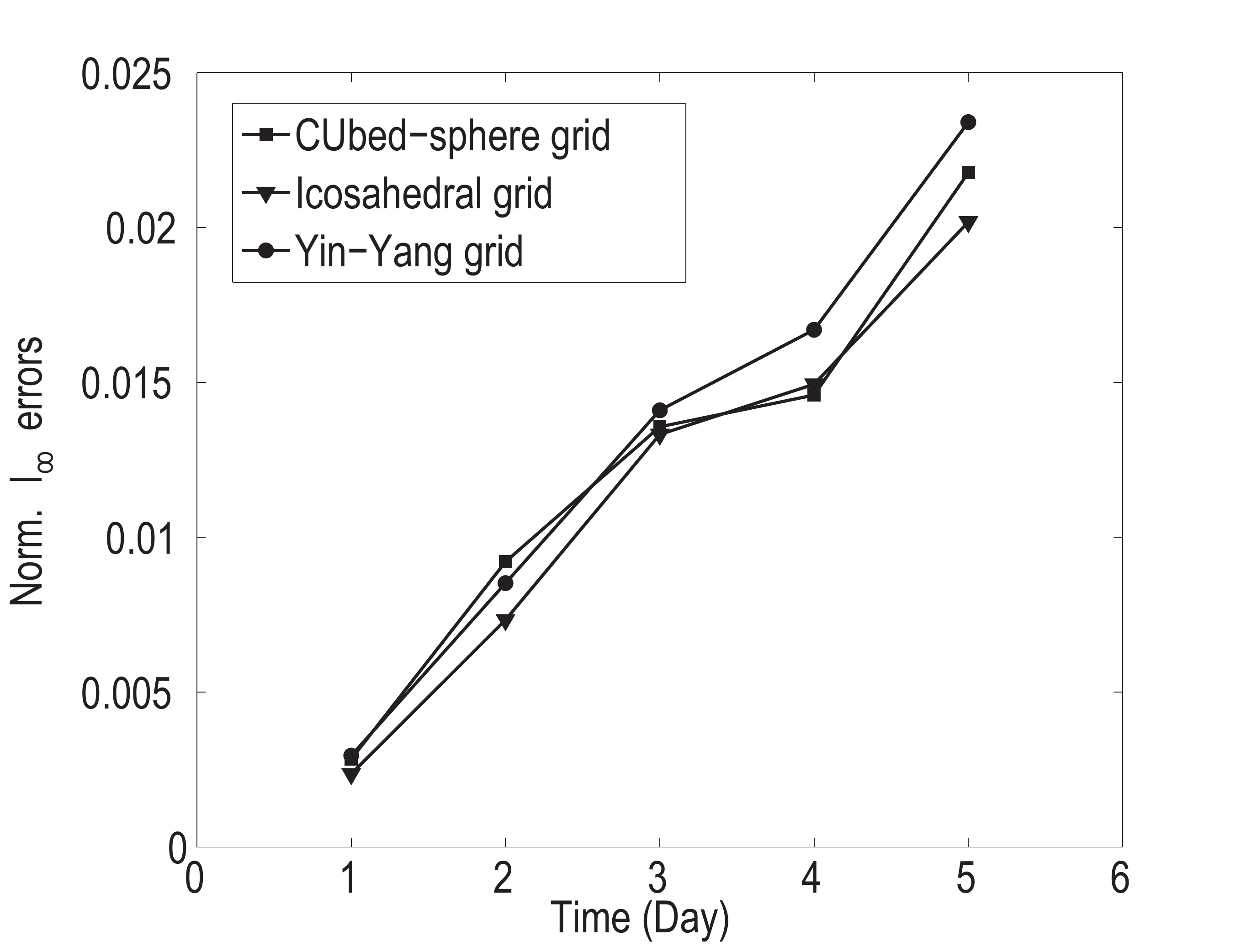}}
\end{center}
\caption{History of the normalized errors of height fields compared with spectral transform solution on T213 grid as the reference solution.}\label{case7error}
\end{figure}

We run this test to day 5 on cubed-sphere and Yin-Yang grids with $N=40$ and icosahedral grid with $N=36$. Contour plots of numerical results of height fields and absolute errors compared with the reference solution of the spectral transform model on T213 grids at day 5 are shown in Fig.\ref{case7h} and  Fig.\ref{case7h-e}. We also calculated the normalized errors by comparing with the spectral transform solution of T213  at each day shown in Fig.\ref{case7error}. The distribution of absolute errors look similar and the magnitude of normalized errors are quite close on three different grids. It reveals that the MCV methodology is robust with regard to different spherical meshes, which indicates the great potential of applications in real cases.

\section{ The compressible non-hydrostatic dynamical core} \index{nonhydrostatic model}

\subsection{The nonhydrostatic governing equations}\index{nonhydromodelequation}

We have developed a 2D compressible non-hydrostatic atmospheric model by using MCV3 and MCV4 schemes \cite{li13}. The governing equations are the Euler conservation laws with the effect of gravity as
\begin{align}\label{eq:2D_org_compre_nonhydra_set1}
\frac{\partial \rho }{\partial t} + \frac{\partial (\rho
u)}{\partial x} + \frac{\partial (\rho w)}{\partial
z} &= 0, \\
\label{eq:2D_org_compre_nonhydra_set2} \frac{\partial (\rho
u)}{\partial t} + \frac{\partial (\rho u^2 +p)}{\partial x} +
\frac{\partial (\rho uw)}{\partial
z} &= 0, \\
\label{eq:2D_org_compre_nonhydra_set3}\frac{\partial (\rho
w)}{\partial t} + \frac{\partial (\rho w u)}{\partial x} +
\frac{\partial (\rho w^2 +p)}{\partial
z} &= -\rho g, \\
\label{eq:2D_org_compre_nonhydra_set4}\frac{\partial (\rho
\theta)}{\partial t} + \frac{\partial (\rho \theta u)}{\partial x} +
\frac{\partial (\rho\theta w)}{\partial z} &= 0
\end{align}
where $\rho$ is the density,
$\mathbf{u}=(u,w)^T$ the vector wind in the Cartesian coordinate,
$p$ the pressure, and $\theta$ the potential
temperature. Since the potential temperature is related to the air
temperature $T$ and pressure $p$ by $\theta=T(p_0/p)^{R_d/c_p}$, the equation of
state is expressed by $p=C_0(\rho\theta)^{\gamma}$, which exactly closes the
equation set where $C_0$ is constant given by
$C_0=R^{\gamma}_{d}p_0^{-R_d/c_v}$.  In the above expressions, the constants are given as
$\gamma=c_p/c_v=1.4$, $R_d=287 \rm J\rm kg^{-1} \rm K^{-1}$,
$c_p=1004.5 \rm J\rm kg^{-1} \rm K^{-1}$, $c_v=717.5\rm J\rm kg^{-1} \rm
K^{-1}$, and $p_0=10^{5}\rm Pa$.

In the presence of topography, the height-based terrain-following coordinate introduced by \citep{gs75}
is utilized to map the physical space $(x,z)$ into the computational domain $(x,\zeta)$ via the transformation relationship $\zeta=\zeta(x,z)$. Put in the component form in the transformed coordinate system, the governing equations read,
\begin{align}\label{eq:terrain_form_equation_set1}
\frac{\partial \rho' }{\partial t} + \frac{1}{\sqrt{G}} \left[ \frac{\partial (\sqrt{G}\rho
u)}{\partial x} + \frac{\partial (\sqrt{G}\rho \tilde{w})}{\partial
\zeta} \right]&= 0, \\
\label{eq:terrain_form_equation_set2}\frac{\partial (\rho
u)}{\partial t} + \frac{1}{\sqrt{G}} \left[ \frac{\partial (\sqrt{G}\rho u^2 +\sqrt{G}p')}{\partial x} +
\frac{\partial (\sqrt{G}\rho u \tilde{w}+\sqrt{G}G^{13} p')}{\partial
\zeta} \right]&= 0, \\
\label{eq:terrain_form_equation_set3}\frac{\partial (\rho
w)}{\partial t} + \frac{1}{\sqrt{G}} \left[ \frac{\partial (\sqrt{G}\rho w u)}{\partial x} +
\frac{\partial (\sqrt{G}\rho w \tilde{w} +p')}{\partial
\zeta} \right] &= -\rho' g, \\
\label{eq:terrain_form_equation_set4}\frac{\partial (\rho
\theta)'}{\partial t} + \frac{1}{\sqrt{G}} \left[ \frac{\partial (\sqrt{G}\rho \theta u)}{\partial x}
+ \frac{\partial (\sqrt{G}\rho\theta \tilde{w})}{\partial \zeta} \right]&= 0,
\end{align}
where $\sqrt{G}=\frac{\partial z}{\partial \zeta}$ is the Jacobian of the transformation and $G^{13}=\frac{\partial \zeta}{\partial x}$ are the contravariant metric, and $\tilde{w}=\frac{d \zeta }{d t}$ is the velocity component in the transformed coordinate. Here, the following relations are utilized by the chain rule \citep{clark77}
\begin{align}\label{eq:chain_rule}
  \frac{\partial \phi}{\partial x}|_{z=\text{constant}}&=\frac{1}{\sqrt{G}} \left[\frac{\partial }{\partial
  x}\left(\sqrt{G}\phi\right)|_{\zeta=\text{constant}}+\frac{\partial }{\partial
  \zeta}\left( \sqrt{G} G^{13}\phi\right) \right] \\
  \frac{\partial \phi}{\partial z}&=\frac{1}{\sqrt{G}} \frac{\partial \phi}{\partial
  \zeta}.
\end{align}
where $\phi$ denotes an arbitrary field variable. The relation between the vertical components of velocity in $z$ coordinate and $\zeta$ coordinate is
\begin{align}\label{eq:w_velocity}
  \tilde{w} = \frac{d \zeta}{d t} = \frac{1}{\sqrt{G}}\left( w +\sqrt{G}G^{13}u \right).
\end{align}

\begin{figure*} [htbp]
\begin{center}
  \includegraphics[scale=1.2,angle=0,clip]{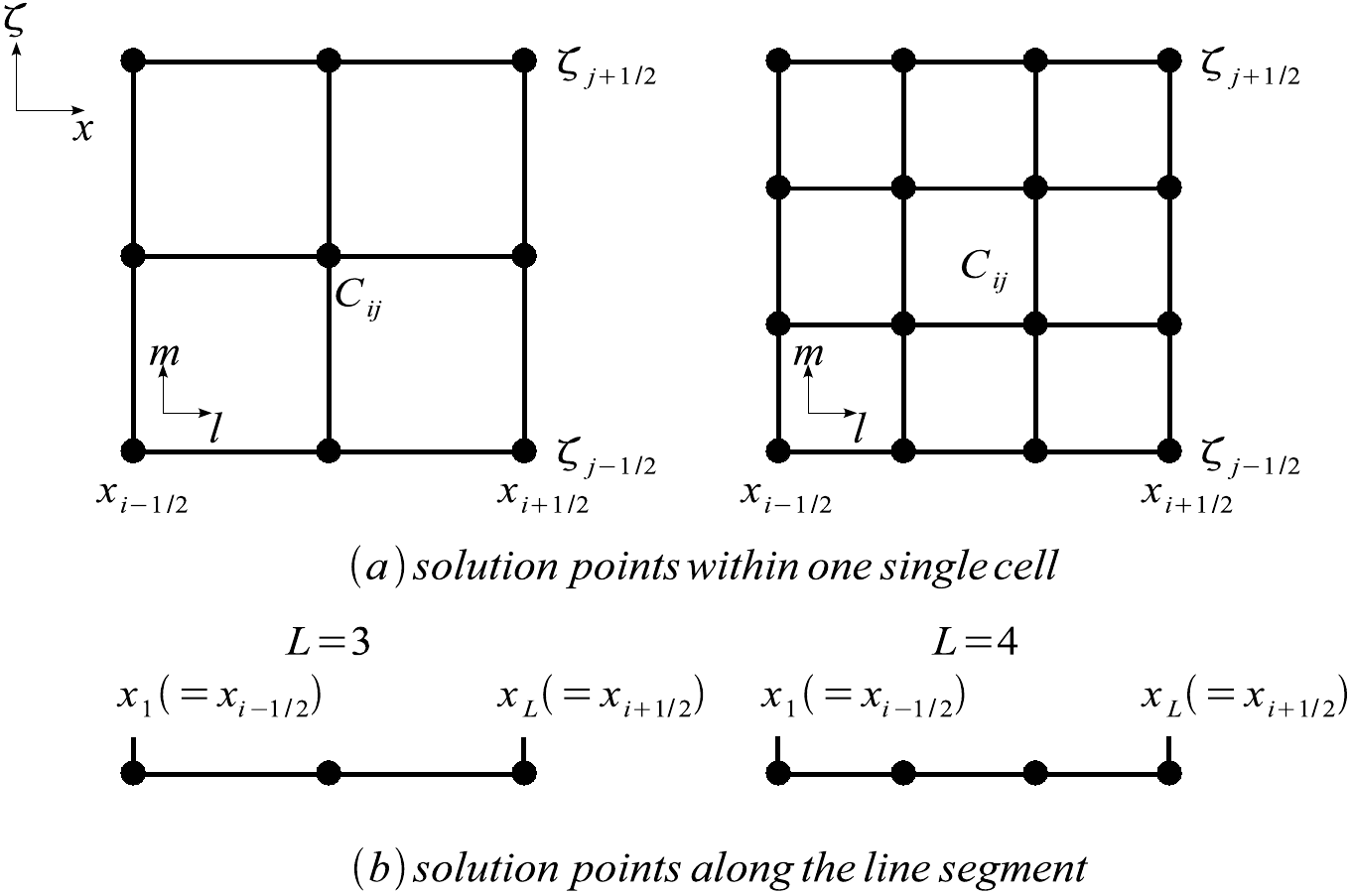}
\end{center}
\vskip -\lastskip \vskip -3pt \caption{ (a) The equidistant solution points in control cell ${\mathcal C}_{ij} = [x_{i-1/2},x_{i+1/2}]\otimes[\zeta_{j-1/2},\zeta_{j+1/2}]$ where the $3 \times 3$ black circles are for the 3rd-order scheme (left) and the $4 \times 4$ black circles for the 4th-order scheme (right). (b) The solution points along the line segment $\overline{x_1x_L}\times \zeta_m$ in $x$ direction for the 3rd-order scheme (left) and the 4th-order scheme (right).
}\label{fig:mcv_collocation}
\end{figure*}

\subsection{The MCV3/4 nonhydrostatic model results}\index{nonhydromodelformulation}

For the MCV nonhydrostatic dynamic core, we use the structured grid with cells $\{ \mathcal{C}_{ij} \}$ numbered by $i$ ($i=1,2,\cdots, I$) and $j$  ($j=1,2,\cdots, J$) respectively in $x$ and $\zeta$ directions. 
As shown in Fig. \ref{fig:mcv_collocation}, the unknown DOFs are defined as the point values at the solution points within each mesh cell. For an $L$-th order scheme, the DOFs  in $\mathcal{C}_{ij}$ are denoted by $q_{ijlm}$ and defined at points $(x_{ijl},\zeta_{ijm})$ with $l,m=1,...,L$, where $l$ denotes the index of solution points in $x$ direction and $m$ in $\zeta$ direction. The total number of DOFs used in an $L$-th order MCV scheme is $\left[I\times(L-1)+1\right]\times \left[J\times(L-1)+1\right]$. Our experience shows that in an MCV scheme the numerical result is not sensitive to the location of the solution points. No significant difference is observed in the solutions between the Gauss-point and the equi-distanced-point configurations. We use the equi-distanced point configuration including the cell boundary points in the present model, which is simpler and easy for implementation. More importantly, the solition points can be adaptively located according to the model configuration in real-case applications.

The basic numerical formulations described in section 2 can be applied straightforwardly.  The generalized Riemann problem formulated in \cite{tt02} and \cite{tmn01} is applied to evaluate the numerical flux and its derivative at cell boundaries. 
Conventional approximate Riemann solvers, such as the local Lax-Friedrich (LLF) \cite{so88} solver or Roe solver\cite{roe1981} can be used.  
As mentioned above, the DOFs (or unknowns) $q_{lm}$ defined at the cell boundary $x_1$ or $x_L$ along the grid line $\overline{x_1 x_L}\cap z_m$ are continuous and shared by the neighboring line grid in the present MCV scheme. Therefore, the flux functions can be computed directly from $q_{lm}$, which has a $C^0$ continuity over the computational domain.
The detailed implementation of the MCV schemes to multi-dimension for nonhydrostatic dynamic core, as well as the numerical results  of 3rd and 4th order schemes, can be found in \cite{li13}. Some of numerical results such as internal gravity waves, hydrostatic mountain waves and nonhydrostatic mountain waves are shown next. 

\subsubsection{Internal gravity waves}
\begin{figure*} [htbp]
\begin{center}
  \includegraphics[scale=0.8,angle=0,clip]{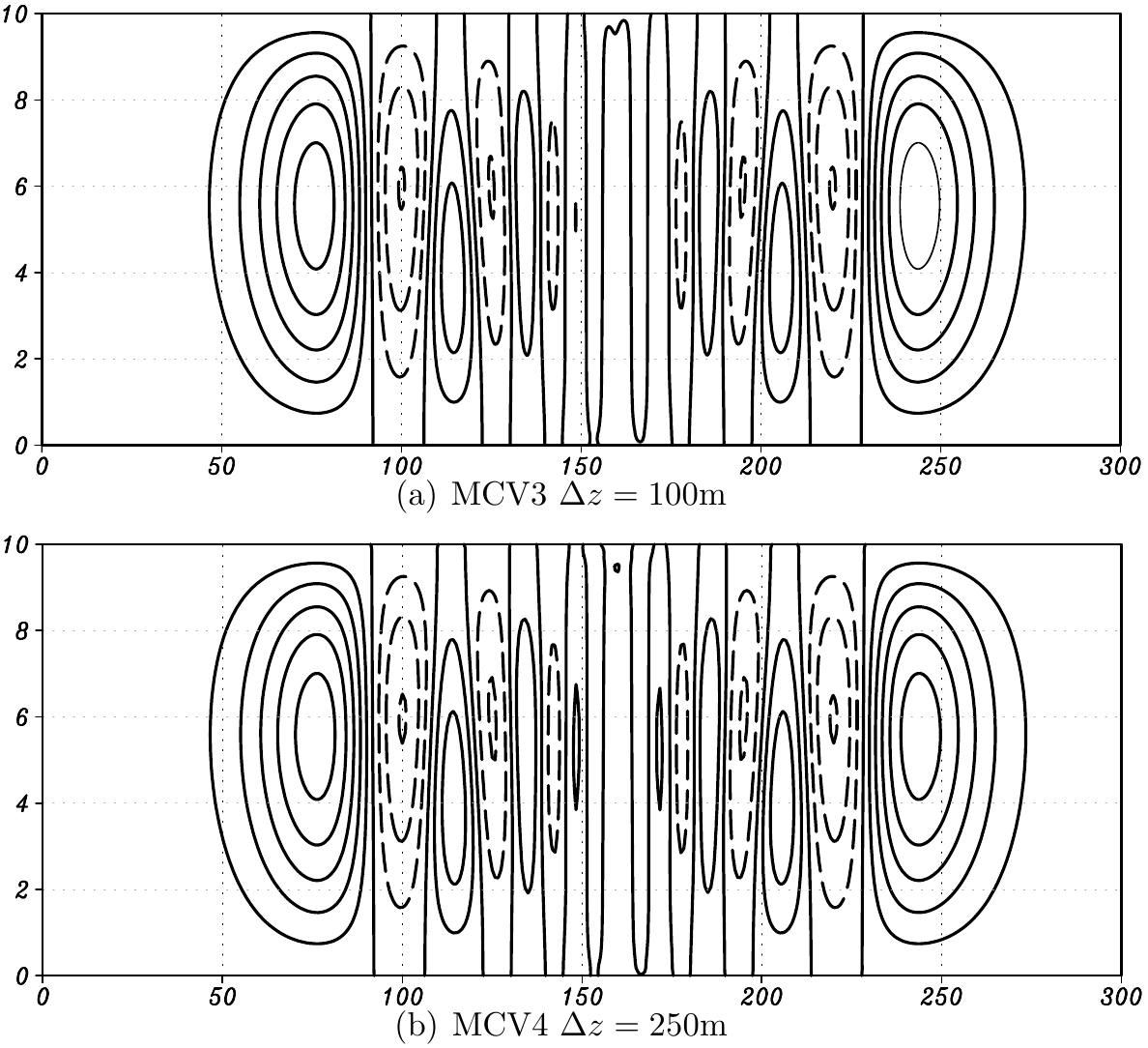}
\end{center}
\vskip -\lastskip \vskip -3pt \caption{Potential temperature
perturbation after $3000$ s using $100$m grid spacing for MCV3
scheme and $250$m grid spacing for MCV4 scheme, with the TVD RK3
time integrator. An aspect ratio of grid spacing, $\Delta x = 10 \Delta z$, is
used for this internal gravity wave test. The
time step is 0.1 s for MCV3 scheme and 0.15 s for MCV4 scheme. The contour values
shown are between $-0.0015$ to $0.003$ with a contour interval of
0.0005. $x$- and $z$-axes are in km and contours of potential temperature
perturbation are in K. }\label{fig:IGWs_mcv34_rk3}
\end{figure*}

We tested the numerical core with the internal gravity wave problem. This test involves the evolution of a potential temperature perturbation in a channel having a periodicity in the horizontal direction. The initial conditions used in this paper are identical to those of  \cite{sk1994}.  Fig. \ref{fig:IGWs_mcv34_rk3} shows the contour plots of the potential
temperature perturbation. The same contouring interval such as \cite{sk1994} and \cite{giraldo08} is adopted for the convenience of comparison. The results of MCV schemes look quite similar with those in \cite{sk1994} and \cite{giraldo08}, as well as other existing studies \cite[e.g.][]{al07,nn11}. It is noted that the error norms measured by \cite{sk1994} can not be adopted here since the analytic solution is available only for the Boussinesq equations but not for the fully compressible equations.

\begin{table*}[h]
\caption{Comparison of the numerical results between the 3rd-order scheme on a $100$ m grid and the 4th-order scheme on a $125$ m grid for the internal-gravity wave test.} {\small
\begin{center}
\begin{tabularx}{\textwidth}{l@{\extracolsep{\fill}}ccccc}
\hline
\hline Schemes  &$w_{max}$ & $w_{min}$ & $\theta'_{max}$ & $\theta'_{min}$ \\
\hline
3rd   & $2.46\times 10^{-3}$ &  $-2.43\times 10^{-3}$ & $2.80\times 10^{-3}$ & $-1.52\times 10^{-3}$ \\
4th   & $2.46\times 10^{-3}$ &  $-2.50\times 10^{-3}$ & $2.80\times 10^{-3}$ & $-1.53\times 10^{-3}$ \\
 \hline \label{table:igws_maxmin}
\end{tabularx}
\end{center}
}
\end{table*}

Table \ref{table:igws_maxmin} shows the maximum and minimum of vertical velocities and potential temperature perturbations for third-order and fourth-order MCV schemes after $3000$ s. It can be seen that the 3rd and 4th-order schemes obtained almost the same numerical outputs though different grid spacings are used here. Specifically, as mentioned by \cite{giraldo08}, the ranges of potential temperature perturbations are $\theta'\in[-1.49\times 10^{-3},2.82\times 10^{-3}]$ in the numerical results of the model based on the technique of flux-based wave decomposition in \cite{al07} and $\theta'\in[-1.51\times 10^{-3},2.78\times 10^{-3}]$ from models based on the spectral element and discontinuous Galerkin in \cite{giraldo08}, while our results are $\theta'\in[-1.52\times 10^{-3},2.80\times 10^{-3}]$ as shown in Table \ref{table:igws_maxmin}. It is observed that these numerical results agree well with each other.

\begin{figure*} [htbp]
\begin{center}
  \includegraphics[scale=0.8,angle=0,clip]{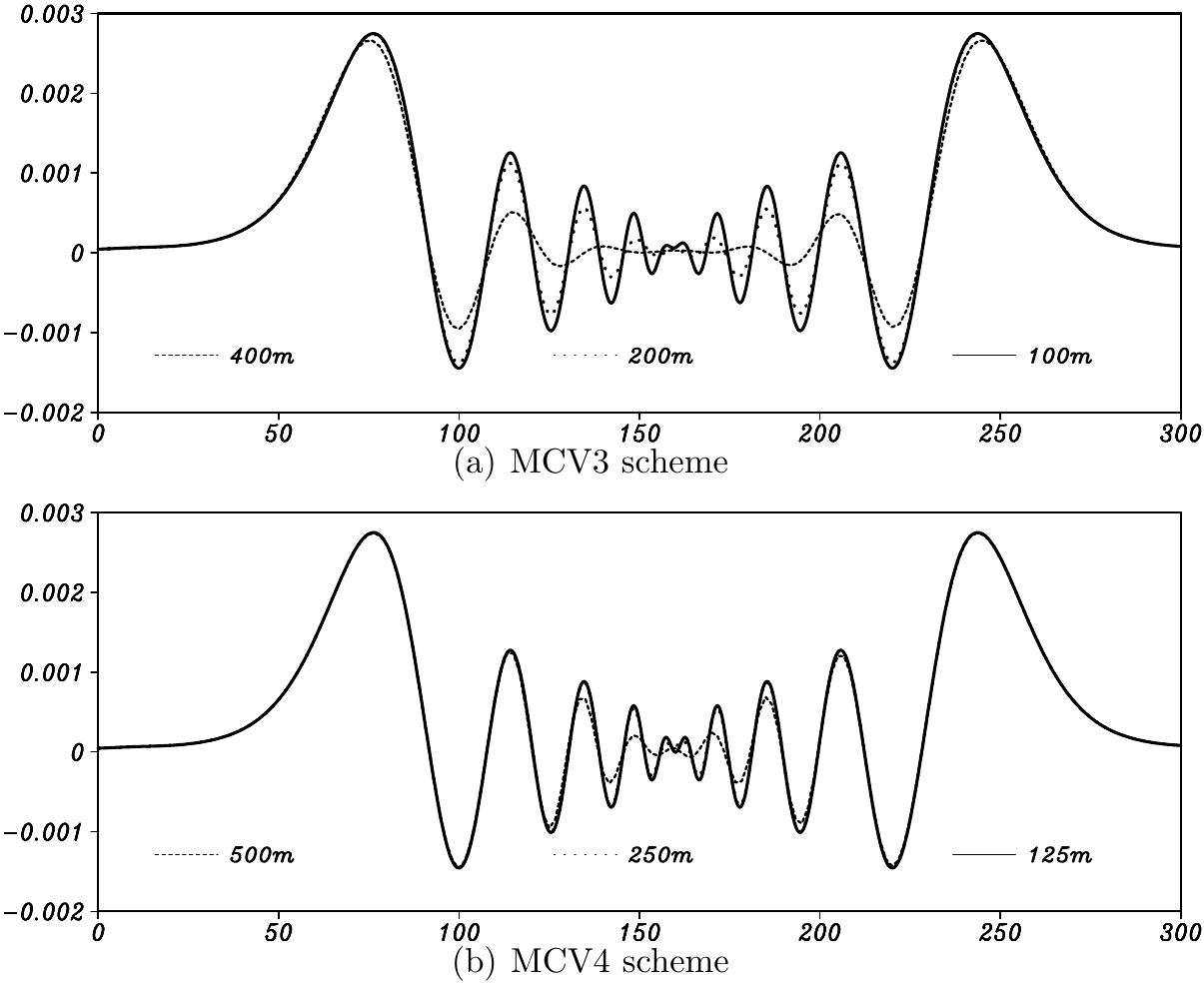}
\end{center}
\vskip -\lastskip \vskip -3pt \caption{Potential temperature
perturbations along the line $z=5$km for the internal gravity wave
test after 3,000 s with different grid resolutions. The cell aspect ratio is $\Delta x = 10 \Delta z$. $x$-axes is in
km and $y$-axes is in K for potential temperature perturbation.
}\label{fig:IGWs_mcv34_z=5km_rk3}
\end{figure*}

Fig. \ref{fig:IGWs_mcv34_z=5km_rk3}
gives the profiles of potential temperature perturbations along the
horizontal cross section $z = 5$ km after 3,000s using the three-order and fourth-order MCV schemes. It is
observed that the profiles are perfectly symmetric about the
position $x=16$ km. Compared with the results of the model using fifth-order WENO\cite{nn11}, the numerical solution of competitive quality are obtained by utilizing
the equivalent DOF resolution with the third-order MCV scheme, as shown in Fig.
\ref{fig:IGWs_mcv34_z=5km_rk3} (a). The fourth-order MCV scheme with a little increase
in the equivalent DOF resolution results in significantly improved numerical outputs as shown in Fig. \ref{fig:IGWs_mcv34_z=5km_rk3}
(b). In addition, the fourth-order MCV scheme also has competitive numerical results with the same grid spacing of $250$ m compared to those using the schemes of spectral element and discontinuous Galerkin methods (see Fig. 2 (b) in \cite{giraldo08}).

\subsubsection{Linear hydrostatic/nonhydrostatic mountain waves}
We have tested with the mountain waves over various bell-shaped mountains in \cite{li16}.  The flows over the mountains of various half widths and heights were simulated with the model. The semi-analytic solutions to the mountain waves through the linear theory are used to check the performance of the MCV nonhydrostatic model. The isolated bell-shaped bottom mountain to trigger the waves is specified as\begin{align} 
   z_s(x)=\frac{h_0}{1+(x-x_0)^2/a^2}
\end{align}
where $h_0$ is the maximum height of the mountain, $x_0$ is the center of physical domain and  $a$ is the half width of the mountain.  Hereafter the numerical results of mountain waves by using the MCV4 scheme are presented. 

\begin{figure*} [htbp]
\begin{center}
  {\includegraphics[width=0.65\textwidth]{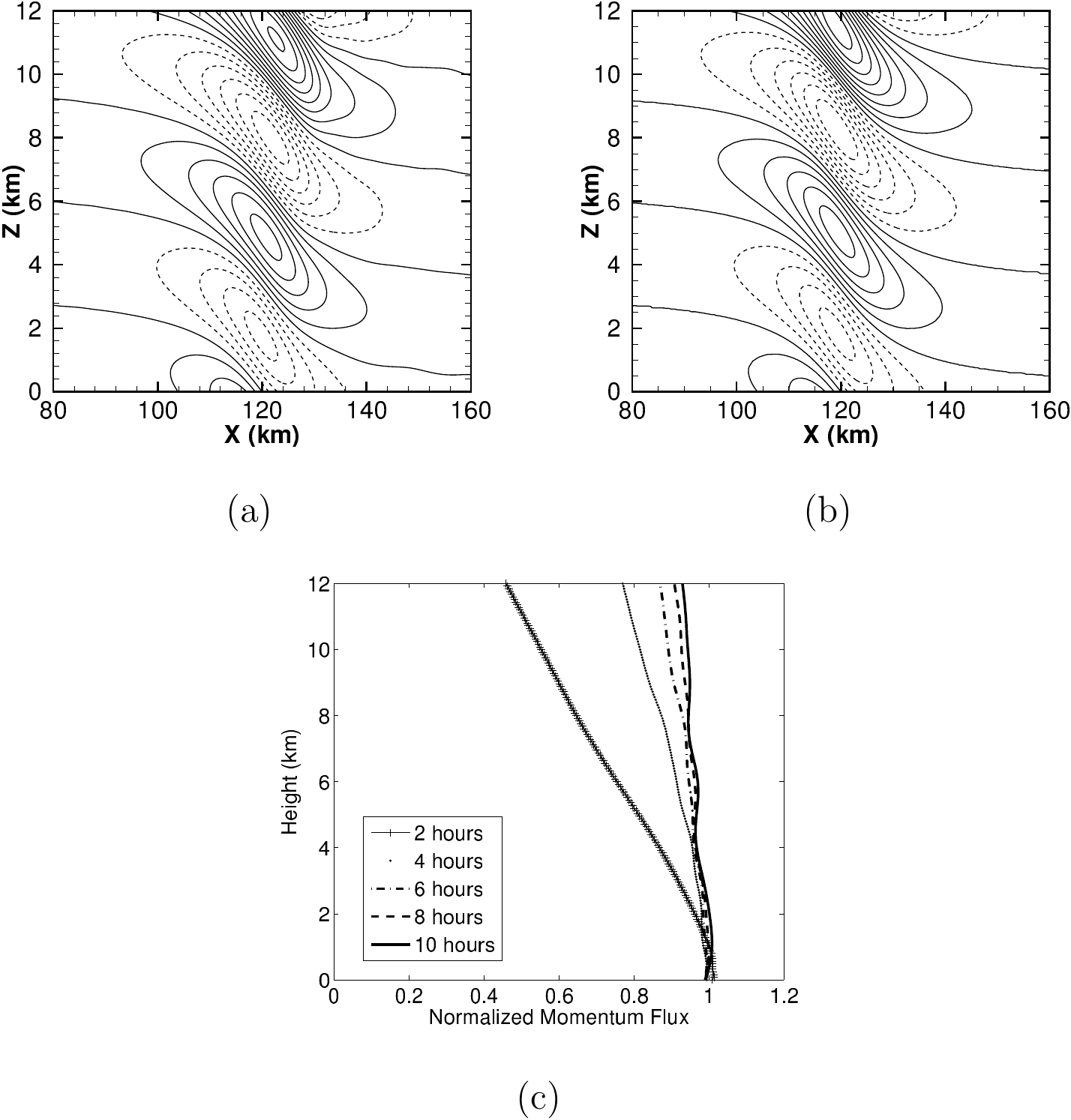}}  
\end{center}
\vskip -\lastskip \vskip -3pt 
\caption{Numerical results of linear hydrostatic mountain waves. Shown are numerical solution (a), semi-analytical solution (b) and normalized momentum flux (c). The contour values vary from -0.005 to 0.005 with an interval of 0.0005. Negative values are denoted by dashed lines.}\label{fig:mcv4hydrowaves}
\end{figure*}

\begin{enumerate}[{-} ]
\item Linear hydrostatic mountain waves
\end{enumerate}

   According to the linear theory, the parameter  $al (\gg1)$ (  $l$ being Scorer parameter) determines the hydrostatically balanced mountain waves in this test. Fig. \ref{fig:mcv4hydrowaves}(a) and (b) show the numerical and the semi-analytical solutions of vertical velocity component $w$ at $t=10$ (nondimensional time $\bar{U}t/a=72$). The MCV model can accurately capture the mountain waves triggered by the gentle slope, except the numerical maxima and minima contours in the vertical velocity field are slightly weaker than the analytic ones. Compared with the existing numerical results obtained by high order schemes such as spectral element (SE) and discontinuous Galerkin (DG), the present results look quite similar to those in \cite{giraldo08}.  

The momentum fluxes at different hours are plotted in Fig.\ref{fig:mcv4hydrowaves}(c). Shown are values normalized by $M_H(z)=-\frac{\pi}{4}\bar{\rho}_0N_0U_0h_0^2$. As the linear hydrostatic mountain waves exist in this test, the vertical variation of the normalized momentum flux is usually used to examine the numerical dissipation of a model. Shown in Fig.\ref{fig:mcv4hydrowaves}(c), as the mountain wave develops into a mature state, the momentum flux gradually approaches to a vertically uniform distribution with a value close to unity of the analytical solution. At  $t=10$, the numerical flux is about 0.99 at the surface and approaches 0.9645 at the height of double vertical wavelength ($z=2\pi/l\approx 6.4 \text{km}$). Present results are competitive to those shown in \cite{xue2000} and \cite{durran1983}, where the flux  is 94\% of its steady value at nondimensional time of 60 in \cite{durran1983} and the flux reaches 0.96 at later time at a height just below their Rayleigh damping layer (12km) in \cite{xue2000}. The results reveal that the high order accuracy of MCV model is much beneficial to improve the simulation of non-dissipative hydrostatic mountain waves.

\begin{enumerate}[{-} ]
\item Linear nonhydrostatic mountain waves
\end{enumerate}
\begin{figure*} [htbp]
\begin{center}
  {\includegraphics[width=0.65\textwidth]{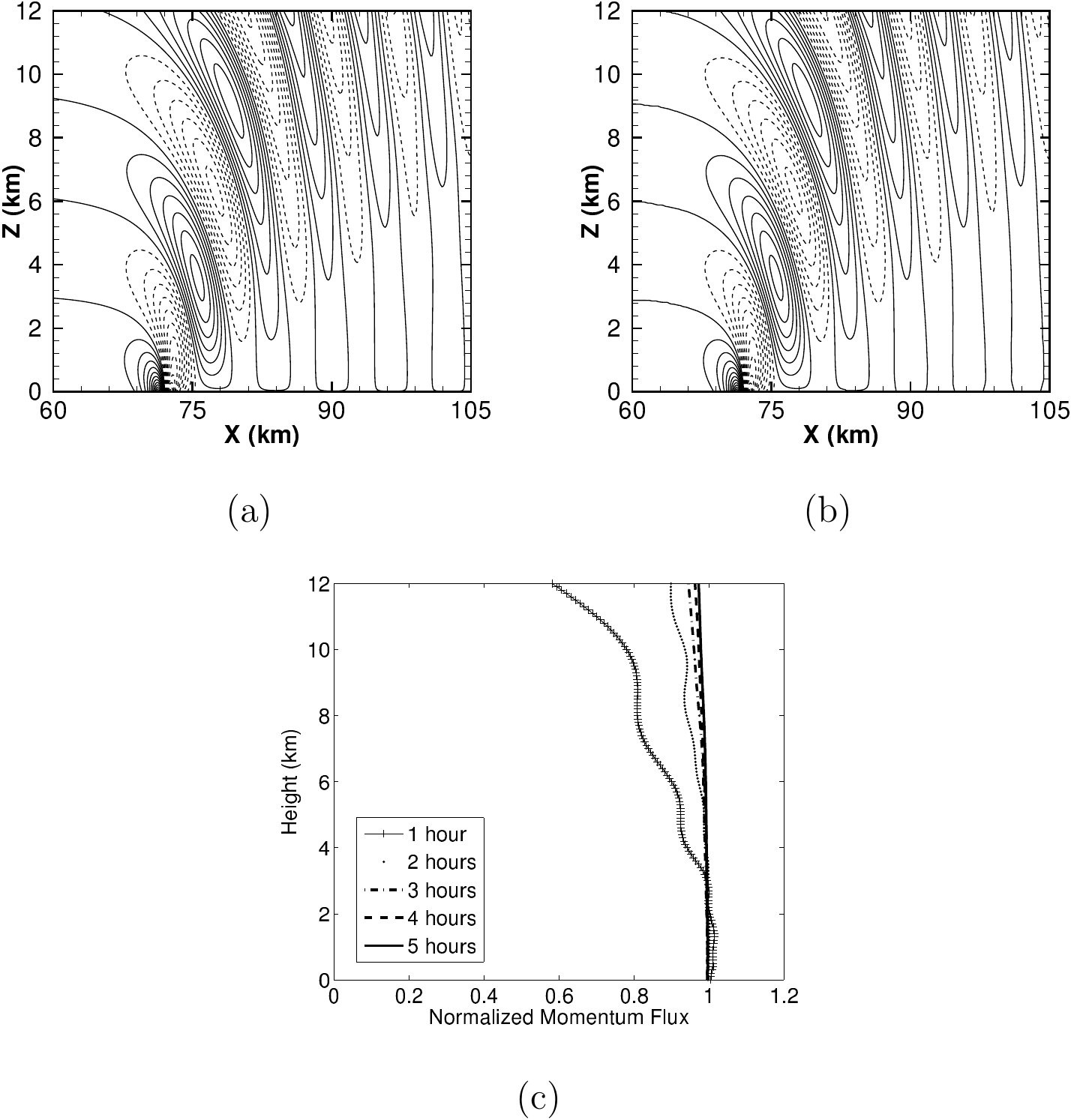}}  
\end{center}
\vskip -\lastskip \vskip -3pt 
\caption{Same as Fig. \ref{fig:mcv4hydrowaves}, but for linear nonhydrostatic mountain }\label{fig:mcv4nonhydrowaves}
\end{figure*}

 Fig. \ref{fig:mcv4nonhydrowaves}(a) and (b) show the numerical and the semi-analytical solutions of vertical velocity after 5 hours (nondimensional time of 180) for linear nonhydrostatic mountain waves. It is observed that the linear nonhydrostatic mountain waves are distinguished from the linear hydrostatic ones by the dispersive character of wave trains behind the mountain peak. The simulated vertical velocity agrees well with the analytical solution and the numerical result of other existing high order schemes such as DG3 in \cite{giraldo08}. 
 
 Similar to previous case, the momentum flux profiles at 1, 2, 3, 4, 5 hours are plotted in Fig. \ref{fig:mcv4nonhydrowaves}(c). It is noted that the momentum flux profiles is normalized by the analytic nonhydrostatic momentum flux $M_{NH}(z)=0.457M_{H}(z)$ which is only valid for $al=1$. It can be seen that the model results have almost reached the steady state at nondimensional time of 180.  This situation is also observed from the convergence of momentum flux at time of 180 approaches 0.97 at the height of 12km. The MCV nonhydrostatic model can represent the linear nonhydrostatic mountain waves quite well as the linear theory predicts.

\subsection{The nonhydrostatic model results with the MCV3-BGS limiter}\index{nonhydromodelformulation}
Next we present some numerical results of the non-hydrostatic dynamical core newly obtained by implementing the MCV scheme with the MCV3-BGS limiter presented in section 2.4.2. It is noted that we have not made use of any artificial diffusion or filtering in the following numericla tests. The intrinsic dissipation in the 4th-order MCV3-BGS limiter effectively eliminates suprious numerical oscillation, which is realized in most existing models by explicit use of diffusion operators or filters.    

\begin{figure*} [htbp]
\begin{center}
\subfigure[]
  {\includegraphics[width=0.4\textwidth]{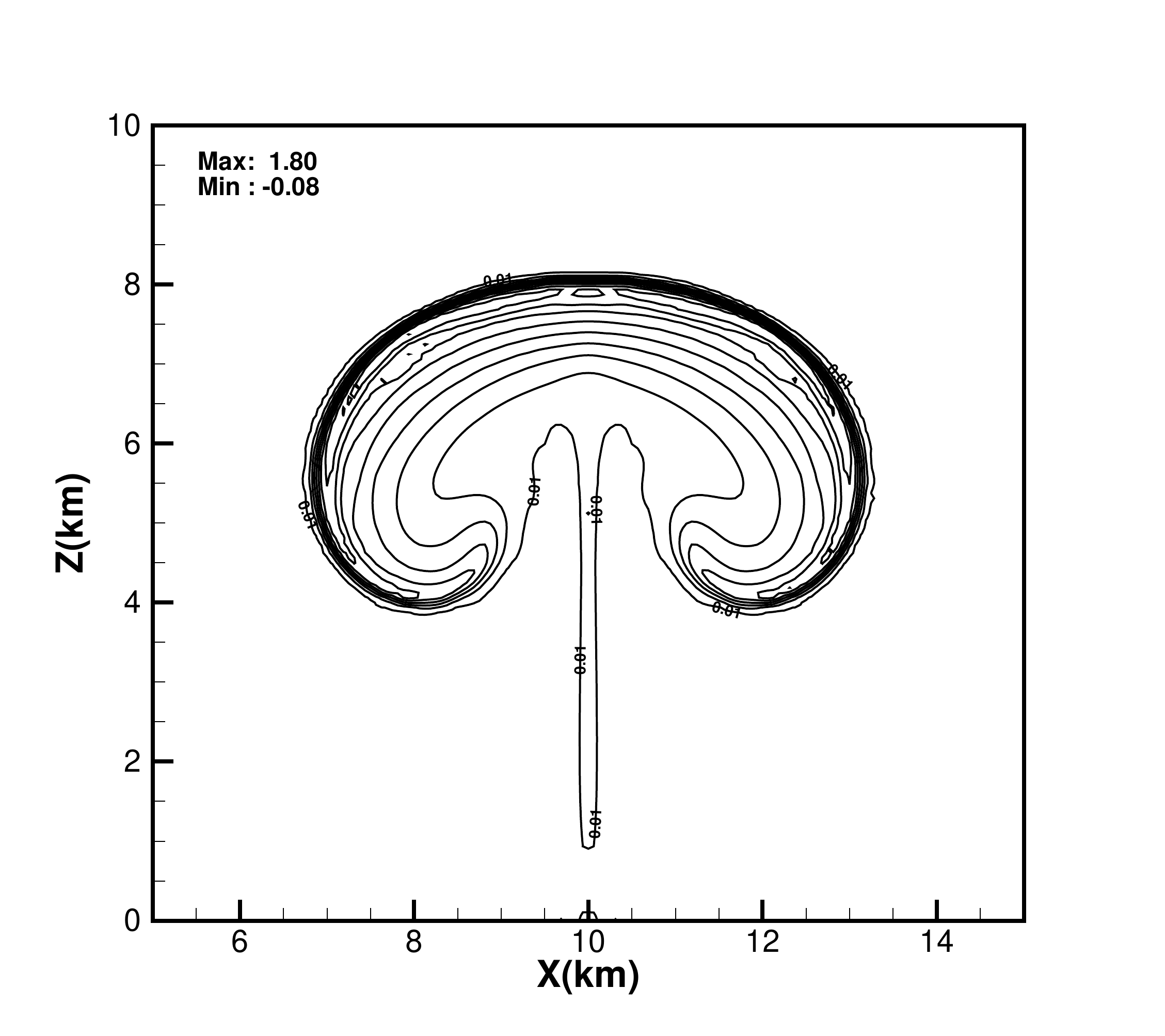}} \hspace{0.5cm}
 \subfigure[]
  {\includegraphics[width=0.4\textwidth]{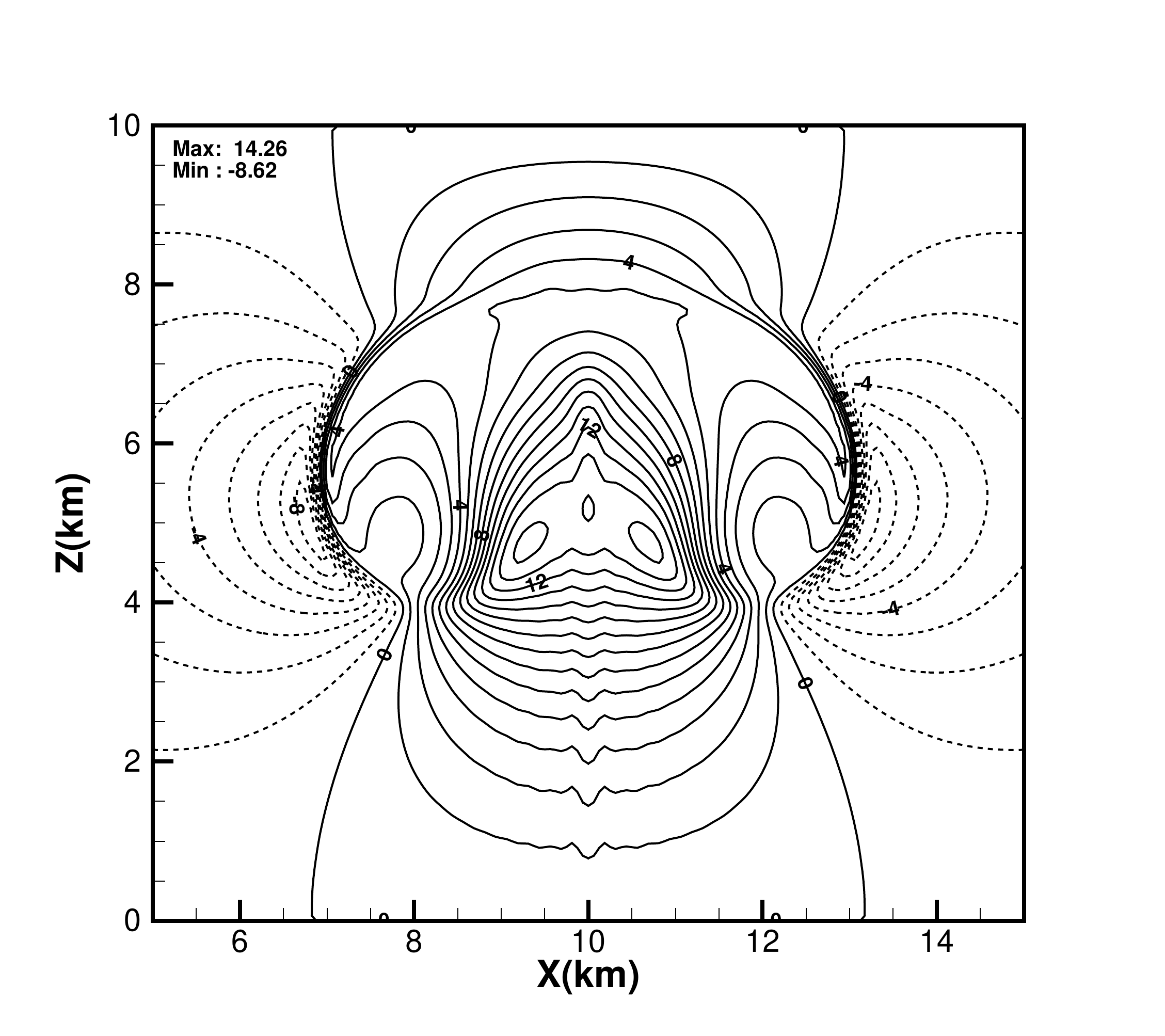}} 
\end{center}
\vskip -\lastskip \vskip -3pt 
\caption{Numerical results of  the convective thermal bubble test.  (a) the potential temperature perturbation and (b) the vertical velocity.}\label{fig:bubble_th_bgs}
\end{figure*}

Fig. \ref{fig:bubble_th_bgs} plots the potential temperature perturbation and vertical velocity in the convective thermal bubble test with a 125-m grid resolution after 1000s using the MCV3-BGS scheme. 

\begin{figure} [htbp]
\begin{center}
  \subfigure[ 400m grid spacing ]{
  \includegraphics[width=0.45\textwidth]{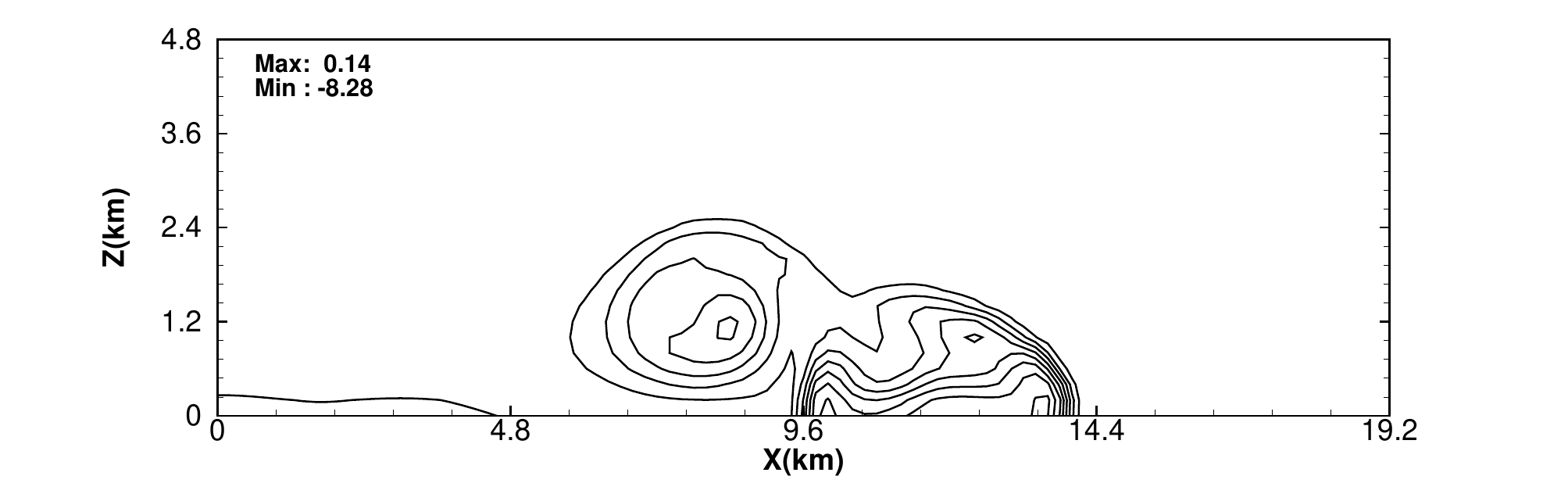}}
  \subfigure[ 200m grid spacing ]{
  \includegraphics[width=0.45\textwidth]{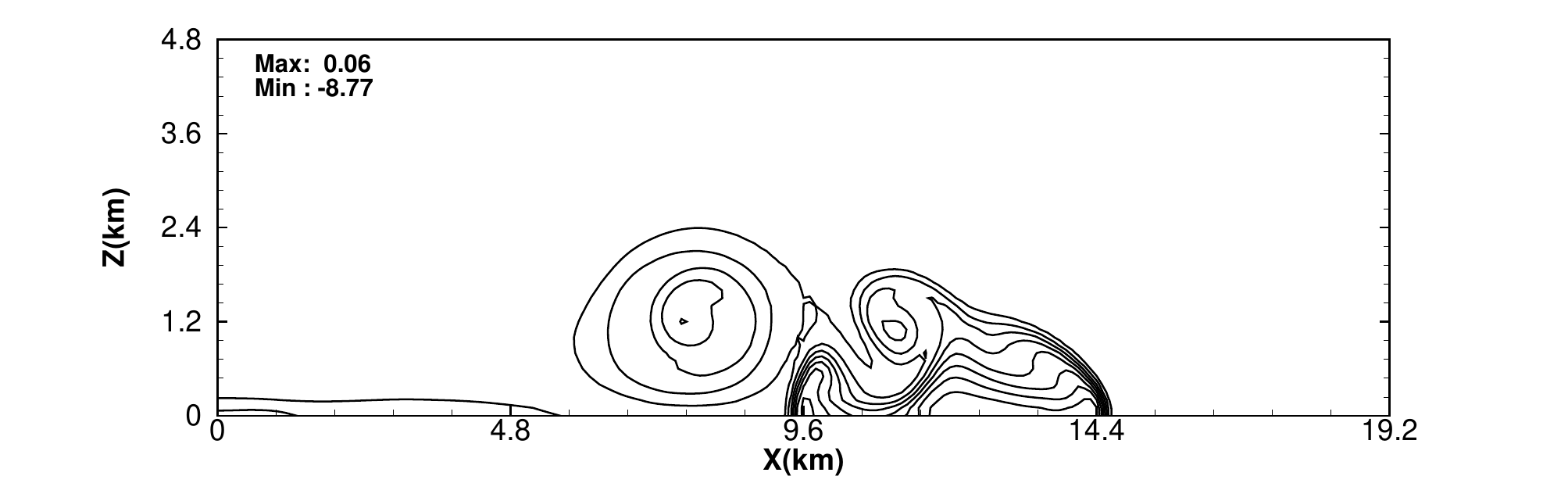}}
  \subfigure[ 100m grid spacing ]{
  \includegraphics[width=0.45\textwidth]{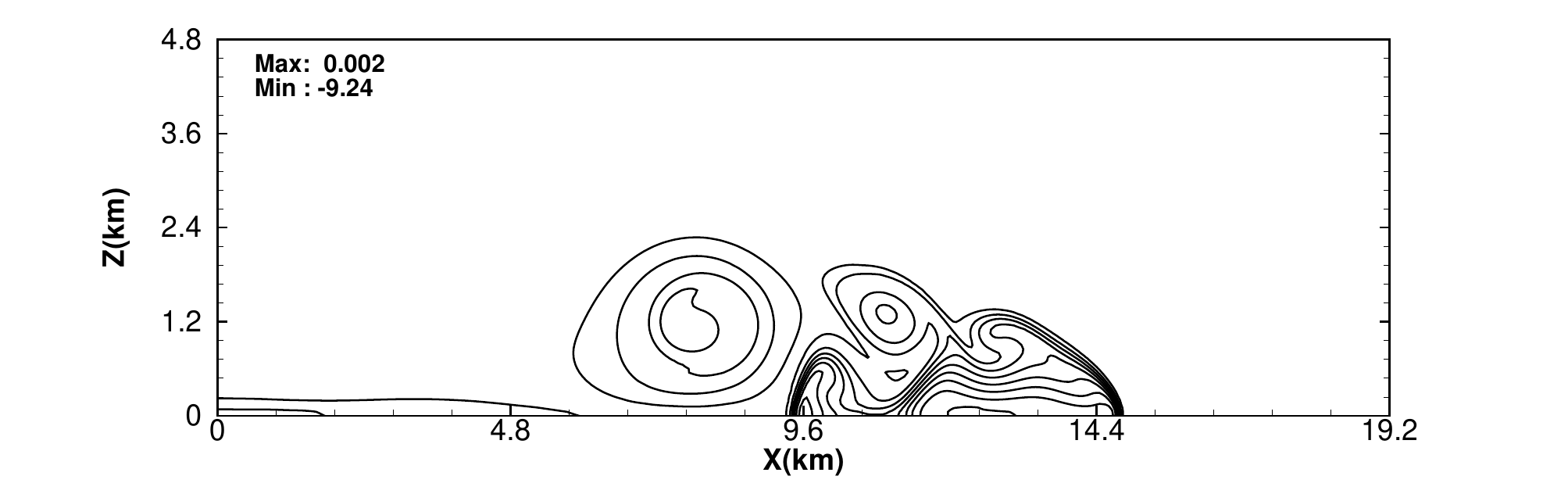}}
  \subfigure[ 50m grid spacing ]{
  \includegraphics[width=0.45\textwidth]{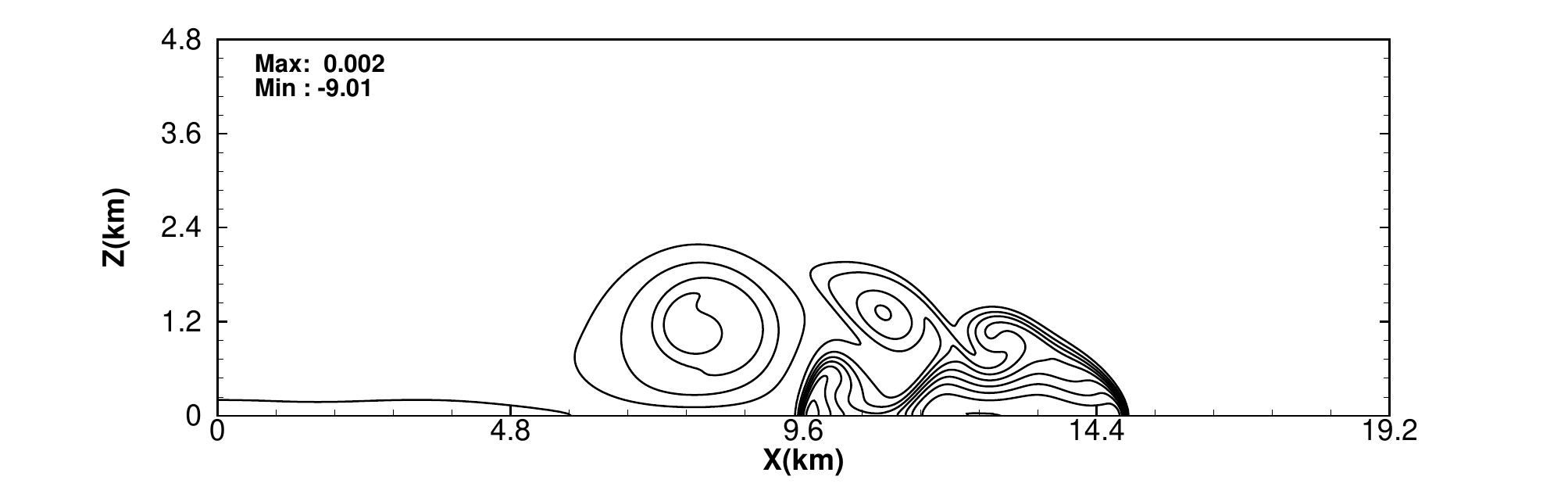}}
\end{center}
\vskip -\lastskip \vskip -3pt \caption{Potential temperature perturbation after 900 s using (a) 400-, (b) 200-, (c) 100- and (d) 50-m grid spacings with the MCV-BGS model.}\label{fig:den_mcv_bgs}
\end{figure}

The numerical results of the density current test on different grid resolutions are shown in Fig. \ref{fig:den_mcv_bgs}. It is found that the MCV-BGS model is able to get the converged numerical solution even with a coarse grid resolution (200m).

\begin{figure} [htbp]
\begin{center}
  \subfigure[ Horizontal wind $u$ ]{
  \includegraphics[width=0.4\textwidth]{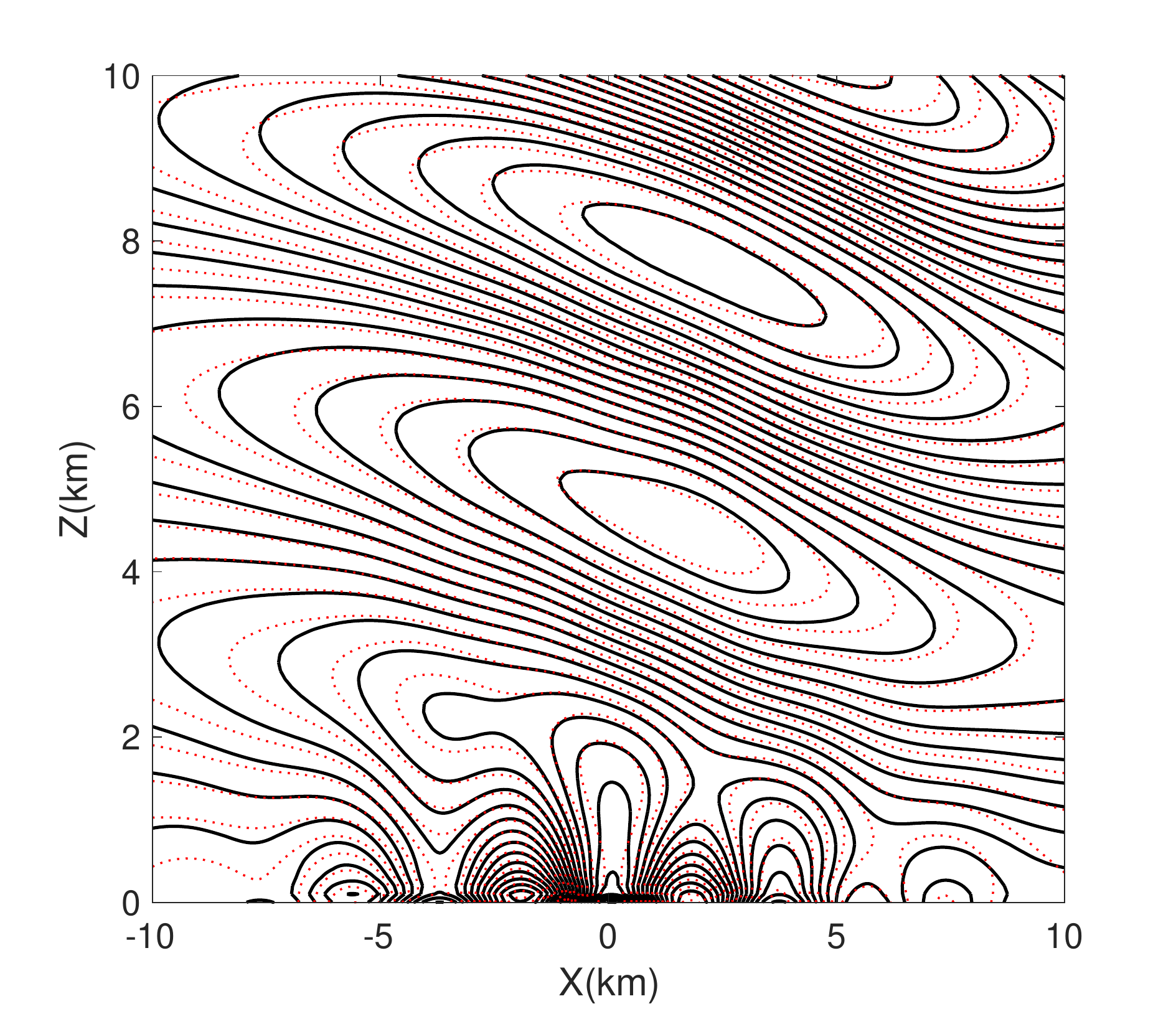}}
  \subfigure[ Vertical wind $w$ ]{
  \includegraphics[width=0.4\textwidth]{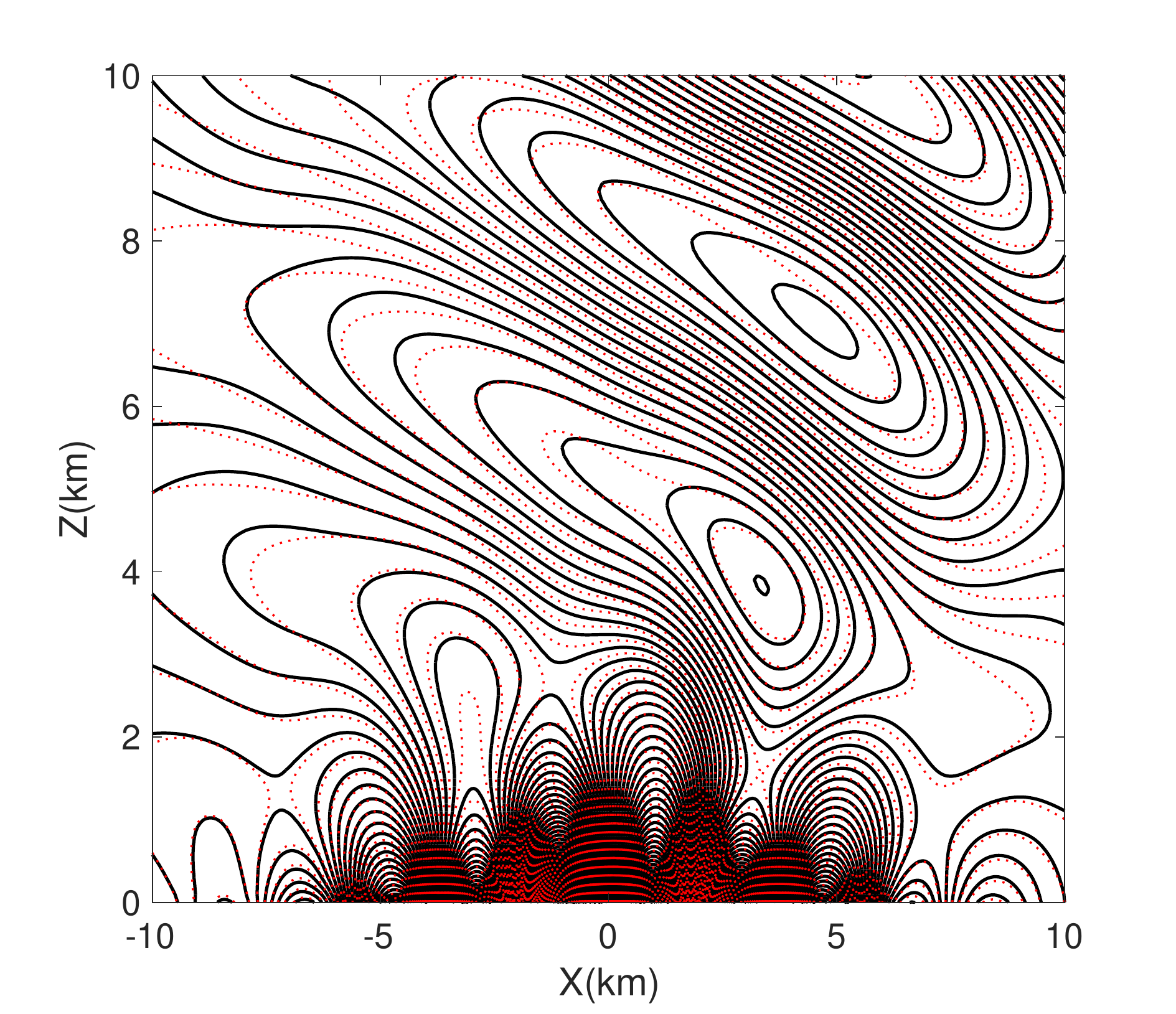}}
\end{center}
\vskip -\lastskip \vskip -3pt \caption{Steady-state flow over the Sch$\ddot{\text{a}}$r mountain computed with the MCV-BGS scheme. (a) Horizontal velocity with contours between -2 and 2 with an interval of 0.2 m/s. (b) Vertical velocity with contours between -2 and 2 with an interval of 0.05 m/s. The numerical solution is represented by a solid line and the analytic solution by a red dashed line.}\label{fig:mcv3-bgs-schaer}
\end{figure}

The Sch$\ddot{\text{a}}$r-mountain wave was simulated to verify the non-hydrostatic dynamical core with MCV-BGS scheme to reproduce the mountain waves generated by complex terrains. Fig. \ref{fig:mcv3-bgs-schaer} shows numerical results of the steady mountain waves computed by the MCV3-BGS4 model after 10 hours with a grid resolution of 250 m ($x$) and 210 m ($\zeta$). 

\subsection{The nonhydrostatic model results with moist thermal dynamics}\index{nonhydromodelformulation}

Before the full atmospheric model that can be run for the operational prediction is accomplished, a key step is that the moist dynamics like moist thermal bubble \cite{bryan2002} and supercell thunderstorm testcases \cite{klemp2015} have been validated by either using direct microphysical feedback or the simple Kessler-type cloud parameterizations. In the past several decades there are many methods for evaluating the performance of a numerical modeling system. In contrast with the dry benchmarks, the suite of test cases that are related physics-dynamics coupling is needed to validate the interplay between the dynamics of the flow and the thermodynamics related to reversible and irreversible moist processes. Bryan and Fritsch (2002) proposed moist thermal bubble test case to check the conversion of reversible processes between the water vapour and cloud water. 

All the following simulations take place in the Cartesian coordinate by neglecting Coriolis force in the nonhydrostatic MCV modelling. For the rising moist thermal bubble, the following processes are ignored: hydrometeor fallout, ice-phase microphysics, the Coriolis force, and subgrid-scale turbulence. The MCV-WENO-FCT \cite{tang2018,li2020} advection scheme is used for transport of water vapour and cloud water. The neutrally stable initial fields in this benchmark is given by following two assumptions: (1) the total water mixing ration is constant at all levels; (2) phase changes are exactly reversible. 

Fig. xxx shows the perturbations of the wet equivalent potential temperature for the moist thermal simulation which ignores the diabatic contribution to the pressure equation, the specific heats of water substances by model configuration labelled as set A in \cite{bryan2002}. Our simulated results agree well with those in the original paper \cite{bryan2002} and those in \cite{neill2014}. The bubble top in the three nonhydrostatic models rises to about 7 km in comparison with the perfect top height of 8.5 km in Fig. 3(a) of \cite{bryan2002} due to different compressible equation set configuration. In this benchmark the vertical velocity is up to 14m/s at 1000s and the nonhydrostatic effects are dominant. It is indicated that the present nonhydrostatic MCV model has the potential to simulate the cloud-resolving atmospheric flows when the vertical acceleration becomes more obvious.

\begin{figure} [htbp]
\begin{center}
  \includegraphics[width=0.8\textwidth]{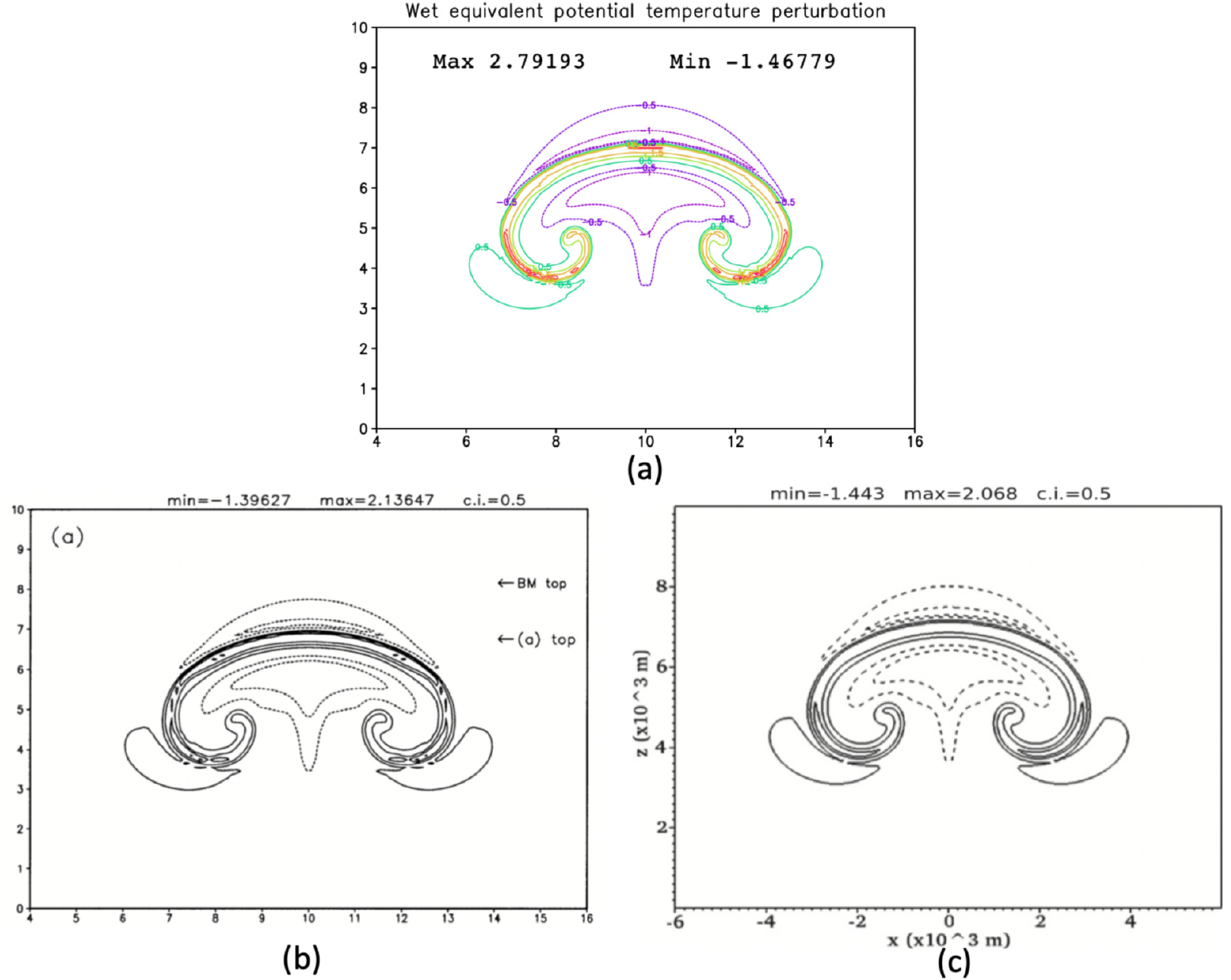}
\end{center}
\vskip -\lastskip \vskip -3pt \caption{ A contour plot of perturbations of the wet equivalent potential temperature $\theta'_e$ at t = 1000 s for the moist bubble simulation with the grid spacing of 100m. (a) $\theta'_e$ in the nonhydrostatic MMFV model; (b) $\theta'_e$ in the Bryan’s nonhydrostatic model; (c) $\theta'_e$ in the pseudo-incompressible model (Neill and Klein, 2014). }\label{fig:moistbubble}
\end{figure}

\section{Development of global nonhydrostatic MCV atmospheric dynamic core on the Cubed sphere}

We have developed a prototype of a global nonhydrostatic atmospheric dynamic core based on the MCV numerical formulations and the cubed sphere grid.   In section 3, we have compared the MCV shallow water models on three quasi-uniform spherical grids.   The numerical results show that the present MCV method works well for all grids with adequate accuracy. Each spherical grid has its own strength and weakness, and the choice of the grid thus should be made according to the prospective applications. We have chosen the cubed-sphere grid as the base because of its rigorious numerical conservativeness and computational efficeincy as a structured grid. Given the numerical components discussed before, the numerical framework of the  global nonhydrostatic dynamic core can be built straighforwardly. 

\subsection{The non-hydrostatic governing equations in cubed-sphere coordinates}
The nonhydrostatic MCV-core model adopts the full compressible non-hydrostatic fluid equations in which the varaibles density $\rho$, momentum $\rho \mathbf{u}$ with the velocity vector $\mathbf{u}=(u^\xi,u^\eta,w)$ and the potential temperature density $\rho \theta$ are solved as the conservative variables. The differential form of the equations of motion under the shallow-atmosphere approximation can be expressed as follows: 
\begin{align} 
   &\frac{\partial \rho}{\partial t} +\nabla \cdot (\rho \mathbf{u}) =0 \label{eq:vector3d-1}   \\
   &\frac{\partial \rho \mathbf{u}}{\partial t} +\nabla \cdot (\rho u^i u^j+G^{ij}p)=-\rho g \mathbf{k} -f\mathbf{k} \times \rho \mathbf{u} \label{eq:vector3d-2}\\
   &\frac{\partial \rho \theta}{\partial t} +\nabla \cdot (\rho \theta \mathbf{u}) =0  \label{eq:vector3d-3}
\end{align}
where the indices $i$ and $j$ are the cyclic of $(1,2,3)=(\xi,\eta,r)$, $G^{ij}$ denotes  the contravariant metric, $\mathbf{k}$ is the basis vector in the radial direction, $g$ is gravity and $f$ is the Coriolis parameter. The pressure $p$ is related to the potential temperature and density by the EOS (equation of state)
\begin{align} 
   p=p_0\left(  \frac{R_d (\rho \theta)}{p_0} \right)^{c_p/c_v}
\end{align}
where $p_0=1000 $hPa is the reference pressure, $R_d$ is the ideal gas constant for dry air, $c_p$ and $c_v$ indicate the specific heat capacity of dry air at constant pressure and constant volume.

Now we denote the vector of prognostic quantities $\mathbf{q}=(\rho',\rho \mathbf{u},(\rho\theta)')$ as the vector of conservative variables. After splitting the thermodynamic variables $(\rho,p,\theta)$ into reference state $(\bar{\rho},\bar{p},\bar{\theta})$ and perturbation $(\rho',p',\theta')$ and assuming the local hydrostatic balance of the reference state, the non-hydrostatic Eqs. \eqref{eq:vector3d-1}-\eqref{eq:vector3d-3} read
\begin{align} 
 \frac{\partial \mathbf{q}}{\partial t}+\frac{\partial \mathbf{e}(\mathbf{q})}{\partial \xi} +\frac{\partial \mathbf{f}(\mathbf{q})}{\partial \eta}+\frac{\partial \mathbf{h}(\mathbf{q})}{\partial r} =\mathbf{S}_m+\mathbf{S}_c+\mathbf{S}_g+\mathbf{S}_r
\end{align}
where
\begin{align} 
   &\mathbf{q}=[\sqrt{G}\rho',\sqrt{G}\rho u^{\xi},\sqrt{G}\rho u^\eta,\sqrt{G}\rho w, \sqrt{G}(\rho\theta)']^T \\
   &\mathbf{e}=[\sqrt{G}\rho u^{\xi},\sqrt{G}\rho (u^{\xi})^2+\sqrt{G}G^{11}p', \sqrt{G} \rho u^\xi u^\eta  +\sqrt{G}G^{21}p', \sqrt{G} \rho u^\xi w +\sqrt{G}G^{31}p', \sqrt{G} \rho \theta u^\xi ]^T  \\   
    &\mathbf{f}=[\sqrt{G}\rho u^{\eta},\sqrt{G}\rho u^{\eta}u^\xi+\sqrt{G}G^{12}p',\sqrt{G}\rho (u^{\eta})^2+\sqrt{G}G^{22}p', \sqrt{G} \rho u^\eta w +\sqrt{G}G^{32}p',\sqrt{G} \rho \theta u^\eta]^T \\
    &\mathbf{h}=[\sqrt{G}\rho w,\sqrt{G}\rho w u^\xi+\sqrt{G}G^{13}p',\sqrt{G}\rho w u^{\eta}+\sqrt{G}G^{23}p', \sqrt{G} \rho w^2+\sqrt{G} G^{33}p',\sqrt{G}\rho \theta w]^T    
\end{align}
with $\mathbf{S}_m$, $\mathbf{S}_c$, $\mathbf{S}_g$ and $\mathbf{S}_r$ denoting the source terms  due to the cubed-sphere geometry, Coriolis forcing, gravity and the vertical non-reflection-boundary damping term, respectively. $\sqrt{G}$ is the Jacobian of the transformation. Note that the contravariant metric $G^{ij}$ is the same as the SWE cubed metrics in the horizontal direction.  

In the presence of topography, the height-based terrain-following coordinate introduced by Gal-Chen and Somerville \cite{gs75} is used to map the physical
space into the computational space. The detailed coordinate transformation can partially be referred to as \cite{li13}. The divergence operator can be easily deduced in the transformed coordinates. The pressure gradient operator should be given explicitly in the transformed curvilinear coordinates during the implementation of the nonhydrostatic MCV-core model.  

\subsection{Time integration procedure}

The explicit time-stepping schemes have the advantage of their simplicity and high parallel efficiency, namely, the minimal interprocessor communication \citep{dennis05,dennis12}.
In the SWE \citep{chen14}  model and  non-hydrostatic model \citep{li13} discussed   in the previous sections,  explicit time integration schemes are used to solve the semi-discretized ODE systems where the spatial discretizations have complished by using the MCV numerical formulations. 
Due to the large difference of the grid spacings in horizontal and vertical directions, a more efficient time integration scheme for global atmospheric models is the horizontally-explicit and vertically-implicit (HEVI) scheme which has been widely adopted in existing dynamic dores of AGEMs.  

The split-explicit and semi-implicit time stepping schemes are two possible alternatives that are widely used in many operational weather forecasting centers to avoid the above time step limitation. Implicit-explicit (IMEX) schemes, a variant of semi-implicit schemes, treat the fast time-scale terms implicitly and the slow time-scale terms explicitly. The HEVI scheme could be viewed as a framework of IMEX time marching schemes. In the HEVI scheme, the time stepping is only limited by the horizontal grid spacing which is ususally acceptable in the real application \citep{sk08}. The additional merit of HEVI scheme is more suitable for data communication in the supercomputer architecture since the horizontal and vertical communication is separate. Weller et al. \cite{weller2013}  present a detailed comparison of popular options of HEVI time stepping schemes. 

 We consider some general semi-discrete system with a finite volume formulation for space discretization  which is represented by the function $\mathcal{R}(\mathbf{q})$ of the vector of all conserved variables $\mathbf{q}$
\begin{align}\label{eq:eq1}     
   \frac{d \mathbf{q}}{dt}+\mathcal{R}(\mathbf{q})=0.
\end{align}
    Different time integrators generally result in the different forms of MOL (method of line) formulations \eqref{eq:eq1}. In our implementation, $\mathcal{R}(\mathbf{q})$ is discretized by one of the MCV family schemes, for example, the MCV3 scheme. 
    
    We have tested two solvers for Eq. \eqref{eq:eq1} for the advection equation. One choice is to locally linearlize the residual $\mathcal{R}(\mathbf{q})$ as
\begin{align} 
   \mathcal{R}^{n+1}=\mathcal{R}^{n}+\frac{\partial \mathcal{R}}{\partial \mathbf{q}} \Delta \mathbf{q}
\end{align}
where $\Delta \mathbf{q} =( \mathbf{q} ^{n+1}-\mathbf{q} ^n )$ and the Jacobian of $\partial \mathcal{R}/\partial \mathbf{q}$ is computed at the time level $n$. Then the linear algebra solver like the Krylov subspace method such as the GMRES approach can be used to advance the solutions.  An alternative to ODE is to utilize the nonlinear iterative solver, so-called Newton's method.
For example, we consider the two-level-time implicit form of Eq. \eqref{eq:eq1}
\begin{align} \label{eq:eq2} 
   \frac{ \mathbf{q}^{n+1}-\mathbf{q}^{n} }{\Delta t}= -\mathcal{R}_{E}(\mathbf{q}^n)-\beta \mathcal{R}_{I}(\mathbf{q}^{n+1})-(1-\beta)\mathcal{R}_{I}(\mathbf{q}^n)
\end{align}
where the subscript $E$ and $I$ denote the explicit and implicit terms respectively,  $\beta$ is an implicit coefficient with interval value of $[0,1]$ and $\mathcal{R}=\mathcal{R}_E+\mathcal{R}_I$. We define the function $\mathbf{F}(\mathbf{q})$ to obtain the nonlinear equation system for the unknown $\mathbf{q}=\mathbf{q}^{n+1}$ from Eq. \eqref{eq:eq2} as follows
\begin{align} 
  \mathbf{F}(\mathbf{q})= \frac{ \mathbf{q}-\mathbf{q}^{n} }{\Delta t}+\mathcal{R}_{E}(\mathbf{q}^n)+\beta \mathcal{R}_{I}(\mathbf{q})+(1-\beta)\mathcal{R}_{I}(\mathbf{q}^n) =0.
\end{align}
The associated Jacobian matrix reads
\begin{align} 
   \mathbf{F}'(\mathbf{q})=\left( \frac{\mathbf{I}}{\Delta t} + \beta \frac{\partial \mathcal{R}_I^n}{\partial \mathbf{q}} \right).
\end{align}
The one Newton step is represented by
\begin{align} 
   \left( \frac{\mathbf{I}}{\Delta t} + \beta \frac{\partial \mathcal{R}_I}{\partial \mathbf{q}} \right)|_{\mathbf{q}^{(k)}} \delta \mathbf{q}=-\mathbf{F}(\mathbf{q}^{(k)})=-\mathcal{R}(\mathbf{q}^{(k)})
\end{align}
where Newton correction $\delta \mathbf{q}=\mathbf{q}^{(k+1)}- \mathbf{q}^{(k)}$ and $k$ is the iterative index. We can solve this linear equation system with system matrix $\mathbf{A}=\left( \frac{1}{\Delta t} + \beta \frac{\partial \mathcal{R}_I}{\partial \mathbf{q}} \right)|_{\mathbf{q}^{(k)}} $ which can use matrix-free methods to obtain the matrix-vector products or be given by analytic matrix expression to compute matrix-vector products.

\begin{figure*} [htbp]
\begin{center}
\subfigure[Linear GMRES solver]
  {\includegraphics[scale=0.3,angle=-90,clip]{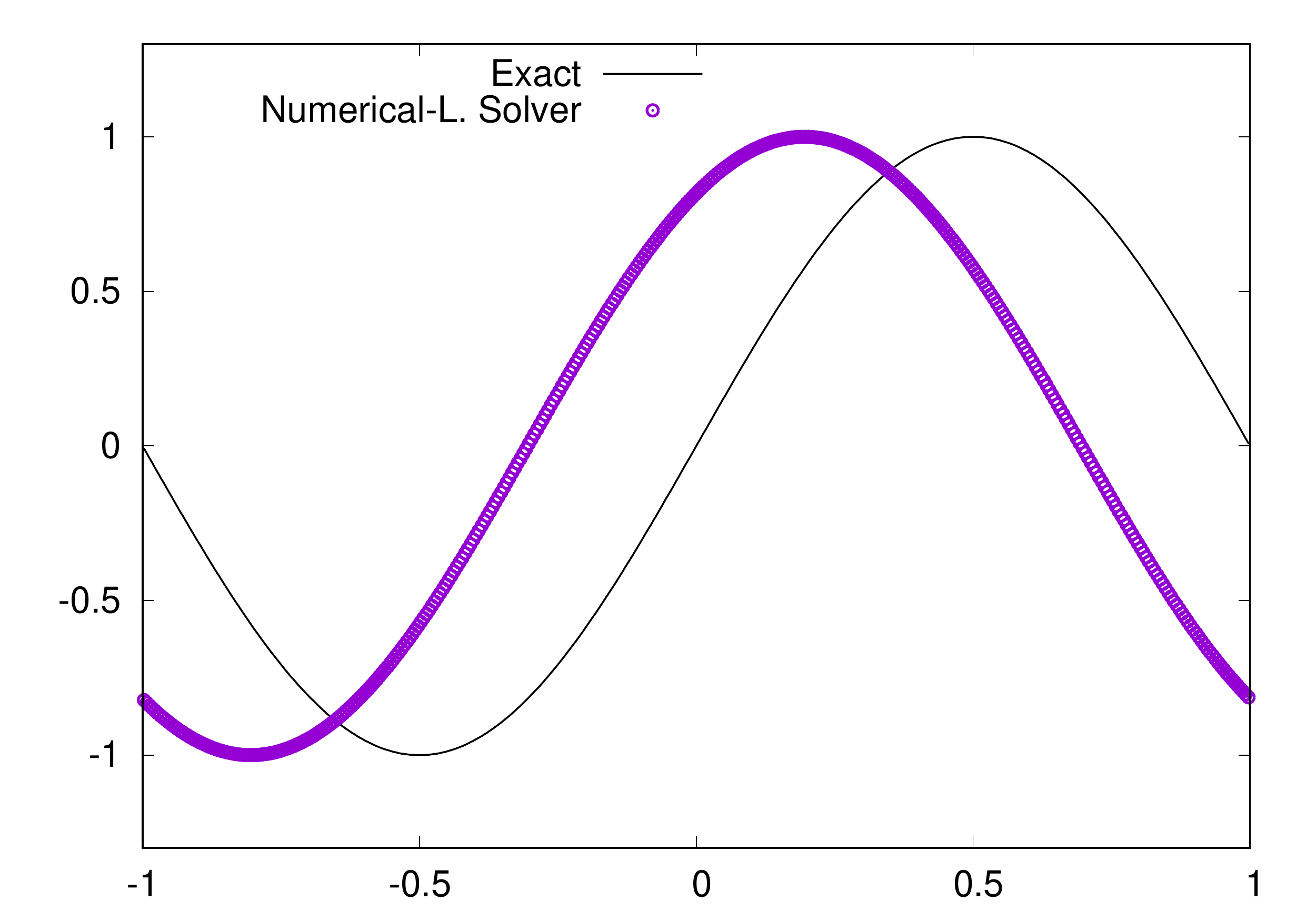}}
 \subfigure[Nonlinear INB solver]
  {\includegraphics[scale=0.3,angle=-90,clip]{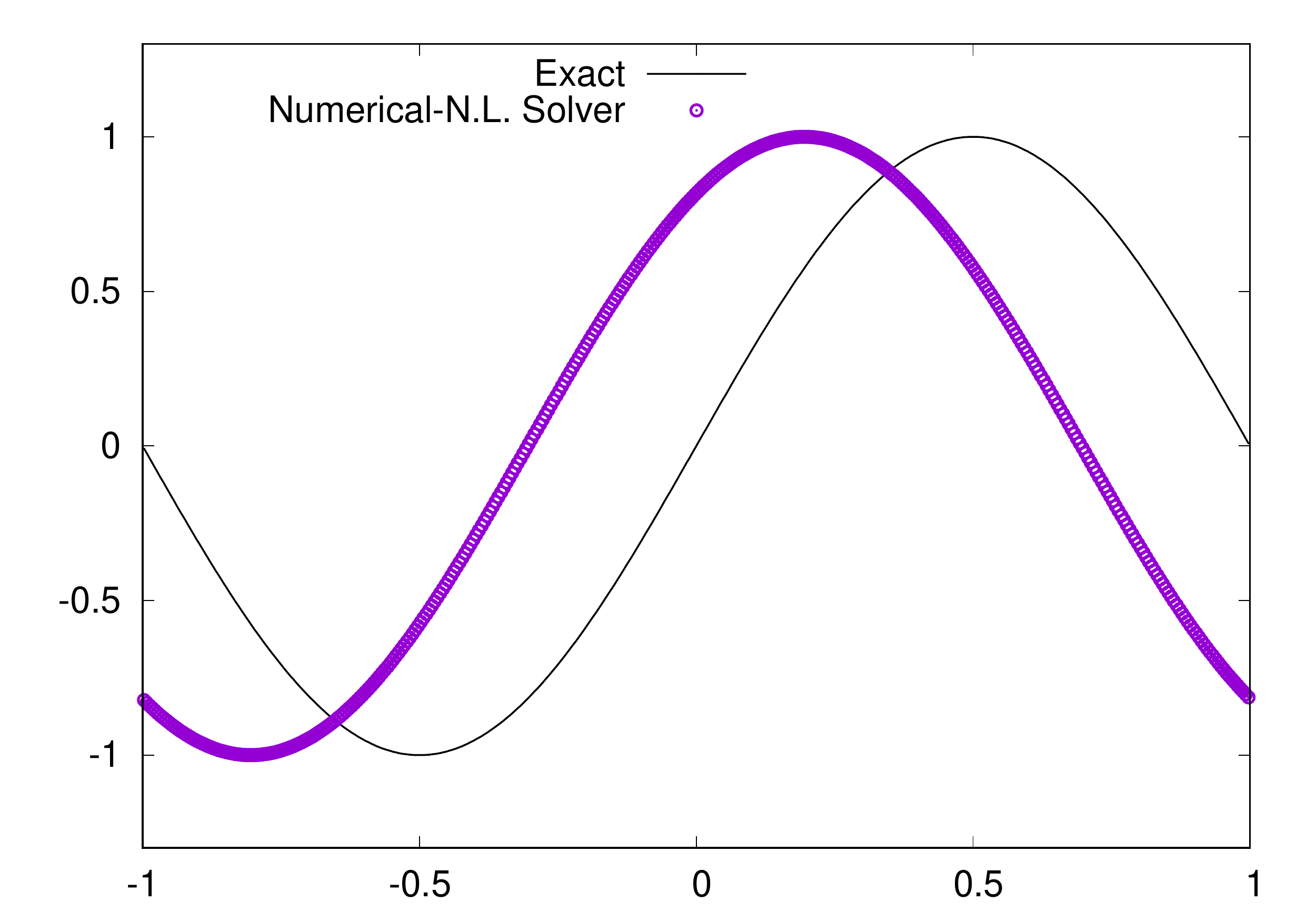}} 
\end{center}
\vskip -\lastskip \vskip -3pt \caption{ Crank-Nicolson time integration scheme for 1D advection test with $\Delta x=0.005$ and $\Delta t=0.5$ in the domain $[-1,1]$. The numerical results at $t=2$ of (a) the GMRES linear solver and (b) the inexact Newton backtracting nonlinear solver.  }\label{fig:lin-nonlin-solver}
\end{figure*}

\begin{figure*} [htbp]
\begin{center}
\includegraphics[scale=0.4,angle=-90,clip]{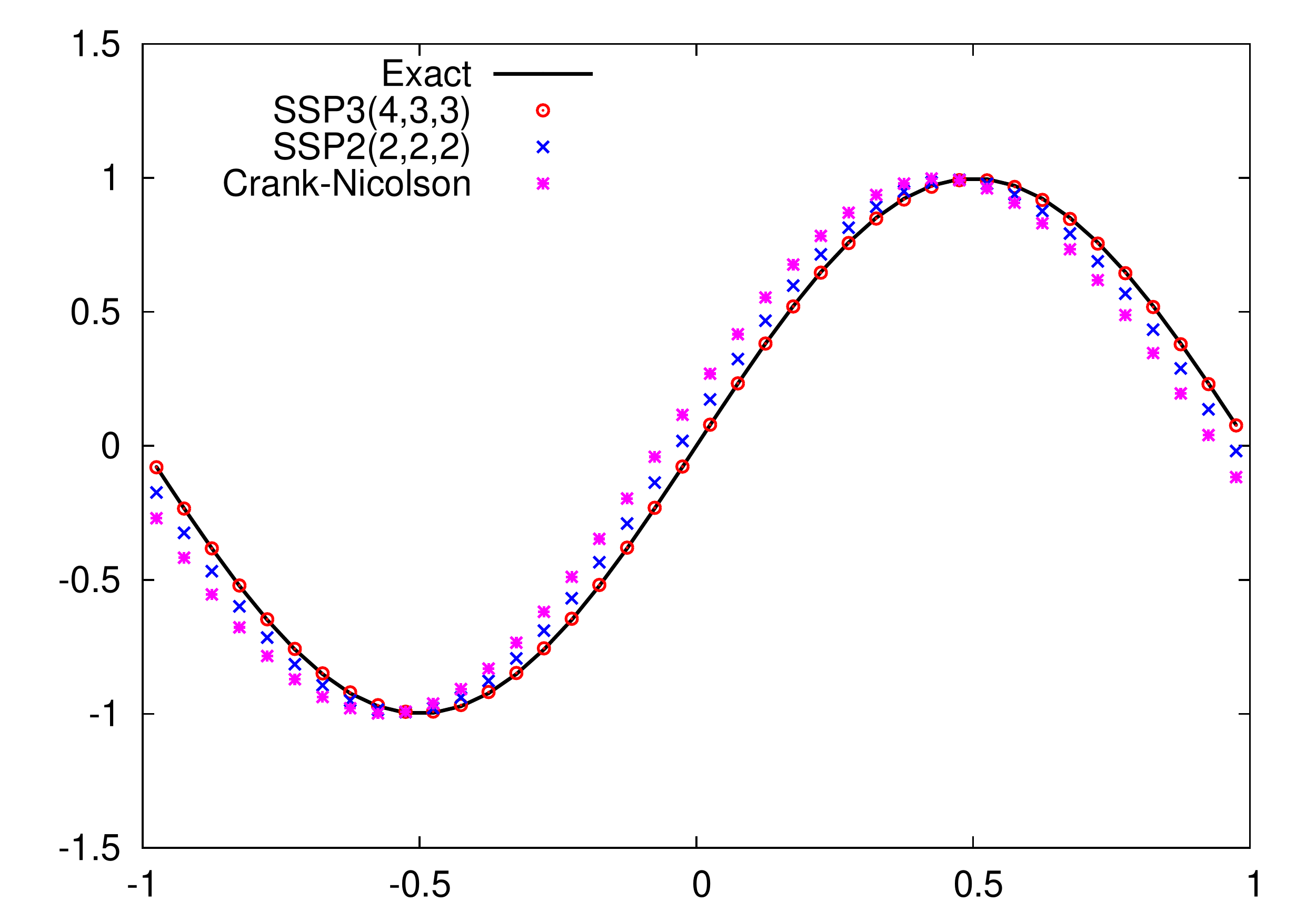}
\end{center}
\vskip -\lastskip \vskip -3pt \caption{ The different time integration schemes for 1D advection test with $\Delta x=0.05$ and $\Delta t=0.2$ in the domain $[-1,1]$ using the INB solver. The numerical results at $t=2$ of Crank-Nicolson method (pink cross-star), IMEX-SSP2 (blue cross) and IMEX-SSP3 (red circle-dot).  }\label{fig:nonlin-solver-comp}
\end{figure*}

We have solved  the advection test by using the linear GMRES solver and the inexact Newton backtracking nonlinear solver with the CFL number equal to 100 as shown in Fig. \ref{fig:lin-nonlin-solver}. It can be seen that for linear advection test the two solvers have produced the same simulations. Generally the nonlinear iterative solvers have better accuracy and robustness for unsteady Navier-Stokes equations although the linear solver is probably popular with the appropriate preconditioned projection for its computational efficiency. Based on the inexact Newton backtracking method, we also compared the different time marching schemes of Crank-Nicolson approach, IMEX-SSP2 and IMEX-SSP3 methods \citep{weller2013} and found improved results in turn for these schemes as indicated in Fig. \ref{fig:nonlin-solver-comp}.  IMEX-SSP3 method overperforms other two and gives the most accurate result. 

These time integration methods can be directly extended to the global nonhydrostatic dynamic core on the cubed sphere in an HEVI fashion.
 According to the stiffness of equation terms, the terms are marked with the underbraces  in the following formulation where the HE terms are advance explicitly via RK3 method and the VI terms are implicitly treated by backward Euler (BE) method,
\begin{align} 
 \frac{\partial \mathbf{q}}{\partial t}+ \underbrace{ \frac{\partial \mathbf{e}(\mathbf{q})}{\partial \xi} +\frac{\partial \mathbf{f}(\mathbf{q})}{\partial \eta}}_{\text{HE terms}}+ \underbrace{ \frac{\partial \mathbf{h}(\mathbf{q})}{\partial r} }_{\text{VI term}}=\underbrace{ \mathbf{S}_m+\mathbf{S}_c}_{\text{HE terms}} +\underbrace{ \mathbf{S}_g+\mathbf{S}_r }_{\text{VI terms}}.
\end{align}

For brevity, we deonte the HE terms explicitly computed as $\mathbf{L}_{he}(\mathbf{q})$ and VI terms implicitly treated as $\mathbf{L}_{vi}(\mathbf{q})$ as follows. Given a time interval of size $\Delta t$ and the solution $\mathbf{q}^n$ at the time level $t=n\Delta t$, the Strang splitting H-V-H steps are given as followed:
\begin{enumerate}[(1)]
\item H-step: advance the solutions explicitly by the TVD 3rd Runge-Kutta method
\begin{align} 
   \mathbf{q}^1=\mathbf{q}^n+RK3\left[  \mathbf{L}_{he}(\mathbf{q}), \frac{1}{2}\Delta t\right],
\end{align}

\item V-step: advance the solutions implicitly by the BE method
\begin{align} 
   \mathbf{q}^2=\mathbf{q}^1+BE\left[  \mathbf{L}_{vi}(\mathbf{q}^2), \Delta t\right],
\end{align}

\item H-step: advance the solutions explicitly one more by the TVD 3rd Runge-Kutta method
\begin{align} 
   \mathbf{q}^{n+1}=\mathbf{q}^2+RK3\left[  \mathbf{L}_{he}(\mathbf{q}^2), \frac{1}{2}\Delta t\right].
\end{align}
Now the solutions $\mathbf{q}^{n+1}$ at the time level $t=(n+1)\Delta t$ are reached.
\end{enumerate}

\subsection{Preliminary results}
The numerical results of the 3D Rossby-Haurwitz wave test case, which is an extension of the 2D shallow-water Rossby-Haurwitz wave \citep{wi92}, are presented. 
The Rossby-Haurwitz wave patterns in the numerical model should be preserved and moves westward. Theoretical analysis \citep{Rossby39,Haurwitz40,zhang1986} has already inidcated that the Rossby-Haurwitz wave is subject to dynamical instability. In the numerical model the numerical solutions are quite sensitive to small perturbations of initial conditions which are then  projected onto the dynamical instability \citep{Hoskins73,thuburn2000}. Generally the first 15 days of the large-scale flow evolution are considered as a dynamically stable period and can be used to verify the ability of a numerical model to keep a 4-wavenumber pattern over the time.

\begin{figure*} [htbp]
\begin{center}
\subfigure[Zonal wind at 850 hPa]
  {\includegraphics[width=0.4\textwidth]{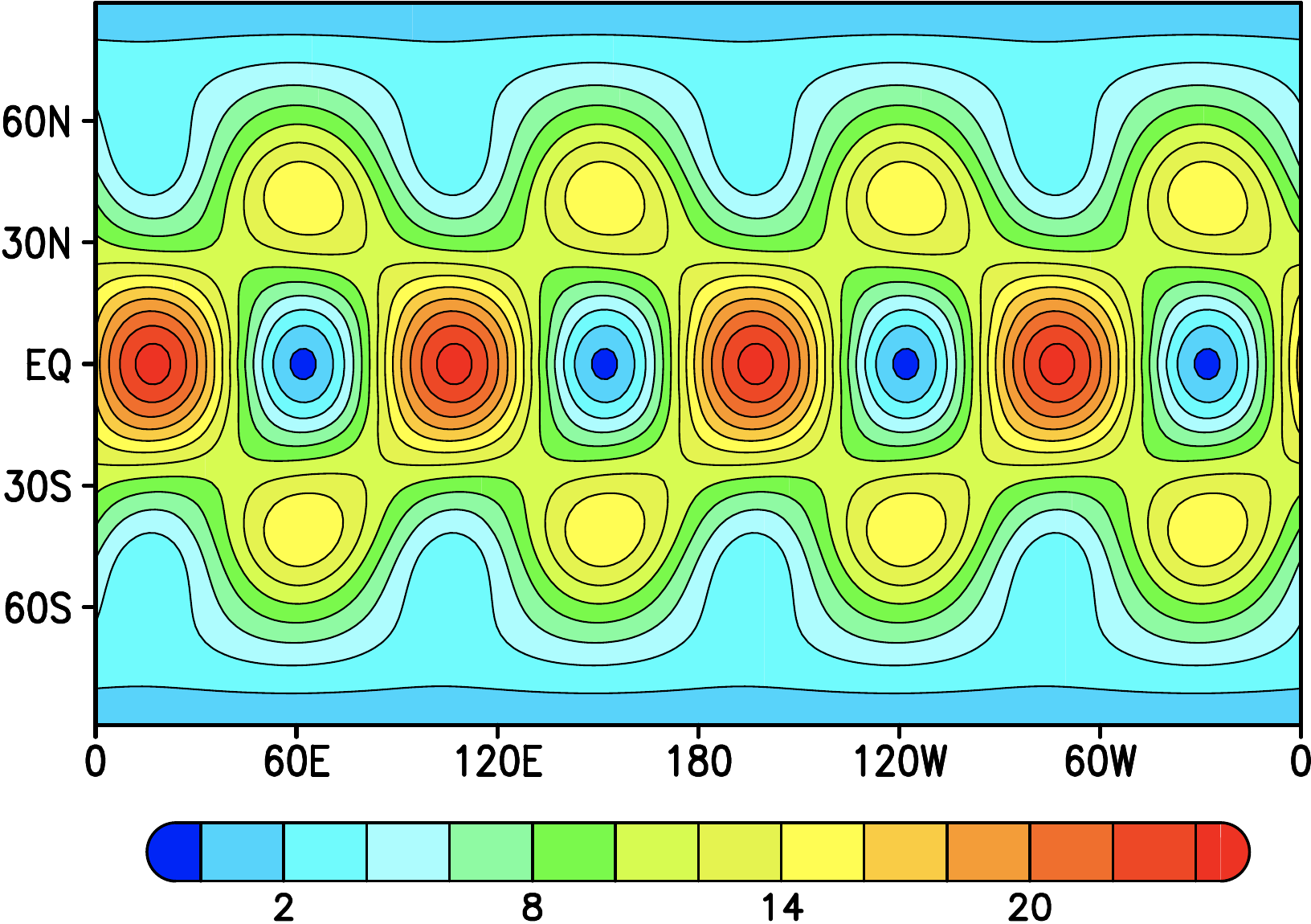}}
 \subfigure[Meridional wind at 850 hPa]
  {\includegraphics[width=0.4\textwidth]{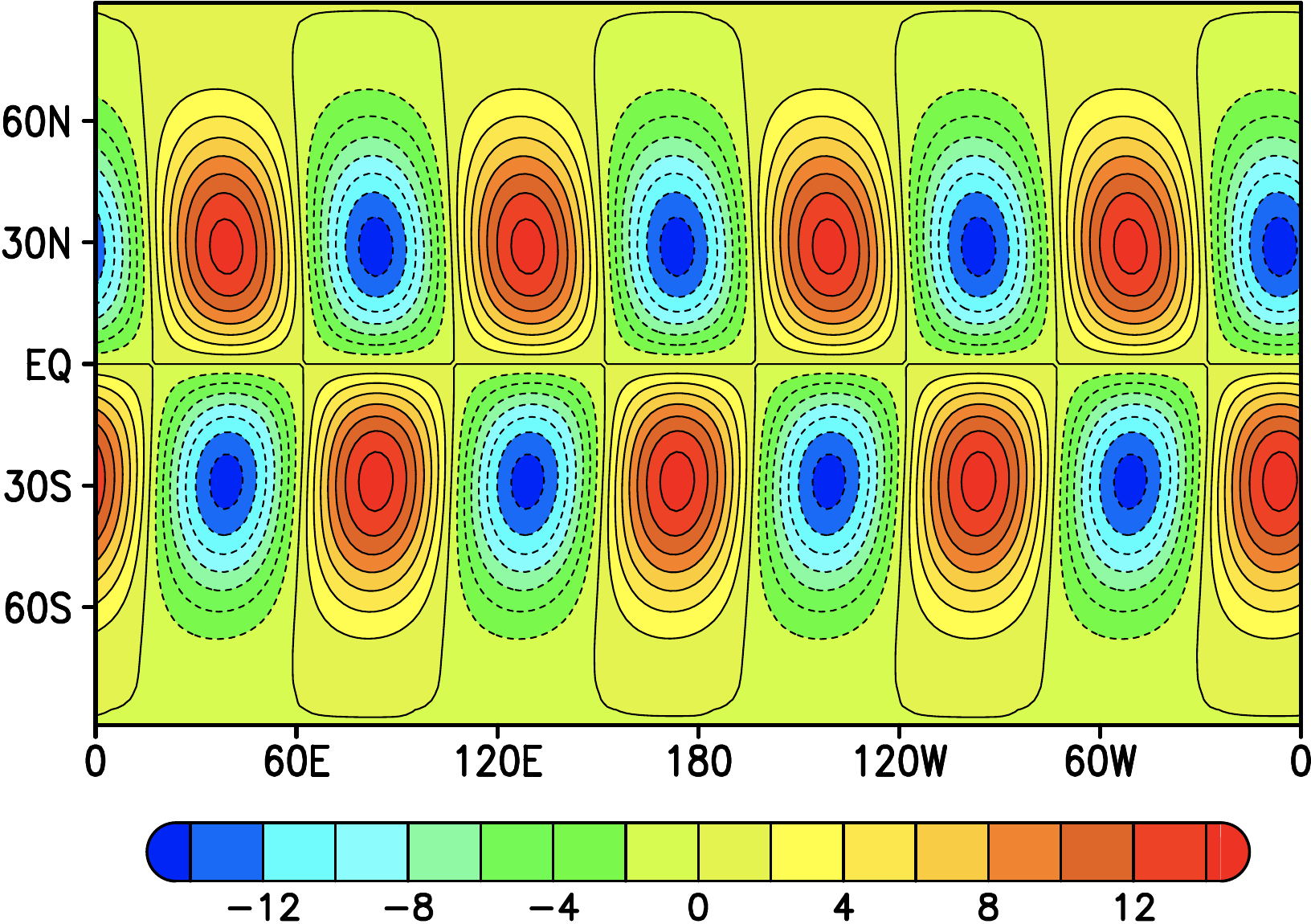}} 
 \subfigure[Surface pressure]
  {\includegraphics[width=0.4\textwidth]{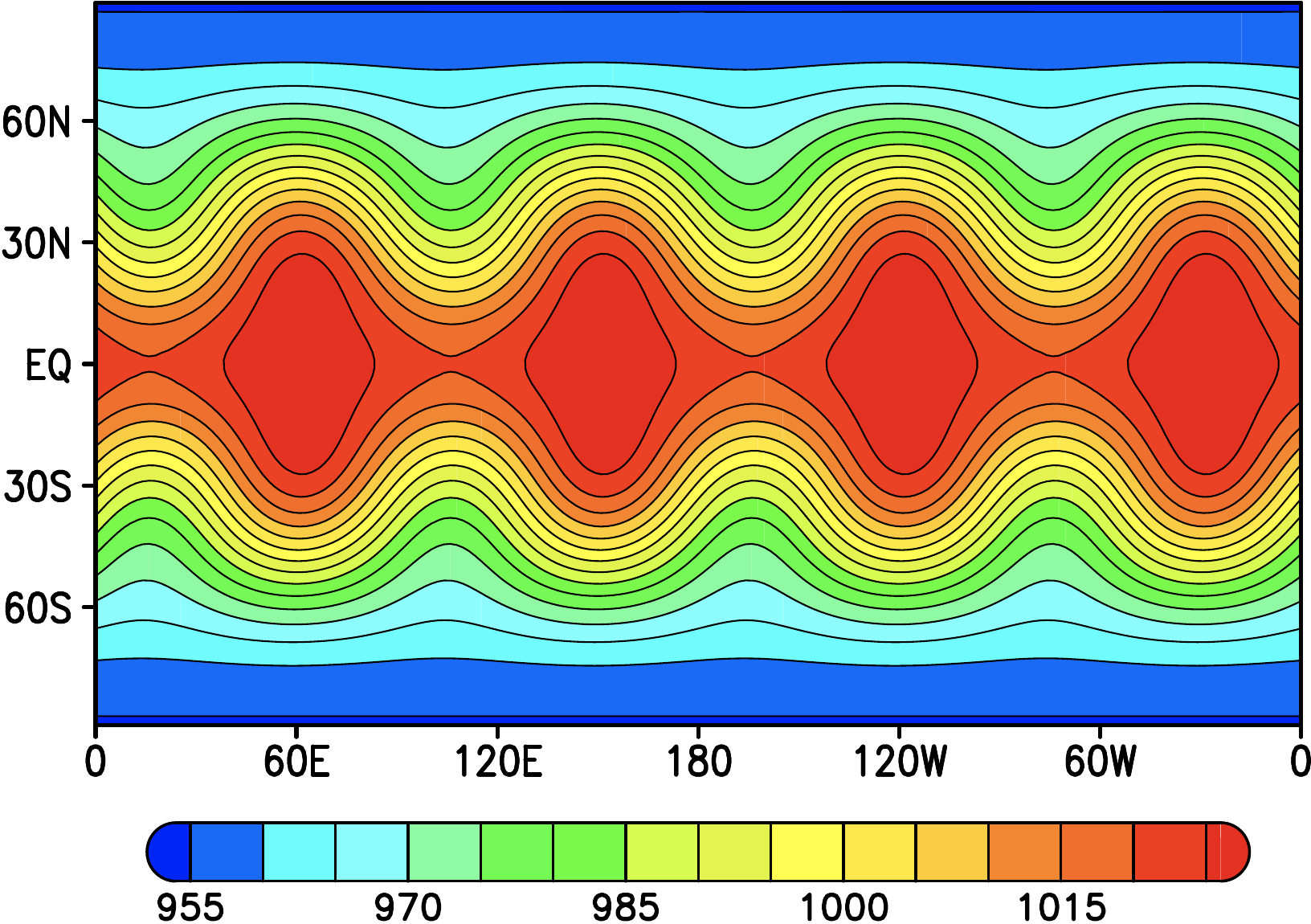}} 
 \subfigure[Temperature at 850 hPa]
  {\includegraphics[width=0.4\textwidth]{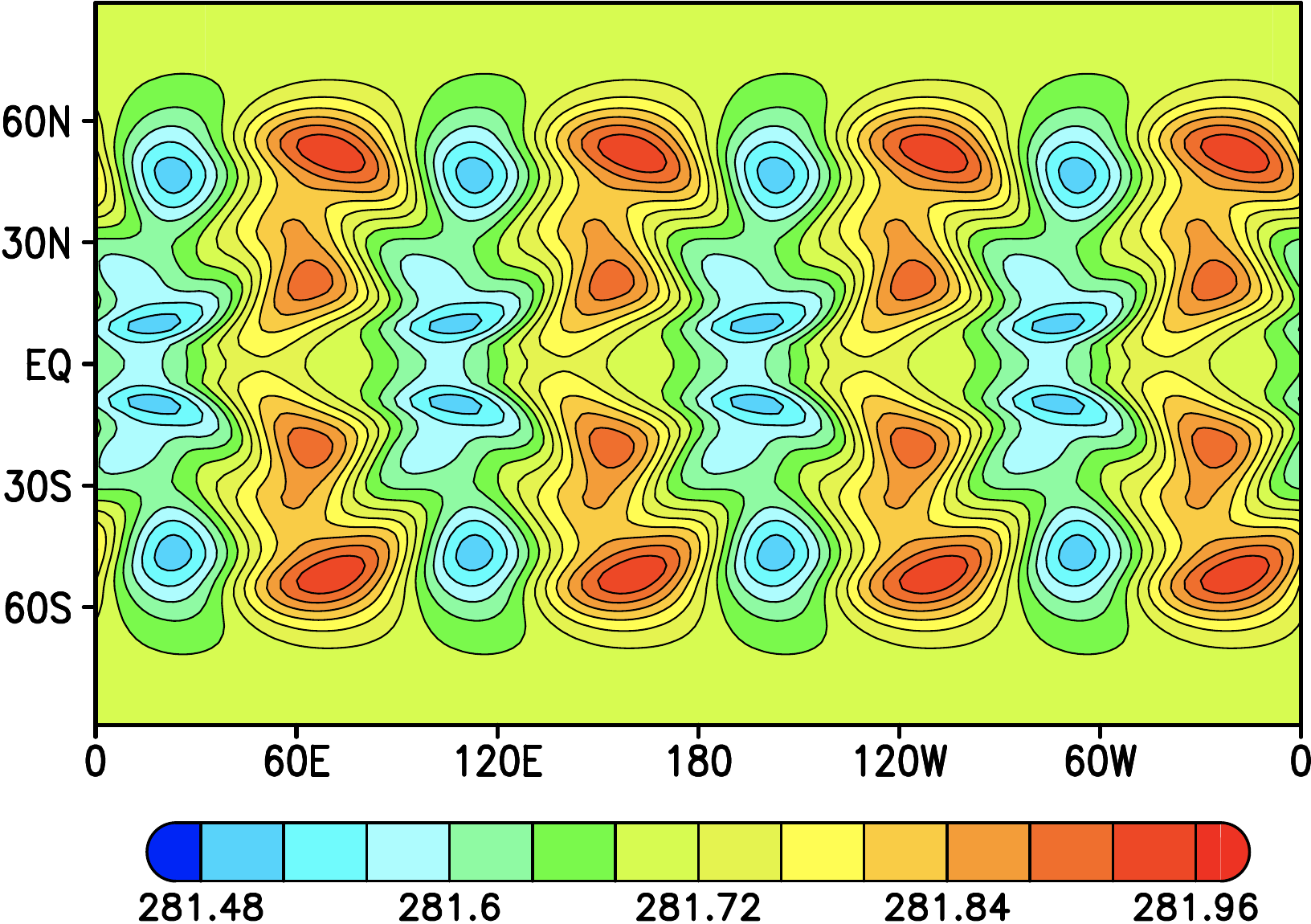}} 
 \subfigure[Geopotential height at 500 hPa]
  {\includegraphics[width=0.4\textwidth]{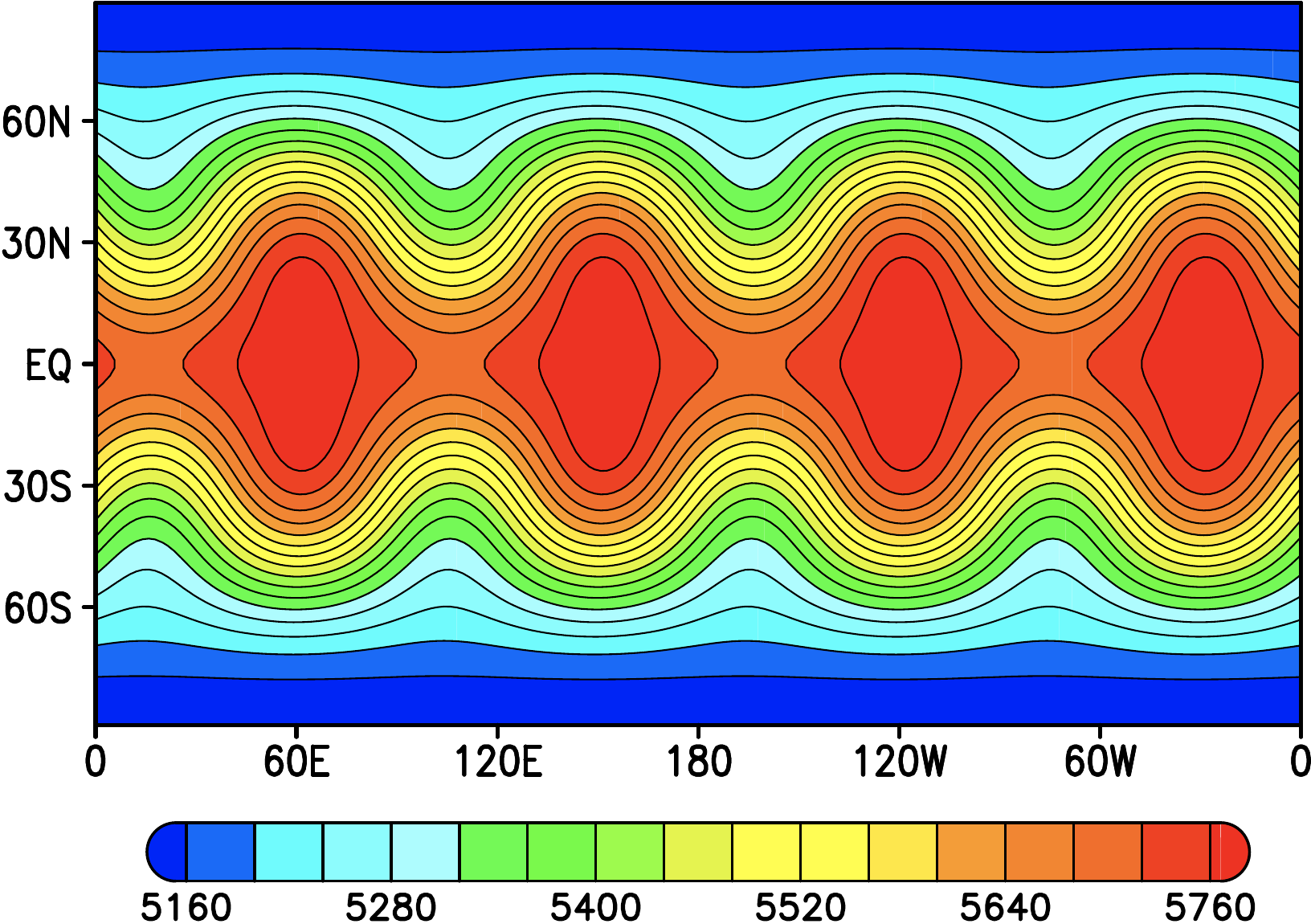}} 
 \subfigure[Vertical velocity at 850 hPa]
  {\includegraphics[width=0.4\textwidth]{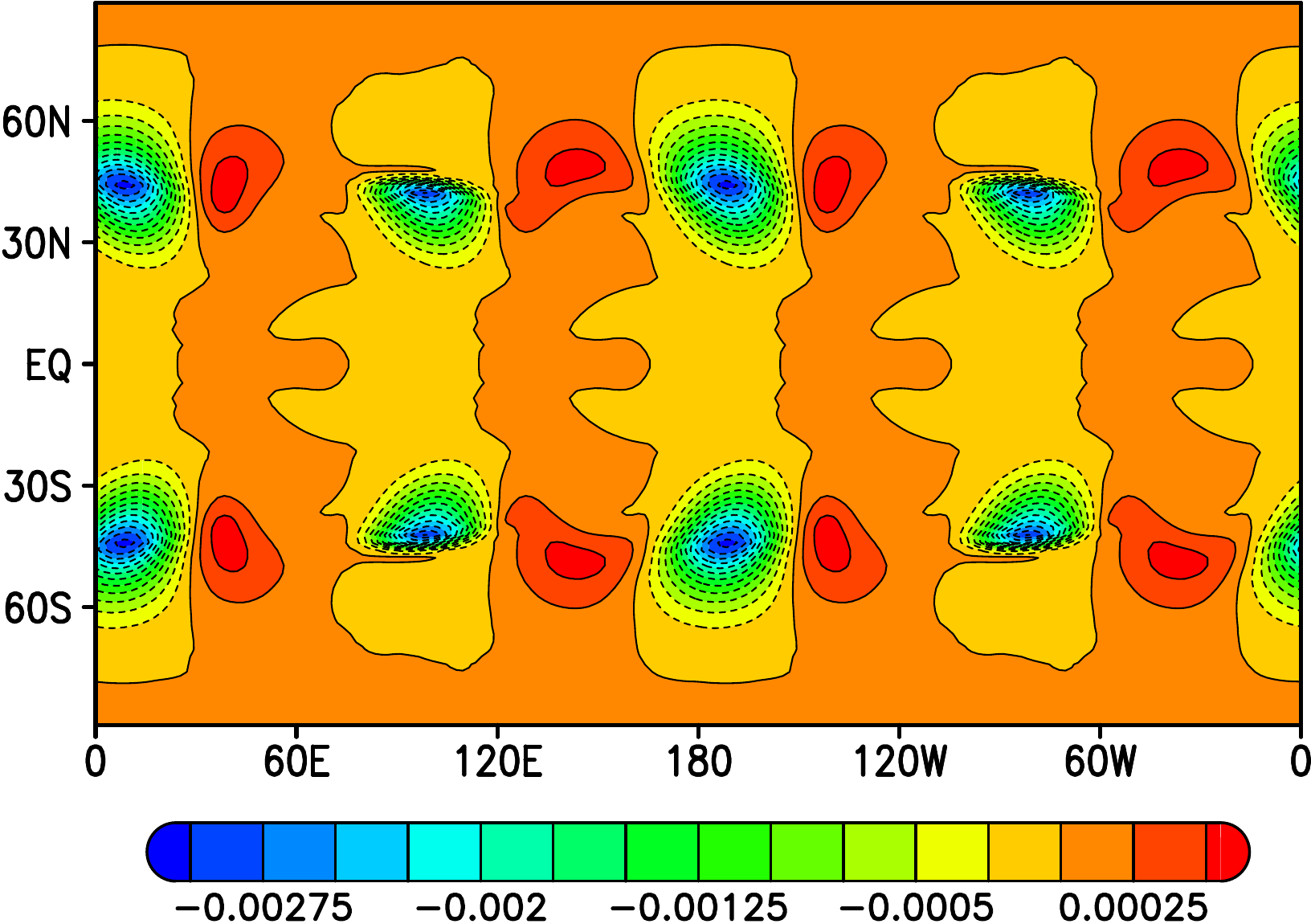}} 
\end{center}
\vskip -\lastskip \vskip -3pt \caption{ Results of the Rossby-Haurwitz wave at day 15 simulated on the horizontal resolution $24\times 24 \times 6$ (equatorial equivalent grid spacing $3.75^{\circ}$) with 13 vertical levels.  Zonal and meridional wind (both at 850 hPa) are plotted in the top row, surface pressure and temperature at 850 hPa are shown in the middle row and 500 hPa geopotential height and 850 hPa vertical velocity are plotted in the bottom row. }\label{fig:rossbywave}
\end{figure*}

Fig. \ref{fig:rossbywave} gives a snapshot of the atmospheric state of Rossby-Haurwitz wave  at day 15 indicating the 850 hPa zonal and meridional wind, surface pressure, 850 hPa temperature, 500 hPa geopotential height and 850 hPa vertical velocity.
The preliminary simulations in the MCV-core nonhydrostatic model show that the wave speed of the Rossby-Haurwitz wave is correctly captured by the MCV3 method with appealing solution quality in comparison with other existing dynamical cores \citep{jltn2008,uj12jcp}.

\section{The parallelization and scalability}

To facilitate the efficient use of the updated supercomputers, it is necessary to develop the spatial and temporal discretization in which the numerical scheme should have computationally desirable local properties such as compact computational stencils, intensive on-processor operations, and minimal communication footprints. Furthermore, in order to fit within the energy limitation while providing tremendous computing power, heterogeneous architectures that use both the multi-core CPU and the many-core accelerator have become a competitive choice to build extreme-scale supercomputing systems.

In global atmospheric simulations, there is an important trend to employ high-order spatial discretizations such as the discontinuous Galerkin (DG) methods, the spectral element (SE) methods, and more generally, the flux reconstruction (FR) methods.  As one of high-order schemes, the MCV scheme is an extension of the traditional FV method by increasing the local degree of freedom (DOF) to achieve local high-order reconstructions in compact stencils, and thus is computationally intensive. 

\begin{figure*} [htbp]
\begin{center}
   \includegraphics[width=0.6\textwidth]{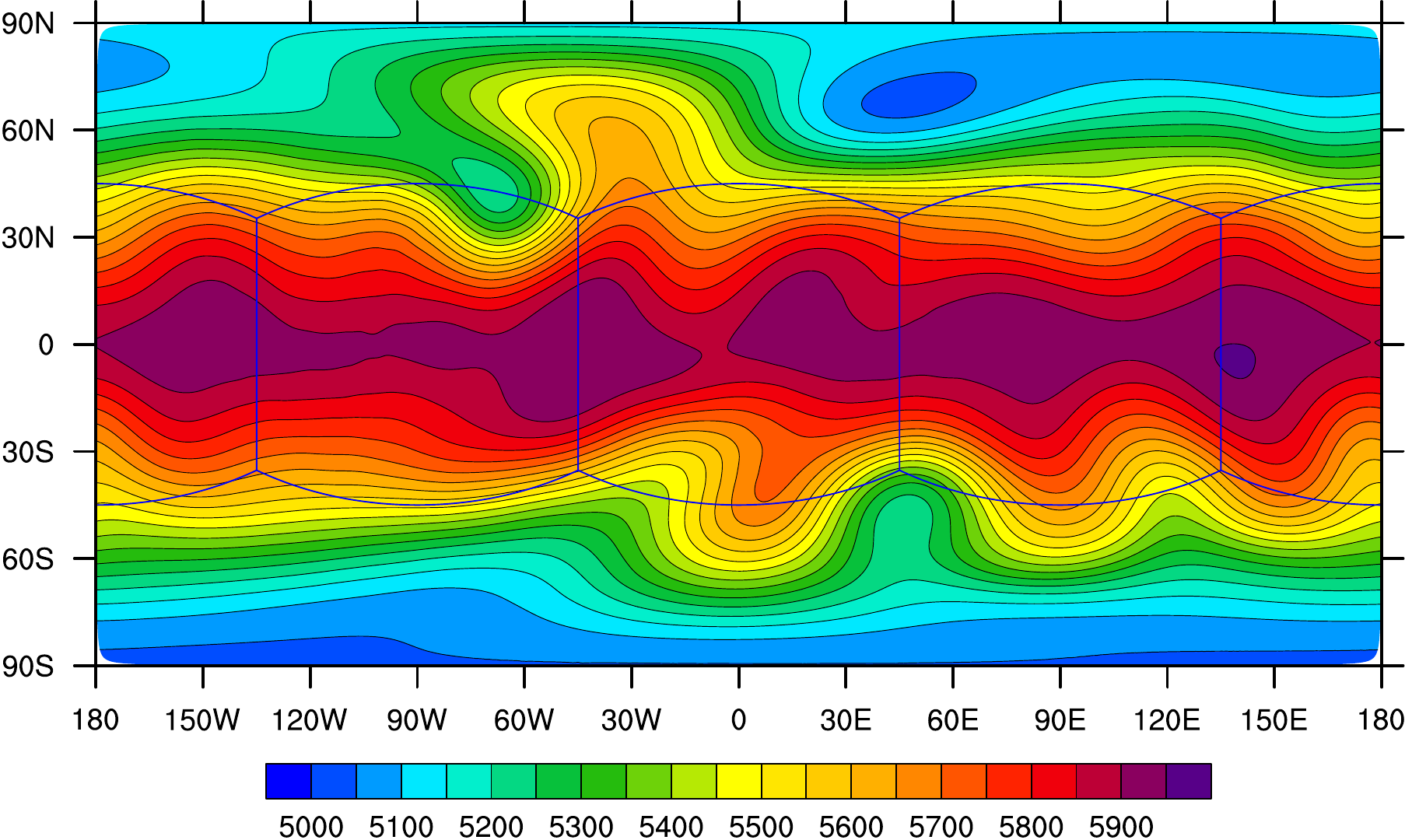}
\end{center}
\vskip -\lastskip \vskip -3pt \caption{ The total height field of mountain wave test (Williamson's test case 5).  }\label{fig:validate_height}
\end{figure*}

Based on a sequential code, Zhang et al. \cite{zhang2017} have developed a hybrid parallel MCV-based global shallow-water model on the cubed-sphere grid through the process-level parallelism using the explicit 3rd TVD Runge-Kutta scheme. The parallel computation was conducted on the described MIC-based heterogenous system with a $512 \times 512 \times 6$ cubed-sphere mesh.  The total height distribution at day 15 is illustrated as a contour plot in Fig. \ref{fig:validate_height} which shows an identical result of the sequential code (Fig. \ref{case5_height}b in \cite{chen14}).

\subsection{Intra-node acceleration results}
\begin{figure*} [htbp]
\begin{center}
\subfigure[]
   {\includegraphics[width=0.4\textwidth]{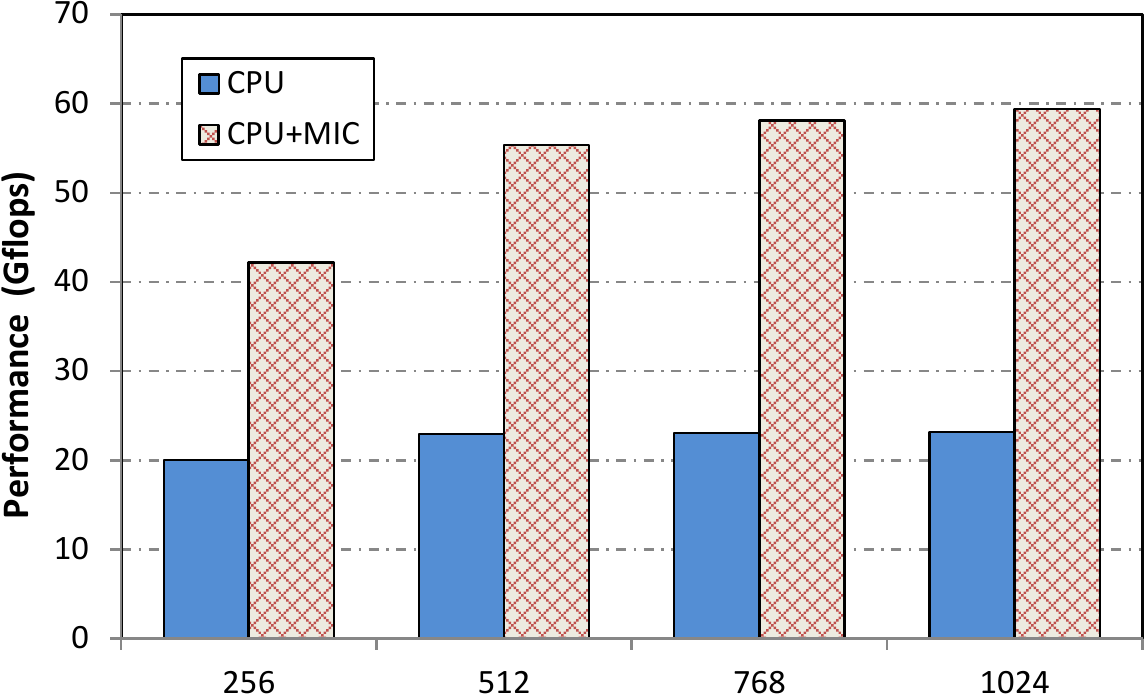}}
 \subfigure[]
  {\includegraphics[width=0.4\textwidth]{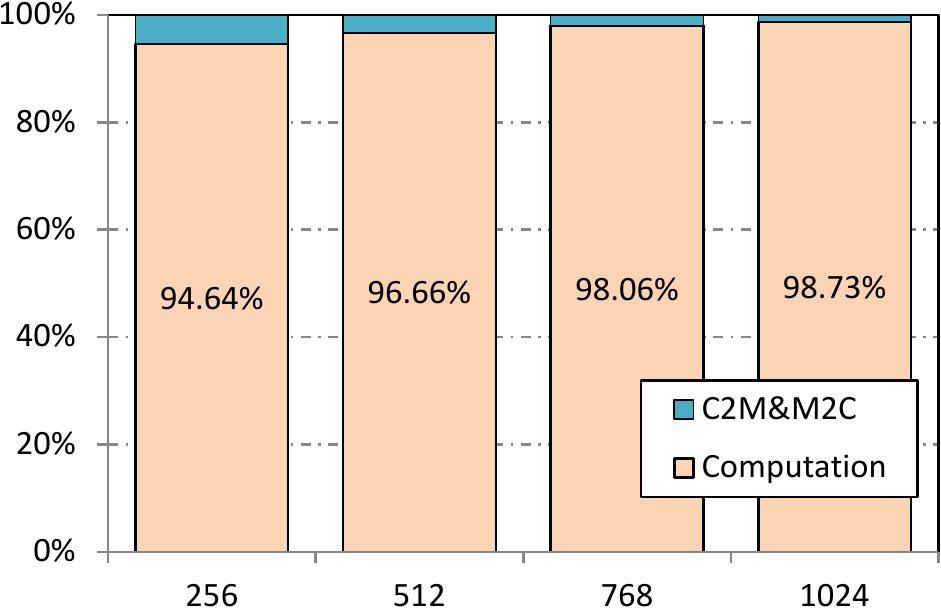}}
\end{center}
\vskip -\lastskip \vskip -3pt \caption{ Performance of the hybrid CPU+MIC code. Four difference sizes of subdomains, i.e., $256 \times 256$, $512 \times 512$, $768 \times 768 $ and $1024 \times 1024$, are tested. (a) Comparison with pure CPU code. (b) Ratio of data transfer cost.  }\label{fig:acceleration}
\end{figure*}

Numerical experiments were carried out to examine the intra-node acceleration. Based on the optimal partition, we measured the performance of the CPU+MIC code with various subdomain sizes and report as shown in Fig. \ref{fig:acceleration}. In the test we not only compared the performance with that of optimized multi-threaded CPU version but also examined specifically the cost of data transfer between the CPU and the MIC. From the two figures, some observations can be made:
\begin{enumerate}[{-}]
\item First, small subdomain size such as $256 \times 256$ is not large enough for the full utilization of the computing capability on both CPU and the MIC. In particular, as the subdomain size is increased, the performance of the CPU-only code reaches the top more quickly than the CPU+MIC version, indicating the higher computation throughput on the MIC accelerator.

\item Second, in terms of the aggregative performance, the hybrid implementation using one CPU and one MIC reaches 59.39 Gflops, outperforming the CPU-only code with a speedup of 2.56$\times$. This clearly shows that the proposed hybrid algorithm can help effectively maximize the performance on both the CPU and MIC sides.

\item Third, in terms of total run time, the ratio of the data transfer between CPU and MIC is kept to a low level, which is only 5.36\% for the $256 \times 256$ subdomain case and is decreased further to only 1.27\% when the subdomain size is $1024 \times 1024$. This indicates the proposed hybrid CPU+MIC algorithm is able to fully exploit the computing capacity of both the CPU and the MIC with very low extra data transfer cost.
\end{enumerate}

\subsection{Inter-node scaling results}
\begin{figure*} [htbp]
\begin{center}
\subfigure[]
   {\includegraphics[width=0.4\textwidth]{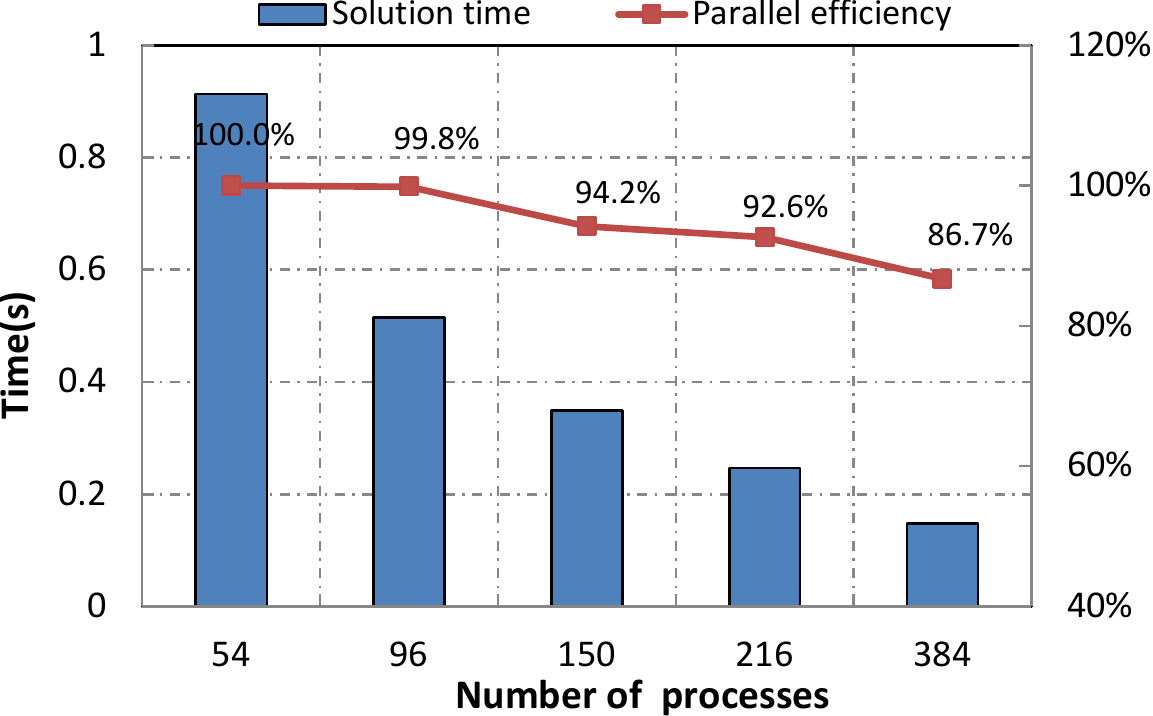}}
 \subfigure[]
  {\includegraphics[width=0.4\textwidth]{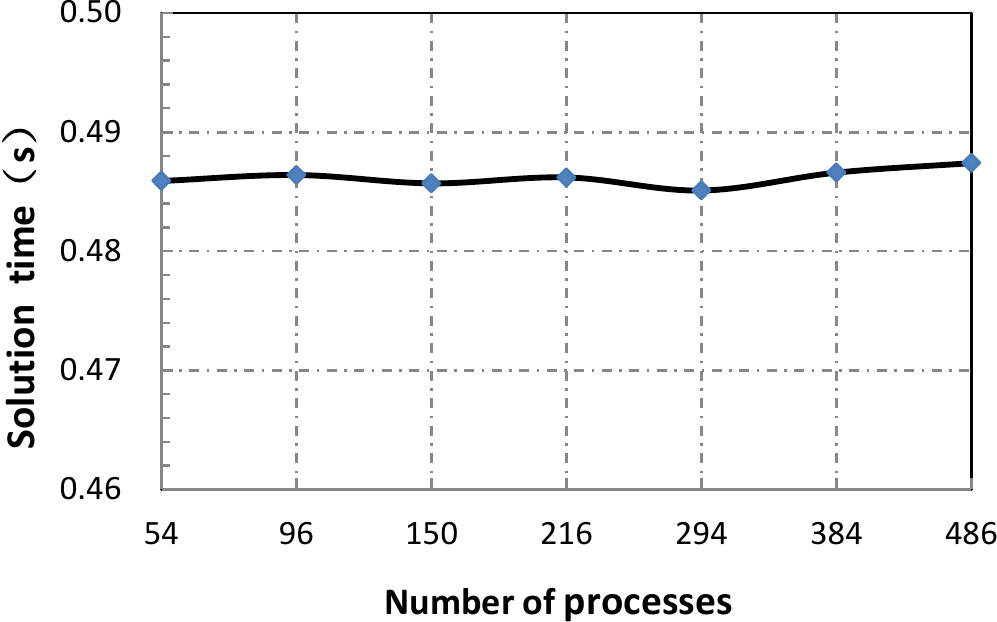}}
\end{center}
\vskip -\lastskip \vskip -3pt \caption{ Results of the inter-node strong and weak scaling tests. (a) Strong scaling. (b) Weak scaling.  }\label{fig:scalemcv3}
\end{figure*}

The hybrid implementation in both the strong and the weak scaling to test the scalability across nodes has been evaluated. In the strong scaling tests, we fix the total problem size to be $4320 \times 4320 \times 6 $ and gradually increase the number of processes from 54 to 384, corresponding to 3726 and 26, 496 cores, respectively. The testing results are provided in Fig. \ref{fig:scalemcv3} (a) , from which we observe a parallel efficiency of  86.7 \% with 384 processes. This slight drop of parallel efficiency in the strong scaling tests is reasonable considering that the workload of each process decreases as more processes are used. For the weak scaling tests, the size of the subdomain for each process is fixed to be $1024 \times 1024$ and we measure the averaged simulation time per time step by increasing the number of processes from 54 to 486. The weak scaling results are provided in Fig. \ref{fig:scalemcv3} (b), which clearly shows that the simulation time is kept to be stable in a very narrow range and the parallel efficiency is 99.7 \% when using 486 processes. 

\section{Summary and future works}

The multi-moment constrained finite volume method is a novel high-order numerical formulation for computational fluid dynamics. It can be viewed as an extension of the traditional finite volume method. All the computational variables (unknowns) are point values defined within each mesh element, and high-order reconstructions with limiting projection can be built over compact mesh stencils on both structured and unstructured grids. The evolution equations to update the unknowns are derived from the constraints on different moments which can be chosen in a more physical and flexible way. Among the constrained conditions is  the finite volume formulation that works as the constraint on cell-integrated average and guarantees the numerical conservativeness. The  multi-moment finite volume method is a numerical framework lies between the conventional finite volume methods and other high-order methods using local spectrum-like reconstructions, such as DG (discontinuous Galerkin) and SE (spectral element) methods.  It well balances among different requirements, such as solution quality, robustness, computational efficiency, parallelization, flexibility and adaptibility to different model configurations, and thus provides a very promising platform of great practical significance and potential to construct next generation atmospheric and oceanic GCMs.  
 
We presented in this report the fundamental researches toward the devlopment of next generation atmospheric GCM. The models  for both global  shallow water equations on three quasi-uniform spherical grids and nonhydrostatic compressible atmospheric flow have been developed. These models have been systematically and extensively verified by widely used bechmark tests. All  numerical results justify that the MCV-based models have competitive numerical accuracy and efficiency in comparison to other existing numerical models. The  3D prototype of  global  nonhydrostatic compressible atmospheric model has been also developed by using the IMEX/HEVI time integration scheme as the base to develop a scalable high-order dynamic core for weather and climate predictions and simulations. 

Next, we will focus on developing continuously the fully compressible nonhydrostatic atmospheric MCV dynamical cores on cubed sphere grid in presence of topography, and examine and improve the dynamic core with more benchmark tests in  3D.  Another big task is to incorporate the physical packages and establish a new system for real-case applications.

\color{black}

\bibliographystyle{elsarticle-num}
\bibliography{<your-bib-database>}



\end{document}